\documentclass[journal]{IEEEtran}

\usepackage{cite}
\usepackage{amsfonts}
\usepackage{graphicx}
\usepackage{textcomp}

\usepackage{xcolor}
\usepackage{tikz}
\usetikzlibrary{positioning}
\usetikzlibrary{arrows}
\usetikzlibrary{circuits.ee.IEC}
\usetikzlibrary{arrows.meta}
\usetikzlibrary{hobby}
\usetikzlibrary{patterns}
\usetikzlibrary{shadows.blur}
\usepackage{booktabs}
\usepackage[font=footnotesize]{caption}
\usepackage[font=footnotesize]{subcaption}
\usepackage[shortlabels]{enumitem}
\usepackage{comment}

\definecolor{backgroundblue}{rgb}{0.6, 0.8, 1}
\definecolor{backgroundorange}{rgb}{1, 0.752, 0}
\definecolor{backgroundcolor}{rgb}{0.4660, 0.6740, 0.1880}
\definecolor{backgroundcolor2}{rgb}{1, 0.4,0.5}

\definecolor{matlab1}{rgb}{0.0000, 0.4470, 0.7410}   
\definecolor{matlab2}{rgb}{0.8500, 0.3250, 0.0980}   
\definecolor{matlab3}{rgb}{0.9290, 0.6940, 0.1250}   
\definecolor{matlab4}{rgb}{0.4940, 0.1840, 0.5560}   
\definecolor{matlab5}{rgb}{0.4660, 0.6740, 0.1880}   
\definecolor{matlab6}{rgb}{0.3010, 0.7450, 0.9330}   
\definecolor{matlab7}{rgb}{0.6350, 0.0780, 0.1840}   
\definecolor{matlab8}{rgb}{0.7500, 0.7500, 0.0000}   

\let\proof\relax
\let\endproof\relax
\usepackage{amsthm}
\usepackage[cmex10]{amsmath}

\newtheorem{remark}{Remark}

\newtheorem{gridcode}{NGGC}

\usepackage{amssymb} 
\usepackage{amsfonts}
\usepackage{mathtools}
\allowdisplaybreaks

\allowdisplaybreaks
\usepackage[bottom]{footmisc}

\usepackage[capitalize,nameinlink,compress]{cleveref}
\crefname{figure}{Fig.}{Figures} 
\crefname{line}{line}{lines} 
\crefname{claim}{Claim}{Claims} 
\crefname{equation}{}{} 
\crefname{problem}{Problem}{Problems}
\crefname{assumption}{Assumption}{Assumptions}

\usepackage{soul}
\usepackage{framed}
\colorlet{shadecolor}{yellow}
\soulregister\cite7
\soulregister\cref7
\soulregister\ref7
\soulregister\pageref7
\soulregister\eqref7

\renewcommand{\baselinestretch}{0.96}

\begin{document}

\title{\fontdimen2\font=0.5ex Next-Generation Grid Codes: Towards a New Paradigm for Dynamic Ancillary Services}

\author{Verena~Häberle,~\IEEEmembership{Member,~IEEE,}
        Kehao Zhuang,~\IEEEmembership{Member,~IEEE,}
        Xiuqiang~He,~\IEEEmembership{Member,~IEEE,}
        Linbin~Huang,~\IEEEmembership{Member,~IEEE,}
        Gabriela Hug,~\IEEEmembership{Senior Member,~IEEE,}
        and~Florian~Dörfler,~\IEEEmembership{Senior Member,~IEEE}

\thanks{This work was supported by the prize funds of the Hans-Eggenberger Award. Verena Häberle, Gabriela Hug and Florian Dörfler are with the Department of Electrical Engineering, ETH Zurich, 8092 Zurich, Switzerland. Email: \{verenhae,hug,dorfler\}@ethz.ch. Xiuqiang He is with the Department of Automation, Tsinghua University, Beijing 100084, China. Email: hxe@tsinghua.edu.cn. Linbin Huang and Kehao Zhuang are with the College of Electrical Engineering, Zhejiang University, Hangzhou 310027, China. Email: \{hlinbin,zhuangkh\}@zju.edu.cn.}
}

\maketitle
\begin{abstract} \fontdimen2\font=0.6ex This paper introduces a conceptual foundation for \emph{Next Generation Grid Codes (NGGCs)} based on stability and performance certificates, enabling the provision of dynamic ancillary services such as fast frequency and voltage regulation through decentralized frequency-domain criteria. The NGGC framework offers two key benefits: (i) rigorous closed-loop stability guarantees, and (ii) explicit performance guarantees for frequency and voltage dynamics in power systems. Regarding (i) \emph{stability}, we employ loop-shifting and passivity-based techniques to derive local frequency-domain stability certificates for individual device dynamics. These certificates ensure the closed-loop stability of the entire interconnected power system through fully decentralized verification. Concerning (ii) \emph{performance}, we establish quantitative bounds on critical time-domain indicators of system dynamics, including the average-mode frequency and voltage nadirs, the rate-of-change-of-frequency (RoCoF), steady-state deviations, and oscillation damping capabilities. The bounds are obtained by expressing the performance metrics as frequency-domain conditions on local device behavior. The NGGC framework is non-parametric, model-agnostic, and accommodates arbitrary device dynamics under mild assumptions. It thus provides a unified, decentralized approach to certifying both stability and performance without requiring explicit device-model parameterizations. Moreover, the NGGC framework can be directly used as a set of specifications for control design, offering a principled foundation for future stability- and performance-oriented grid codes in power systems.
\end{abstract}

\section{Introduction}
\fontdimen2\font=0.6ex \IEEEPARstart{C}{urrent} grid codes for dynamic ancillary services (e.g., fast frequency and voltage regulation) with converter-based generation are usually defined through prescribed time-domain step responses, response-time limits, and steady-state specifications. The European network code \cite{european2016commission}, for example, specifies frequency containment and voltage regulation via piecewise linear active/reactive power curves. National codes in Finland \cite{oyj2021technical} and Germany \cite{GermanGridCode} likewise define fast frequency reserves and synthetic inertia through power injection profiles or time constants, enabling converter support in low inertia grids. Although simple in concept, these specifications are often applied ad hoc through open-loop commands or look-up tables, making them rigid and inefficient. They focus mainly on grid-following controls, with only recent exceptions such as Germany’s code \cite{GermanGridCode} adding grid-forming requirements, and they provide no guarantees for closed-loop stability or overall system performance. Consequently, even devices that satisfy current grid codes can exhibit poor performance or lead to system instabilities \cite{gross2022compensating,geng2017small,ying2018impact}.

Nordic grid-code proposals \cite{NordicGridCodes} have begun to target \emph{closed-loop stability and performance guarantees} in the frequency domain. However, they remain limited: they address only single-device-to-grid connections, and focus on qualitative disturbance rejection rather than explicit bounds on metrics such as frequency nadir or RoCoF. Early ideas for embedding passivity requirements in grid-forming control design \cite{NERCdocument} face similar issues, as power system networks are not inherently passive. These limitations motivate the formulation of a more general, quantitative, and practically implementable framework for specifying dynamic ancillary services provision in future grid codes.

This work introduces the concept of \emph{Next Generation Grid Codes (NGGCs)} as a first impetus toward a new grid-code paradigm for dynamic ancillary services via decentralized frequency-domain criteria. The NGGC concept offers two key benefits: (i) guaranteed closed-loop stability, and (ii) explicit performance guarantees for the overall power system. Regarding \emph{(i) closed-loop stability}, we employ loop-shifting and passivity-based techniques to derive decentralized stability certificates for the interconnected system. The certificates translate directly into local tuning rules for individual device controllers, ensuring that each device contributes to global stability in a modular fashion. Concerning \emph{(ii) performance guarantees}, we establish quantitative bounds on critical time-domain performance indicators of the system’s small-signal frequency and voltage dynamics, including both local voltage responses at the individual device buses as well as the average-mode frequency response. The considered indicators include frequency and voltage nadirs, rate-of-change-of-frequency (RoCoF), steady-state deviations, and oscillation damping capabilities. The bounds are derived by linking system-level time-domain metrics to local frequency-domain conditions imposed on each device. Both the decentralized stability and performance requirements can be expressed through Nyquist-plot envelopes that characterize local converter dynamics in terms of their transfer functions, thereby allowing them to be directly used as specifications for control design. The NGGC framework accommodates devices with a grid-forming signal causality, i.e., mapping active and reactive power injections to frequency and voltage outputs, and integrates with existing converter and synchronous-generator models.

In contrast to existing approaches for decentralized closed-loop stability certificates that rely on explicit device-model parameterizations \cite{siahaan2024decentralized,gross2022compensating,haberle2025decentralized,subotic2021lyapunov}, the proposed framework is model-agnostic and accommodates arbitrary local device dynamics, provided that a set of mild regularity conditions on the device behavior is satisfied. This generality allows the interconnection of heterogeneous device models, independent of specific control architectures or tuning methods, thereby offering greater design flexibility to power plant operators. While related generic stability certificates have been proposed in \cite{huang2024gain,niehues2025small,huang2025geometric}, typically based on small-gain, small-phase, or sector-bounded conditions, our work extends these results by also providing explicit performance guarantees for the system frequency and voltage dynamics. In particular, we derive quantitative bounds on average-mode frequency and voltage nadirs, steady-state deviations, frequency RoCoF, and oscillation damping capabilities, applicable to arbitrary device models without explicit parameterization. Although previous studies have addressed such performance specifications for particular device parameterizations \cite{jiang2021grid}, to the best of our knowledge, no existing work provides a unified framework that ensures \emph{both} closed-loop stability and performance under general, non-parametric device dynamics.

The paper is structured as follows. Section \ref{sec:problem_setup} presents the power-system model used to develop the proposed grid-code framework. Section \ref{sec:framework} introduces the device-level certification methodology that guarantees closed-loop stability and performance, which forms the basis for the grid-code formulation in Section \ref{sec:NGGC}. Section \ref{sec:case_study} evaluates the grid codes in numerical case studies. Section \ref{sec:conclusion} summarizes the main findings.

\section{Power System Model}\label{sec:problem_setup}
\subsection{Small-Signal Network Dynamics}
\fontdimen2\font=0.6ex  We study the stability and dynamic performance of an interconnected power system composed of three-phase generation units, including grid-forming voltage-source converters (VSCs) and synchronous generators (SGs). These units are interconnected through a dynamic transmission network modeled by resistive–inductive lines, as illustrated in \cref{fig:multi_conv_system}. The network is assumed to be balanced and is represented in its Kron-reduced form, consisting of $n \in \mathbb{N}$ generator nodes indexed by ${1,\dots,n}$.

The associated network dynamics are captured by a \emph{quasi-stationary small-signal} model, formulated in the global per-unit system, which relates frequency and voltage-magnitude deviations, $\Delta f_i(s)$ and $\Delta |v|_i(s)$, at each node $i$ to the corresponding active and reactive power injections, $\Delta p_i(s)$ and $\Delta q_i(s)$. In particular, the transmission-line dynamics can be compactly described by a $2n \times 2n$ transfer matrix $N(s)$ (see \cite{haberle2025decentralized} for a detailed derivation), yielding
\begin{align}\label{eq:network_polar}
   \hspace{-1mm} \underset{\eqqcolon\,{\tiny\begin{bmatrix}
       \Delta p(s)\\\Delta q(s)
   \end{bmatrix}}}{\underbrace{\begin{bmatrix}
       \Delta p_1(s)\\ \Delta q_1(s)\\\vdots\\ \Delta p_n(s)\\ \Delta q_n(s)
   \end{bmatrix}}}
  \hspace{-0.5mm} = \hspace{-0.5mm}\underset{\eqqcolon \, {N}(s)}{\underbrace{
   \begin{bmatrix}
       N_{11}(s)&\cdots&N_{1n}(s)\\\vdots & \ddots& \vdots\\ N_{n1}(s)& \cdots & N_{nn}(s)
   \end{bmatrix}}}
   \underset{\eqqcolon\,{\tiny\begin{bmatrix}
       \Delta f(s)\\\Delta |v|(s)
   \end{bmatrix}}}{\underbrace{
   \begin{bmatrix}
       \Delta f_1(s)\\\Delta {|v|}_{1} (s)\\ \vdots\\ \Delta f_n(s)\\\Delta {|v|}_{n} (s)
   \end{bmatrix}}},\hspace{-1mm} 
\end{align}
with the $2\times2$ transfer matrix blocks $N_{ii}(s)$ and $N_{ij}(s)$
\begin{align}\label{eq:full_network_polar_matrix_blocks_level1}
    \begin{split}
    \hspace{-2mm}N_{ii}(s) \hspace{-0.5mm}&=\hspace{-0.5mm} \textstyle\sum_{j\neq i}^n\hspace{-0.5mm} b_{ij}\tfrac{1}{1+\rho^2}\hspace{-0.8mm}\begin{bmatrix}
        \tfrac{2\pi|v|_{0,i}|v|_{0,j}}{s}\hspace{-2mm}&0\\0 \hspace{-2mm}& 2|v|_{0,i}\hspace{-0.5mm}-\hspace{-0.5mm}|v|_{0,j}
    \end{bmatrix}\hspace{-2mm}\\
    \hspace{-2mm}N_{ij}(s)\hspace{-0.5mm} &=\hspace{-0.5mm} b_{ij}\tfrac{|v|_{0,i}|v|_{0,j}}{1+\rho^2}\hspace{-0.5mm}\begin{bmatrix}
        -\tfrac{2\pi}{s}&0\\0&-\tfrac{1}{|v|_{0,j}}
    \end{bmatrix}\hspace{-0.5mm}.\hspace{-2mm}
\end{split}
\end{align}
Here, $|v|_{0,i}$ denotes the steady-state voltage magnitude at node $i$, $b_{ij}=1/l_{ij}$ the line susceptance, and $\rho_{ij}=r_{ij}/l_{ij}$ the resistance-to-inductance ratio of line $ij$. The derivation of \cref{eq:full_network_polar_matrix_blocks_level1} uses the standard small-angle approximation $\delta_{0,i}\approx\delta_{0,j}$. The network is assumed to be dominantly inductive, with a small and uniform ratio $\rho_{ij}=\rho\ll1$ for all $i,j\in\{1,\dots,n\}$. If no line connects nodes $i$ and $j$, we set $b_{ij}=\rho_{ij}=0$. Under these assumptions, \cref{eq:full_network_polar_matrix_blocks_level1} reveals a decoupling of active-power and frequency dynamics from reactive-power and voltage-magnitude dynamics.
\begin{figure}[t!]
    \centering
    \vspace{-1mm}
    \resizebox{0.45\textwidth}{!}{

\tikzstyle{roundnode}=[circle,draw=black!60, fill=black!5,scale=0.5]
\begin{tikzpicture}[scale=1,every node/.style={scale=0.65}]
\draw (-2.2,1.4) node (v4) {} -- (0.2,0.4) node (v1) {} -- (-1.7,2.2) node (v3) {} --  (1.6,2)  -- (0.2,0.4) node (v5) {} -- (1.6,1) node (v2) {} -- (1.6,2) node (v6) {} -- (1.6,1) -- (1.6,1) node (v7) {};

\node [scale=1.8] at (0,1.6) {$\ddots$};
\node at (-1.7,2.4) {$1$};
\draw (-1.7,2.2) -- (-2.7,2.2); 
\draw (-2.2,1.4) -- (-3.02,1.4); 
\draw (0.2,0.4) -- cycle;
\draw (0.2,0.4) -- (-1.7,0.4); 
\draw (1.6,2) -- (2.1,2); 
\draw (1.6,1) -- (2.5,1);
\draw  [rounded corners = 3,fill=gray!30](-3.4,2.5) rectangle (-2.7,1.9);
\draw  [rounded corners = 3,fill=gray!30](-2.9,1.7) rectangle (-3.6,1.1);
\draw  [rounded corners = 3,fill=gray!30](-1.7,0.7) rectangle (-2.4,0.1);
\node  [scale=1.8] at (-2.6,1.1) {$\ddots$};
\draw  [rounded corners = 3,fill=gray!30](2.092,2.3) rectangle (2.792,1.7);
\draw  [rounded corners = 3,fill=gray!30](2.5,1.3) rectangle (3.2,0.7);

\node at (2.85,0.55) {VSC};
\node at (-3.24,0.95) {SG};
\node at (-2.3,1.6) {2};
\node at (0.3,0.2) {$i$};
\node at (1.7,0.8) {$n-1$};
\node at (1.7,2.2) {$n$};

\draw (-2.28,0.39) -- (-2.18,0.39); 
\draw (-2.18,0.54) -- (-2.18,0.24); 
\draw (-2.13,0.54) -- (-2.13,0.24); 
\draw (-2.13,0.44) -- (-2.03,0.49) -- (-2.03,0.64); 
\draw (-2.13,0.34) -- (-2.03,0.29); 
\draw (-2.03,0.29) node (v9) {} -- (-2.03,0.14); 
\draw (-2.03,0.19) -- (-1.93,0.19) --  (-1.93,0.34);
\draw  (-1.93,0.44) -- (-1.93,0.59) -- (-2.03,0.59);
\draw (-1.98,0.44) -- (-1.88,0.44); 
\draw (-1.93,0.44) node (v8) {} -- (-1.98,0.34) -- (-1.88,0.34) -- (-1.93,0.44);
\draw (-2.06,0.33) --  (-2.03,0.29) -- (-2.08,0.29);

\draw (2.62,1.01) -- (2.72,1.01); 
\draw (2.72,1.16) -- (2.72,0.86); 
\draw (2.77,1.16) -- (2.77,0.86); 
\draw (2.77,1.06) -- (2.87,1.11) -- (2.87,1.26); 
\draw (2.77,0.96) -- (2.87,0.91); 
\draw (2.87,0.91) node (v9) {} -- (2.87,0.76); 
\draw (2.87,0.81) -- (2.97,0.81) --  (2.97,0.96);
\draw  (2.97,1.06) -- (2.97,1.21) -- (2.87,1.21);
\draw (2.92,1.06) -- (3.02,1.06); 
\draw (2.97,1.06) node (v8) {} -- (2.92,0.96) -- (3.02,0.96) -- (2.97,1.06);
\draw (2.84,0.95) --  (2.87,0.91) -- (2.82,0.91);

\draw (-3.26,2.2) -- (-3.16,2.2); 
\draw (-3.16,2.35) -- (-3.16,2.05); 
\draw (-3.11,2.35) -- (-3.11,2.05); 
\draw (-3.11,2.25) -- (-3.01,2.3) -- (-3.01,2.45); 
\draw (-3.11,2.15) -- (-3.01,2.1); 
\draw (-3.01,2.1) node (v9) {} -- (-3.01,1.95); 
\draw (-3.01,2) -- (-2.91,2) --  (-2.91,2.15);
\draw  (-2.91,2.25) -- (-2.91,2.4) -- (-3.01,2.4);
\draw (-2.96,2.25) -- (-2.86,2.25); 
\draw (-2.91,2.25) node (v8) {} -- (-2.96,2.15) -- (-2.86,2.15) -- (-2.91,2.25);
\draw (-3.04,2.14) --  (-3.01,2.1) -- (-3.06,2.1);

\draw[fill=gray!10,color=gray!10,opacity=0.5]  (-0.2,1.5) ellipse (2 and 1.1);
\draw[-latex,gray] (-2.55,2.3) -- (-2,2.3);
\node at (-2.35,2.7) {$\begin{bmatrix}\Delta p_{1}\\ \Delta q_1\end{bmatrix}$};
\draw[dashed,gray]  (-0.2,1.5) ellipse (2 and 1.1);
\node [color=gray] at (-0.2,2.35) {power network};
\node [roundnode] at (-1.7,2.2) {};
\node [roundnode] at (-2.2,1.4) {};
\node [roundnode] at (0.2,0.4) {};
\node [roundnode] at (1.6,1) {};
\node [roundnode] at (1.6,2) {};
\draw[-latex,gray] (-1.8,2.2) -- (-1.8,2.9);
\node at (-1.3,2.9) {$\begin{bmatrix}\Delta f_{1}\\\Delta |v|_{1}\end{bmatrix}$};
\fill[gray] (-1.9,2.2) circle(0.2mm);

\draw  (-3.24,1.4) ellipse (0.225 and 0.225);
\draw (-3.36,1.462) arc (180:0:0.12);
\draw (-3.12,1.342) node (v10) {} arc (0:-180:0.12); 
\draw (-3.36,1.462) -- (-3.28,1.462) -- (-3.28,1.342) -- (-3.36,1.342);
\draw (-3.12,1.462)--(-3.2,1.462) -- (-3.2,1.342)--(-3.12,1.342);

\draw (2.44,2) ellipse (0.225 and 0.225);
\draw (2.32,2.062) arc (180:0:0.12);
\draw (2.56,1.942) node (v10) {} arc (0:-180:0.12); 
\draw (2.32,2.062) -- (2.4,2.062) -- (2.4,1.942) -- (2.32,1.942);
\draw (2.56,2.062)--(2.48,2.062) -- (2.48,1.942)--(2.56,1.942);

\end{tikzpicture}
}
    \vspace{-9mm}
    \caption{Multi-device transmission system with SGs and grid-forming VSCs connected by resistive–inductive transmission lines.}
    \label{fig:multi_conv_system}
    \vspace{-4mm}
\end{figure}

\subsection{Small-Signal Device Dynamics}
\fontdimen2\font=0.6ex For the network in \cref{fig:multi_conv_system}, we consider three-phase generation devices with \emph{grid-forming signal causality}, i.e., devices that regulate frequency and voltage outputs based on measured active and reactive power deviations. These devices may include grid-forming VSCs and conventional SGs.

The linearized small-signal behavior of the $i$-th device (in the global per unit system) is captured by a diagonal $2\times 2$ transfer matrix $D_i(s)$, which maps the device's active and reactive power injections, $\Delta p_i(s)$ and $\Delta q_i(s)$, to the resulting frequency and voltage deviations, $\Delta f_i(s)$ and $\Delta |v|_i(s)$, i.e.,
\begin{align}\label{eq:local_conv_model}
    -\begin{bmatrix}
        \Delta f_i(s)\\ \Delta |v|_i(s)
    \end{bmatrix}= \underset{=D_i(s)}{\underbrace{\begin{bmatrix}
        D_i^\mathrm{pf}(s)&0\\0&D_i^\mathrm{qv}(s)
    \end{bmatrix}}}\begin{bmatrix}
        \Delta p_i(s)\\ \Delta q_i(s)
    \end{bmatrix}.
\end{align}
In this work, $D_i(s)$ is retained as an arbitrary diagonal rational transfer function that captures any decoupled active-power-frequency ($\mathrm{pf}$) and reactive-power-voltage ($\mathrm{qv}$) control behavior. This highlights a key advantage of the NGGC framework: it is \emph{model-agnostic} and does not depend on specific device parametrizations. While $D_i(s)$ can recover standard control architectures, such as droop or virtual synchronous generator (VSG) implementations for converter outer-loop control \cite{tayyebi2020frequency}, as well as governor-turbine and AVR dynamics for synchronous generators, the framework is not limited to such models. It can accommodate general and flexible device dynamics and control strategies, including novel transfer-function based control designs recently proposed in \cite{haberle2021control,haberle2022control}. This generality enables stability and performance guarantees for a wide class of devices, without restricting the analysis to a particular parametrization.

By stacking the dynamics \cref{eq:local_conv_model} of all $n$ devices, we obtain the overall device dynamics as
\begin{align}\label{eq:full_converter_polar}
     \hspace{-1.5mm}- \hspace{-1mm}
   \begin{bmatrix}
       \Delta f_1(s)\\\Delta {|v|}_{1} (s)\\ \vdots\\ \Delta f_n(s)\\\Delta {|v|}_{n} (s)
   \end{bmatrix}\hspace{-0.7mm}= \hspace{-0.7mm} \underset{\eqqcolon\,D(s)}{\underbrace{\begin{bmatrix}
       D_1(s) \hspace{-0.6mm}& \hspace{-0.6mm}0_{2\times 2} \hspace{-0.6mm}& \hspace{-0.6mm} \dots  \hspace{-0.6mm}& \hspace{-0.6mm} 0_{2\times 2}\\0_{2\times2}  \hspace{-0.6mm}&  \hspace{-0.6mm}D_2(s) \hspace{-0.6mm} &  \hspace{-0.6mm}\dots  \hspace{-0.6mm}&  \hspace{-0.6mm}0_{2\times 2}\\
       \vdots  \hspace{-0.6mm}&  \hspace{-0.6mm}\vdots
        \hspace{-0.6mm}& \hspace{-0.6mm}\ddots \hspace{-0.6mm}& \hspace{-0.6mm}\vdots\\0_{2\times 2} \hspace{-0.6mm}& \hspace{-0.6mm}0_{2\times 2} \hspace{-0.6mm}& \hspace{-0.6mm}\dots \hspace{-0.6mm}& \hspace{-0.6mm}D_n(s)
   \end{bmatrix}}}
    \begin{bmatrix}
       \Delta p_1(s)\\ \Delta q_1(s)\\\vdots\\ \Delta p_n(s)\\ \Delta q_n(s)
   \end{bmatrix} \hspace{-1mm}.\hspace{-0.5mm}
\end{align}

\subsection{Interconnected Power System Model}
\fontdimen2\font=0.6ex 
The closed-loop dynamics of the power system are modeled as the feedback interconnection of the device dynamics in \cref{eq:full_converter_polar} and the network dynamics in \cref{eq:network_polar}, as illustrated in the top of \cref{fig:power_sys_model_orig}. We denote this feedback interconnection by ${D}\#{N}$. In \cref{sec:framework}, we establish the internal feedback stability\footnote{Internal feedback stability refers to the stability of all four closed-loop input-output transfer functions of the feedback interconnection in \cref{fig:power_sys_model_orig}, meaning that their poles lie in the open left half-plane \cite{skogestad2005multivariable}.} of ${D}\#{N}$ and derive quantitative bounds on key time-domain performance metrics of the output frequency and voltage responses under mild, decentralized conditions on the device dynamics $D_i(s)$. As discussed in \cref{sec:NGGC}, these decentralized conditions form a practical and implementable specification set for \emph{next-generation grid code (NGGC)} design.

Because both the device dynamics \cref{eq:full_converter_polar} and the network equations \cref{eq:network_polar} decouple into active-power–frequency ($\mathrm{pf}$) and reactive-power–voltage ($\mathrm{qv}$) channels, the closed-loop system in the top of \cref{fig:power_sys_model_orig} can be partitioned into two corresponding subsystems as indicated in the bottom of \cref{fig:power_sys_model_orig}, where $D^\mathrm{pf}(s)=\mathrm{diag}(D_1^\mathrm{pf}(s),...,D_n^\mathrm{pf}(s))$ and $D^\mathrm{qv}(s)=\mathrm{diag}(D_1^\mathrm{qv}(s),...,D_n^\mathrm{qv}(s))$. Importantly, although the $\mathrm{pf}$- and $\mathrm{qv}$-channels are decoupled at the device level and in the network equations, the network dynamics within each channel couple the device nodes. More specifically, the active-power flow dynamics couple all bus-frequency responses, i.e., 
\begin{align*}
    \hspace{-2mm}N^\mathrm{fp}(s) \hspace{-0.5mm}=\hspace{-0.5mm} \tfrac{1}{s}\hspace{-1mm}\underset{=\,L^\mathrm{fp}}{\underbrace{\begin{bmatrix}
        \hspace{-0.5mm}\textstyle\sum_{j\ne 1}^n\hspace{-1mm}\tfrac{2\pi b_{1j} |v|_{0,1}|v|_{0,j}}{1+\rho^2} & \hspace{-0.5mm}\dots \hspace{-0.5mm}& -\tfrac{2\pi b_{1n}|v|_{0,1}|v|_{0,n}}{1+\rho^2} \\\vdots&\hspace{-0.5mm}\ddots\hspace{-0.5mm}&\vdots\\-\tfrac{2\pi b_{n1}|v|_{0,n}|v|_{0,1}}{1+\rho^2} &\hspace{-0.5mm}\dots\hspace{-0.5mm}& \textstyle\sum_{j\ne n}^n \hspace{-1mm}\tfrac{2\pi b_{nj} |v|_{0,n}|v|_{0,j}}{1+\rho^2}
    \hspace{-1mm}\end{bmatrix}}}\hspace{-1mm},\hspace{-2mm}
\end{align*}
where $L^\mathrm{fp}$ is an undirected weighted Laplacian matrix of the interconnection network with nonnegative and real-valued eigenvalues $0=\lambda_1\leq \lambda_2 \leq ... \lambda_n$. Likewise, the reactive-power flow dynamics couple all bus-voltage responses, i.e., 
\begin{align*}
    \hspace{-2mm} N^\mathrm{vq}\hspace{-0.1mm}\hspace{-0.1mm} = \hspace{-0.5mm}\hspace{-1mm}\begin{bmatrix}\hspace{-0.5mm}
        \textstyle\sum_{j\ne 1}^n\hspace{-1.5mm}\tfrac{b_{1j} (2|v|_{0,1}-|v|_{0,j})}{1+\rho^2}\hspace{-1mm} & \dots & -\tfrac{b_{1n}|v|_{0,1}}{1+\rho^2} \\\vdots&\ddots&\vdots\\-\tfrac{b_{n1}|v|_{0,n}}{1+\rho^2} &\dots& \hspace{-1mm}\textstyle\sum_{j\ne n}^n \hspace{-1.5mm}\tfrac{b_{nj} (2|v|_{0,n}-|v|_{0,j})}{1+\rho^2}\hspace{-0.5mm}
    \end{bmatrix}\hspace{-1mm}.\hspace{-2mm}
\end{align*}

For stability analysis, we retain the full multi-node feedback interconnection. For performance analysis, we adopt reduced models, since performance requirements are typically imposed on local or aggregate quantities rather than full network interactions \cite{ENTSOE_Policy1_2009,ENTSOE_Inertia_RoCoF_2017}. In particular, we consider the average-mode (center-of-inertia) frequency response for the $\mathrm{pf}$-subsystem (see \cref{sec:avg_mode}) and local open-loop voltage responses at each node for the $\mathrm{qv}$-subsystem.
\begin{figure}[t!]
\vspace{-1mm}
    \centering
   \resizebox{0.5\textwidth}{!}{
\tikzstyle{roundnode}=[circle,draw=black!60,fill=black!5,scale=0.7]
\begin{tikzpicture}[scale=1,every node/.style={scale=0.8}]

\draw [rounded corners = 3,fill=gray!30] (-0.3,4.3) rectangle (1.3,3.5);
\node [scale=1.2] at (0.5,3.9) {${D}(s)$};
\draw [rounded corners = 3,fill=gray!10,opacity=0.5] (-0.3,2.7) rectangle (1.3,1.9);
\node [scale=1.2] at (0.5,2.3) {${N}(s)$};
\draw [-latex](1.3,3.9) -- (2.5,3.9);
\draw (-0.3,2.3) -- (-1.1,2.3);

\draw[-latex] (1.8,2.3) -- (1.3,2.3);
\draw [-latex](2.6,2.3) -- (2,2.3); 
\draw [-latex](1.9,3.9) -- (1.9,2.4);
\node [roundnode] at (1.9,2.3) {};

\draw [-latex](-1,3.9) -- (-0.3,3.9);
\draw [-latex](-1.8,3.9) -- (-1.2,3.9); 
\draw [-latex](-1.1,2.3) node (v1) {} -- (-1.1,3.8);
\node [roundnode] at (-1.1,3.9) {};
\node at (-0.9,3.6) {$-$};
\node at (-1.8,4.35) {$\begin{bmatrix}\Delta p_\mathrm{d}\\ \Delta q_\mathrm{d}\end{bmatrix}$};
\node at (2.5,4.35) {$\begin{bmatrix} \Delta f\\\Delta |v|\end{bmatrix}$};
\node at (-0.65,4.35) {$\begin{bmatrix}\Delta p\\ \Delta q\end{bmatrix}$};
\node at (2.6,2.75) {$\begin{bmatrix} \Delta f_\mathrm{d} \\\Delta |v|_\mathrm{d}\end{bmatrix}$};
\node at (0.5,3.3) {device dynamics};
\node at (0.5,1.7) {network dynamics};
\node at (-1.05,4.35) {$-$};
\draw[-latex] (-1.1,2.3) -- (-1.8,2.3);
\node at (-1.8,2.75) {$\begin{bmatrix}\Delta p_\mathrm{e}\\ \Delta q_\mathrm{e}\end{bmatrix}$};

\draw [rounded corners = 3,fill=gray!30] (2.2,0.3) rectangle (3.6,-0.5);
\node [scale=1.2] at (2.9,-0.1) {${D}^\mathrm{qv}(s)$};
\draw [rounded corners = 3,fill=gray!10,opacity=0.5] (2.2,-1.3) rectangle (3.6,-2.1);
\node [scale=1.2] at (2.9,-1.7) {${N}^\mathrm{qv}(s)$};
\draw [-latex](3.6,-0.1) -- (4.7,-0.1);
\draw (2.2,-1.7) -- (1.5,-1.7);

\draw[-latex] (4,-1.7) -- (3.6,-1.7);
\draw [-latex](4.8,-1.7) -- (4.2,-1.7); 
\draw [-latex](4.1,-0.1) -- (4.1,-1.6);
\node [roundnode] at (4.1,-1.7) {};

\draw [-latex](1.6,-0.1) -- (2.2,-0.1);
\draw [-latex](0.8,-0.1) -- (1.4,-0.1); 
\draw [-latex](1.5,-1.7) node (v1) {} -- (1.5,-0.2);
\node [roundnode] at (1.5,-0.1) {};
\node at (1.7,-0.4) {$-$};
\node at (1,0.1) {$\Delta q_\mathrm{d}$};
\node at (4.6,0.1) {$ \Delta |v|$};
\node at (1.8,0.1) {$-\Delta q$};
\node at (4.6,-1.5) {$\Delta|v|_\mathrm{d}$};
\node at (2.9,-0.7) {device dynamics};
\node at (2.9,-2.3) {network dynamics};

\draw[-latex] (1.5,-1.7) -- (0.8,-1.7);
\node at (1,-1.5) {$\Delta q_\mathrm{e}$};
\draw[-latex,dashed] (-0.8,1.4) -- (-1.7,0.6);

\draw [rounded corners = 3,fill=gray!30] (-2.5,0.3) rectangle (-1.1,-0.5);
\node [scale=1.2] at (-1.8,-0.1) {${D}^\mathrm{pf}(s)$};
\draw [rounded corners = 3,fill=gray!10,opacity=0.5] (-2.5,-1.3) rectangle (-1.1,-2.1);
\node [scale=1.2] at (-1.8,-1.7) {${N}^\mathrm{pf}(s)$};
\draw [-latex](-1.1,-0.1) -- (0,-0.1);
\draw (-2.5,-1.7) -- (-3.2,-1.7);

\draw[-latex] (-0.7,-1.7) -- (-1.1,-1.7);
\draw [-latex](0.1,-1.7) -- (-0.5,-1.7); 
\draw [-latex](-0.6,-0.1) -- (-0.6,-1.6);
\node [roundnode] at (-0.6,-1.7) {};

\draw [-latex](-3.1,-0.1) -- (-2.5,-0.1);
\draw [-latex](-3.9,-0.1) -- (-3.3,-0.1); 
\draw [-latex](-3.2,-1.7) node (v1) {} -- (-3.2,-0.2);
\node [roundnode] at (-3.2,-0.1) {};
\node at (-3,-0.4) {$-$};
\node at (-3.7,0.1) {$\Delta p_\mathrm{d}$};
\node at (-0.1,0.1) {$ \Delta 
f$};
\node at (-2.9,0.1) {$-\Delta p$};
\node at (-0.1,-1.5) {$\Delta f_\mathrm{d}$};
\node at (-1.8,-0.7) {device dynamics};
\node at (-1.8,-2.3) {network dynamics};

\draw[-latex] (-3.2,-1.7) -- (-3.9,-1.7);
\node at (-3.7,-1.5) {$\Delta p_\mathrm{e}$};
\draw[-latex,dashed] (-0.8,1.4) -- (-1.7,0.6);
\node at (0.5,0.9) {PARTITIONING};
\draw[-latex,dashed] (1.8,1.4) -- (2.7,0.6);
\end{tikzpicture}
}
    \vspace{-9mm}
    \caption{Closed-loop power system dynamics where $\Delta p_\mathrm{d}$ and  $\Delta q_\mathrm{d}$ are the active and reactive power disturbances, $\Delta f_\mathrm{d}$ is the frequency disturbance, and $\Delta |{v}|_{\mathrm{d}}$ is the voltage magnitude disturbance.}
        \label{fig:power_sys_model_orig}
            \vspace{-4mm}
\end{figure}

\subsection{Average-Mode System Frequency Response Approximation}\label{sec:avg_mode}
\fontdimen2\font=0.6ex We now derive an approximation of the closed-loop system frequency response corresponding to the \emph{average (coherent) mode}. Consider the closed-loop feedback interconnection of the $\mathrm{pf}$-subsystem shown in the bottom-left of \cref{fig:power_sys_model_orig}, and assume the interconnection is internally stable (which is guaranteed under the stability conditions established in \cref{sec:stability_certificates}).

The closed-loop transfer function from the disturbance input $\Delta p_{\mathrm{d}}$ to the frequency deviation $\Delta f$ is given by
\begin{align}\label{eq:cl_w_1}
    \Delta f(s) = D^\mathrm{pf}(s)\left( I + N^\mathrm{fp}(s)D^\mathrm{pf}(s) \right)^{-1} \Delta p_\mathrm{d}(s),
\end{align}
where $D^\mathrm{pf}(s)=\mathrm{diag}(D_1^\mathrm{pf}(s),...,D_n^\mathrm{pf}(s))$ and $N^\mathrm{fp}(s)=\tfrac{1}{s}L^\mathrm{fp}$. Applying an eigenvalue decomposition of the Laplacian matrix,
$L^\mathrm{fp} = V \Lambda V^\top$, where $V$ is orthonormal, i.e., $VV^\top = V^\top V = I$ and $V^\top = V^{-1}$, \cref{eq:cl_w_1} can be written as
\begin{align}\label{eq:cl_w_2}
    \hspace{-2.5mm}\Delta f(s)
    \hspace{-0.5mm}= \hspace{-0.5mm}\mathrm{diag}\big( \hspace{-0.5mm}D_i^\mathrm{pf}\hspace{-0.5mm}(s) \hspace{-0.5mm}\big) \hspace{-0.5mm}
    \left(\hspace{-0.5mm}
    I
     \hspace{-1mm}+  \hspace{-1mm}\tfrac{1}{s}
    V \Lambda V^\top \hspace{-1mm}
    \mathrm{diag}\big(\hspace{-0.5mm} D_i^\mathrm{pf}\hspace{-0.5mm}(s)\hspace{-0.5mm}\big)\hspace{-1mm}
    \right)^{-1} \hspace{-2.5mm}
    \Delta p_\mathrm{d}(s).\hspace{-2mm}
\end{align}

The average-mode frequency response corresponds to the dominant mode of \cref{eq:cl_w_2} on long time-scales \cite{paganini2019global,min2019dynamics,jiang2021grid}. More specifically, in the quasi-steady-state limit $s \rightarrow 0$, one obtains
\begin{align}\label{eq:cl_w_3}
     \hspace{-1mm}\lim_{s \rightarrow 0} \Delta f(s)
    =
    \lim_{s \rightarrow 0}
    \left(\sum_{i=1}^{n} \left(D_i^\mathrm{pf}(s)\right)^{-1}\right)^{-1} \hspace{-1mm}
    \mathbf{1}_n \mathbf{1}_n^\top
    \Delta p_\mathrm{d}(s),
\end{align}
where $\mathbf{1}_n$ is the eigenvector of $L^\mathrm{fp}$ associated with the dominant zero eigenvalue $\lambda_1 = 0$. A detailed derivation of \cref{eq:cl_w_3} is provided in Appendix \ref{appendix:avg_mode}, aligning with classical slow-coherency theory in which the coherent frequency captures the dominant low-frequency behavior \cite{paganini2019global,min2019dynamics,jiang2021grid}. Finally, the average-mode system frequency is obtained by left-multiplying \cref{eq:cl_w_3} with $\tfrac{1}{n}\mathbf{1}_n^\top$, where we define $\tfrac{1}{n}\sum_{i=1}^n\Delta f_i(s)=\Delta f_{\mathrm{avg}}(s)$, yielding
\begin{align}\label{eq:avg_f_response}
   f_{\mathrm{avg}}(s)=
    \underbrace{\left(\textstyle\sum_{i=1}^{n} \left(D_i^\mathrm{pf}(s)\right)^{-1}\right)^{-1}}_{=:~D_{\mathrm{avg}}(s)}
   \underbrace{\textstyle\sum_{i=1}^{n} \Delta p_{\mathrm{d},i}(s)}_{=:~\Delta p_\mathrm{d}^{\sum}(s)}.
\end{align}
As an example, \cite{haberle2022control,he2026aggregate,domingo2024dynamic} leverage this approximation to design aggregate dynamic control strategies for collections of distributed energy resources.

\section{Device-Level Certification Methods}\label{sec:framework}
\subsection{Stability Certification Criteria}\label{sec:stability_certificates}
\fontdimen2\font=0.6ex By considering the feedback interconnection in \cref{fig:power_sys_model_orig} and applying loop-shifting and passivity arguments as in \cite{haberle2025decentralized}, we obtain decentralized stability conditions that act as local tuning rules and guarantee internal feedback stability of $D \# N$.

Recall the stability result in \cite[Corollary~1]{haberle2025decentralized}, derived by analyzing the feedback interconnection via loop shifting as indicated in \cref{fig:loop_shifting}. The key idea is to apply the passivity theorem \cite[Theorem 1]{haberle2025decentralized} to the loop-shifted system in \cref{fig:loop_shifted_power_system}. Since $N(s)$ is not inherently passive, the theorem ensures internal feedback stability of the loop-shifted system provided that $N'(s)$ is passive, $D'(s)$ is strictly passive, and the additional small-gain condition $\bar{\sigma}(N'(\mathrm{j}\infty))\bar{\sigma}(D'(\mathrm{j}\infty)) < 1$ is satisfied at $\omega = \infty$. More specifically, for the block-diagonal loop-shifting matrix $\Gamma = \mathrm{diag}(\Gamma_1,\ldots,\Gamma_n)$, where each block is defined as
\begin{align}\label{eq:gamma}
    \Gamma_i = 
    \begin{bmatrix}
        0 & 0\\
        0 & c_i
    \end{bmatrix},
    \qquad 
    c_i = \textstyle\sum_{j\neq i}^n b_{ij}\tfrac{0.8}{1+\rho^2}\tfrac{1}{|v|_{0,i}},
\end{align}
it can be shown that $N'(s)$ is passive. To ensure strict passivity of $D'(s)$ for $\Gamma$ defined in \cref{eq:gamma}, the loop-shifted device dynamics
\begin{align}\label{eq:D_dash}
    \hspace{-1mm}D_i'(s)
    \hspace{-0.5mm}= \hspace{-0.5mm}D_i(s)\big(I\hspace{-0.5mm}-\hspace{-0.5mm}\Gamma_i D_i(s)\big)^{-1}\hspace{-0.5mm}
    = \hspace{-0.5mm}\begin{bmatrix}
        D_i^{\mathrm{pf}}(s)\hspace{-1mm} & 0\\[1mm]
        0 & \hspace{-1mm}\tfrac{D_i^{\mathrm{qv}}(s)}{1-c_i D_i^{\mathrm{qv}}(s)}
      \end{bmatrix}\hspace{-1mm}
\end{align}
must be strictly passive for all $i \in \{1,\ldots,n\}$. If, in addition, the condition $\bar{\sigma}(N'(\mathrm{j}\infty)),\bar{\sigma}(D'(\mathrm{j}\infty)) < 1$ is satisfied, then stability of the original system $D\#N$ in \cref{fig:power_sys_model} follows from the stability of the loop-shifted system $D'\#N'$ in \cref{fig:loop_shifted_power_system}. 

Because the device dynamics in \cref{eq:D_dash} are decoupled, the stability conditions separate into independent constraints for active-power frequency and reactive-power voltage control.

\subsubsection{Frequency Control}
For the $\mathrm{pf}$-control, strict passivity of $D'(s)$ and $\bar{\sigma}(N'(\mathrm{j}\infty))\bar{\sigma}(D'(\mathrm{j}\infty))<1$ is equivalent to satisfying, for each unit~$i$,
\begin{align}
    \tag*{{\color{red}(1-i)}}\label{cond:1i}
    & {\color{red} D_i^{\mathrm{pf}}(s)\ \text{is stable and strictly proper},}\\[-0.5mm]
    \tag*{{\color{matlab3}(1-ii)}}\label{cond:1ii}
    & {\color{matlab3}\mathrm{Re}[D_i^{\mathrm{pf}}(\mathrm{j}\omega)] > 0,\quad \forall\,\omega\in[0,\infty).}
\end{align}
The conditions \ref{cond:1i} and \ref{cond:1ii} match the strict passivity criteria \cite{haberle2025decentralized} and also ensure $\bar{\sigma}(D_i^{\mathrm{pf}}(\mathrm{j}\infty))=0$.

\begin{figure}[t!]
    \centering
    \vspace{-2mm}
    \begin{subfigure}{0.21\textwidth}
        \centering
        \resizebox{1.1\textwidth}{!}{
\tikzstyle{roundnode}=[circle,draw=black!60,fill=black!5,scale=0.7]
\begin{tikzpicture}[scale=1,every node/.style={scale=0.8}]

\draw [rounded corners = 3,fill=gray!30] (-0.3,4.3) rectangle (1.3,3.5);
\node [scale=1.2] at (0.5,3.9) {${D}(s)$};
\draw [rounded corners = 3,fill=gray!10,opacity=0.5] (-0.3,2.7) rectangle (1.3,1.9);
\node [scale=1.2] at (0.5,2.3) {${N}(s)$};
\draw [-latex](1.3,3.9) -- (2.5,3.9);
\draw (-0.3,2.3) -- (-1.1,2.3);

\draw[-latex] (1.8,2.3) -- (1.3,2.3);
\draw [-latex](2.6,2.3) -- (2,2.3); 
\draw [-latex](1.9,3.9) -- (1.9,2.4);
\node [roundnode] at (1.9,2.3) {};

\draw [-latex](-1,3.9) -- (-0.3,3.9);
\draw [-latex](-1.8,3.9) -- (-1.2,3.9); 
\draw [-latex](-1.1,2.3) node (v1) {} -- (-1.1,3.8);
\node [roundnode] at (-1.1,3.9) {};
\node at (-0.9,3.6) {$-$};
\node at (-1.8,4.35) {$\begin{bmatrix}\Delta p_\mathrm{d}\\ \Delta q_\mathrm{d}\end{bmatrix}$};
\node at (2.5,4.35) {$\begin{bmatrix} \Delta f\\\Delta |v|\end{bmatrix}$};
\node at (-0.65,4.35) {$\begin{bmatrix}\Delta p\\ \Delta q\end{bmatrix}$};
\node at (2.6,2.75) {$\begin{bmatrix} \Delta f_\mathrm{d} \\\Delta |v|_\mathrm{d}\end{bmatrix}$};
\node at (0.5,3.3) {device dynamics};
\node at (0.5,1.7) {network dynamics};
\node at (-1.05,4.35) {$-$};
\draw[-latex] (-1.1,2.3) -- (-1.8,2.3);
\node at (-1.8,2.75) {$\begin{bmatrix}\Delta p_\mathrm{e}\\ \Delta q_\mathrm{e}\end{bmatrix}$};

\end{tikzpicture}
}
        \vspace{-8.5mm}
        \caption{Original feedback system.}
        \label{fig:power_sys_model}
            \vspace{-1mm}
    \end{subfigure}
    \hspace{0.05cm}
    \begin{subfigure}{0.24\textwidth}
        \centering
        \resizebox{1.1\textwidth}{!}{
\tikzstyle{roundnode}=[circle,draw=black!60,fill=black!5,scale=0.75]
\begin{tikzpicture}[scale=1,every node/.style={scale=0.8}]
\draw [rounded corners = 3,dashed] (2,3.7) rectangle (-1.1,1.6);
\draw [rounded corners = 3, dashed] (2,4.4) rectangle (-1.1,6.5);
\draw [rounded corners = 3,fill=gray!30] (-0.3,6.2) rectangle (1.3,5.4);
\draw [rounded corners = 3,fill=gray!10,opacity=0.5] (-0.3,2.5) rectangle (1.3,1.7);
\node [scale=1.2] at (0.5,5.8) {${D}(s)$};
\node [black,scale=1.2] at (1.65,6.15) {${D}'$};
\node [scale=1.2] at (0.5,2.1) {${N}(s)$};
\node [black,scale=1.2] at (1.65,3.35) {${N}'$};
\draw [-latex](1.3,5.8) -- (3,5.8);
\draw (-0.9,2.1) -- (-1.4,2.1);

\draw[-latex] (2.2,2.1) -- (1.3,2.1);

\draw [-latex](2.3,5.8) -- (2.3,2.2);

\draw [-latex](-1.3,5.8) -- (-0.9,5.8);
\draw [-latex](-2.1,5.8) -- (-1.5,5.8); 
\draw [-latex](-1.4,2.1) node (v2) {} -- (-1.4,5.7);
\node [roundnode] at (-1.4,5.8) {};
\node at (-1.25,5.5) {$-$};

\draw [rounded corners = 3, fill = black!45] (-0.3,3.4) rectangle (1.3,2.6);
\node [scale=1.2] at (0.5,3) {$\Gamma$};
\draw [rounded corners = 3, fill = black!45] (-0.3,5.3) rectangle (1.3,4.5);
\node [scale=1.2] at (0.5,4.9) {$\Gamma$};
\draw [-latex](1.8,2.1) -- (1.8,3) -- (1.3,3);
\draw [-latex](-0.3,2.1) -- (-0.7,2.1); 
\draw [-latex](-0.3,3) -- (-0.8,3) -- (-0.8,2.2);
\node [roundnode] at (-0.8,2.1) {}; 

\node[roundnode] at (-0.8,5.8) {}; 
\draw [-latex](-0.7,5.8) -- (-0.3,5.8);
\draw [-latex](-0.3,4.9) -- (-0.8,4.9) -- (-0.8,5.7);
\draw[-latex] (1.8,5.8) -- (1.8,4.9) -- (1.3,4.9);

\node at (-2.1,6.25) {$\begin{bmatrix}\Delta p_\mathrm{d}\\ \Delta q_\mathrm{d}\end{bmatrix}$};
\node at (3,6.25) {$\begin{bmatrix} \Delta f\\\Delta |{v}|\end{bmatrix}$};
\node at (0.5,4.2) {device dynamics};
\node at (0.5,1.4) {network dynamics};
\node [roundnode] (v1) at (2.3,2.1) {};

\draw [-latex](3,2.1) --  (2.4,2.1);
\draw [-latex](-1.4,2.1) -- (-2.1,2.1);
\node at (-2.1,2.55) {$\begin{bmatrix}\Delta p_\mathrm{e}\\ \Delta q_\mathrm{e}\end{bmatrix}$};
\node at (3,2.55) {$\begin{bmatrix} \Delta f_\mathrm{d} \\\Delta |{v}|_\mathrm{d}\end{bmatrix}$};
\end{tikzpicture}
}
        \vspace{-8.5mm}
        \caption{Loop-shifted system with $\Gamma$.}
        \label{fig:loop_shifted_power_system}
           \vspace{-1mm}
    \end{subfigure}
    \caption{\footnotesize Closed-loop interconnection used for stability analysis.}
    \label{fig:loop_shifting}
    \vspace{-4mm}
\end{figure}
\subsubsection{Voltage Control}
For the $\mathrm{qv}$-control, strict passivity of $D'(s)$ and $\bar{\sigma}(N'(\mathrm{j}\infty))\bar{\sigma}(D'(\mathrm{j}\infty))<1$ requires, for each unit~$i$,
\begin{align}
    \tag*{{\color{red}(2-i)}}\label{cond:2i}
    & {\color{red}D_i^{\mathrm{qv}}(s)\ \text{is stable and strictly proper},}\\[-0.5mm]
    \tag*{{\color{matlab3}(2-ii)}}\label{cond:2ii}
    & {\color{matlab3} c_i < \mathrm{Re}[D_i^{\mathrm{qv}}(\mathrm{j}\omega)^{-1}],\quad \forall\,\omega\in[0,\infty).}
\end{align}
The conditions in \ref{cond:2i} and \ref{cond:2ii} follow from phase constraints in the complex plane. A detailed proof is stated in Appendix \ref{appendix:stability_qv}.

\begin{figure*}
    \centering
    \vspace{-2mm}
    \resizebox{1.01\textwidth}{!}{
\begin{tikzpicture}[scale=1,every node/.style={scale=1.5}]

\draw [-latex](4.35,-2.05) -- (4.35,2.25);
\draw[-latex] (1.95,0) -- (6.75,0);
\node [scale=0.7,anchor=west] at (4.4,2.15) {$\text{Im}[D_i^\mathrm{pf}(\mathrm{j}\omega)]$};
\node [scale=0.7,anchor =east] at (6.95,-0.3) {$\text{Re}[D_i^\mathrm{pf}(\mathrm{j}\omega)]$};
\node[scale=0.7,color=red] at (4.35,-2.3) {cond (1-i)};
\node[scale=0.7,color=red] at (3.45,0.35) {$|D_i^\mathrm{pf}(\mathrm{j}\infty)| = 0$};

\draw[fill=red,color=red]  (4.35,0) node (v7) {} ellipse (0.06 and 0.06);

\draw [fill=matlab3!30,color=matlab3!30,opacity=0.5] (9.4,2.1) node (v6) {} rectangle (11.55,-2.05) node (v1) {};
\draw [-latex](9.35,-2.05) -- (9.35,2.25);
\draw[-latex] (6.95,0) -- (11.75,0);
\node [scale=0.7,anchor=west] at (9.4,2.15) {$\text{Im}[D_i^\mathrm{pf}(\mathrm{j}\omega)]$};
\node [scale=0.7,anchor=east] at (11.95,-0.3) {$\text{Re}[D_i^\mathrm{pf}(\mathrm{j}\omega)]$};

\node[scale=0.7,color=matlab3] at (9.35,-2.3) {cond (1-ii)};

\node [scale=0.7,color=matlab3] at (10.95,1.45) {$\text{Re}[D_i^\mathrm{pf}(\mathrm{j}\omega)]>0,$};
\node [scale=0.7,color=matlab3] at (11.3,0.95) {$\forall\omega\in[0,\infty)$};

\draw [dotted,dash pattern=on 1pt off 1pt,color=matlab3,thick] (9.4,2.1) -- (9.4,-2.05);

\draw[fill=orange!20, color=orange!20,opacity = 0.7]  (14.35,0) node (v2) {} rectangle  (16.55,-2.05);

\draw[fill=orange!20, color=orange!20,opacity = 0.5] (14.35,0) -- (16.55,1.5) node (v8) {} -- (16.55,0) --(14.35,0);
\draw [-latex](14.35,-2.05) -- (14.35,2.25);
\draw[-latex] (11.95,0) -- (16.75,0);
\node [scale=0.7,anchor=west] at (14.4,2.15) {$\text{Im}[D_i^\mathrm{pf}(\mathrm{j}\omega)]$};
\node [scale=0.7,anchor=east] at (16.95,-0.3) {$\text{Re}[D_i^\mathrm{pf}(\mathrm{j}\omega)]$};

\node[scale=0.7,color=orange] at (14.35,-2.3) {cond (1-iii)};
\node  [scale=0.7,color=orange] at (15.25,-1) {$\angle (D_i^\mathrm{pf}(\mathrm{j}\omega))\in[-\frac{\pi}{2},\frac{\pi}{6}],$};
\node  [scale=0.7,color=orange] at (16.4,-1.5) {$\forall\omega \geq 0$};

\draw[color=orange,dotted,dash pattern=on 1pt off 1pt,thick](14.35,0) node (v10) {} --(16.55,1.5);

\draw[color=matlab4,ultra thin] (19.35,0) node (v4) {} -- (19.1,0.15);
\node[scale=0.7,color=matlab4] at (18.8,0.15) {$\varepsilon_\mathrm{f}$};
\node[scale=0.7,color=matlab4] at (18.1,1.2) {$|D_i^\mathrm{pf}(\mathrm{j}\omega)|\leq \varepsilon_\mathrm{f},$};
\node[scale=0.7,color=matlab4] at (18.45,0.7) {$\forall \omega\geq \omega_\mathrm{bw}$};

\draw [fill=matlab4!30,opacity=0.5,draw=none] (19.35,0) node (v3) {} ellipse (0.3 and 0.3);
\draw [-latex](19.35,-2.05) -- (19.35,2.25);
\draw[-latex] (16.95,0) -- (21.75,0);
\node [scale=0.7,anchor=west] at (19.4,2.15) {$\text{Im}[D_i^\mathrm{pf}(\mathrm{j}\omega)]$};
\node [scale=0.7,anchor=east] at (21.95,-0.3) {$\text{Re}[D_i^\mathrm{pf}(\mathrm{j}\omega)]$};
\draw [color=matlab4,dotted,dash pattern=on 1pt off 1pt,thick] (19.35,0) node (v3) {} ellipse (0.3 and 0.3);
\node[scale=0.7,color=matlab4] at (19.35,-2.3) {cond (1-iv)};

\draw [fill=matlab1!20,opacity=0.4,draw=none] (24.35,0) ellipse (1.75 and 1.75);
\draw [color=matlab1,dotted,dash pattern=on 1pt off 1pt,thick] (24.35,0) ellipse (1.75 and 1.75);
\draw [-latex](24.35,-2.05) -- (24.35,2.25);
\draw[-latex] (21.95,0) -- (26.75,0);
\node [scale=0.7,anchor=west] at (24.4,2.15) {$\text{Im}[D_i^\mathrm{pf}(\mathrm{j}\omega)]$};
\node [scale=0.7,anchor=east] at (26.95,-0.3) {$\text{Re}[D_i^\mathrm{pf}(\mathrm{j}\omega)]$};
\node[scale=0.7,color=matlab1] at (24.35,-2.3) {cond (1-v)};
\draw[color=matlab1,ultra thin] (24.35,0)  -- (24,1.7);
\node [scale=0.7,color=matlab1] at (23.6,0.5) {$\frac{\Delta f_\mathrm{max}}{2.5\Delta p_\mathrm{d}^{\tikz[baseline=0] \draw (0.05,0.05) -- (0.15,0.05) -- (0.15,0.15) -- (0.25,0.15);}}$};
\node  [scale=0.7,color=matlab1] at (24.1,-0.95) {$||D_i^\mathrm{pf}(\mathrm{j}\omega)||_\infty\leq \frac{\Delta f_\mathrm{max}}{2.5\Delta p_\mathrm{d}^{\tikz[baseline=0] \draw (0.05,0.05) -- (0.15,0.05) -- (0.15,0.15) -- (0.25,0.15);}}$};

\draw [fill=matlab5!20,opacity=0.6,draw=none] (29.35,0) node (v5) {} ellipse (1.1 and 1.1);
\draw [color=matlab5,dotted,dash pattern=on 1pt off 1pt,thick] (29.35,0) node (v5) {} ellipse (1.1 and 1.1);
\draw [-latex](29.35,-2.05) -- (29.35,2.25);
\draw[-latex] (26.95,0) -- (31.75,0);
\node [scale=0.8,anchor=west] at (29.4,2.15) {$\text{Im}[D_i^\mathrm{pf}(\mathrm{j}\omega)]$};
\node [scale=0.7,anchor=east] at (31.95,-0.3) {$\text{Re}[D_i^\mathrm{pf}(\mathrm{j}\omega)]$};
\node[scale=0.7,color=matlab5] at (29.35,-2.3) {cond (1-vi)};
\draw[color=matlab5,ultra thin] (29.35,0)  -- (29.55,1.08);
\node [scale=0.7,color=matlab5] at (30.3,0.45) {$\frac{\Delta f_\mathrm{ss,\,max}}{\Delta p_\mathrm{d}^{\tikz[baseline=0] \draw (0.05,0.05) -- (0.15,0.05) -- (0.15,0.15) -- (0.25,0.15);}}$};
\node  [scale=0.7,color=matlab5] at (30.3,1.35) {$|D_i^\mathrm{pf}(0)|\leq \frac{\Delta f_\mathrm{ss,\,max}}{\Delta p_\mathrm{d}^{\tikz[baseline=0] \draw (0.05,0.05) -- (0.15,0.05) -- (0.15,0.15) -- (0.25,0.15);}}$};

\draw [fill=magenta!30,opacity=0.5,draw=none] (40.35,0) ellipse (1 and 1);
\draw [color=magenta,dotted,dash pattern=on 1pt off 1pt,thick] (40.35,0) ellipse (1 and 1);
\draw[color=magenta,ultra thin] (40.35,0)  -- (40.55,1);
\draw [-latex](39.35,-2.05) -- (39.35,2.25);
\draw[-latex] (36.95,0) -- (41.75,0);
\node [scale=0.7,anchor=west] at (39.4,2.15) {$\text{Im}[D_i^\mathrm{pf}(\mathrm{j}\omega)]$};
\node [scale=0.7,anchor=east] at (41.95,-0.3) {$\text{Re}[D_i^\mathrm{pf}(\mathrm{j}\omega)]$};
\node[scale=0.7,color=magenta] at (39.35,-2.3) {cond (1-viii)};

\node[scale=0.7,color=magenta] at (39.1,-1.3) {$\text{Re}[D_i^\mathrm{pf}(\mathrm{j}\omega)^{-1}] \hspace{-0.5mm}=\hspace{-0.5mm} \tfrac{\text{Re}[D_i^\mathrm{pf}(\mathrm{j}\omega)]}{|D_i^\mathrm{pf}(\mathrm{j}\omega)|^2}\hspace{-0.5mm}\geq\hspace{-0.5mm} \eta_\mathrm{f},$};
\node[scale=0.7,color=magenta] at (40.95,-1.85) {$\forall\omega\geq 0$};

\node [scale=0.7,color=magenta] at (40.2,0.8) {$\tfrac{1}{2\eta_\mathrm{f}}$};

\node [scale=0.7,color=magenta] at (41.1,0.3) {$(\tfrac{1}{2\eta_\mathrm{f}},0)$};
\draw [color=magenta,fill=magenta] (40.35,0) ellipse (0.05 and 0.05);

\draw [-latex](34.35,-2.05) -- (34.35,2.25);
\draw[-latex] (31.95,0) -- (36.75,0);

\node[scale=0.7,color=cyan] at (34.35,-2.3) {cond (1-vii)};
\node [scale=0.7,color=cyan] at (34.9,1.1) {$|\underset{{\omega\rightarrow\infty}}{\text{lim}}\,\mathrm{j}\omega D_i^\mathrm{pf}(\mathrm{j}\omega)|\leq\frac{\Delta\dot{f}_\mathrm{max}}{\Delta p_\mathrm{d}^{\tikz[baseline=0] \draw (0.05,0.05) -- (0.15,0.05) -- (0.15,0.15) -- (0.25,0.15);}}$};
\draw [fill=cyan!30,opacity=0.5,draw=none] (34.35,0) ellipse (0.6 and 0.6);
\draw [color=cyan,dotted,dash pattern=on 1pt off 1pt,thick] (34.35,0) ellipse (0.6 and 0.6);
\draw[color=cyan,ultra thin] (34.35,0)  -- (34.65,0.53);
\node [scale=0.7,color=cyan] at (35.4,0.4) {$\frac{\Delta\dot{f}_\mathrm{max}}{\Delta p_\mathrm{d}^{\tikz[baseline=0] \draw (0.05,0.05) -- (0.15,0.05) -- (0.15,0.15) -- (0.25,0.15);}}$};
\node [scale=0.7,anchor=west] at (34.4,2.15) {$\text{Im}[\mathrm{j}\omega D_i^\mathrm{pf}(\mathrm{j}\omega)]$};
\node [scale=0.7,anchor=east] at (36.95,-0.3) {$\text{Re}[\mathrm{j}\omega D_i^\mathrm{pf}(\mathrm{j}\omega)]$};


\end{tikzpicture}  
}
    \vspace{-9mm}
    \caption{Qualitative graphical illustration of the eight frequency-domain conditions \ref{cond:1i}-\ref{cond:1viii} for each $D_i^\mathrm{pf}(\mathrm{j}\omega)$ in the complex plane.}
    \label{fig:nyquist_pf_elements}
\end{figure*}
\begin{figure*}
    \centering
    \vspace{-3mm}
    \resizebox{0.79\textwidth}{!}{
\begin{tikzpicture}[scale=1,every node/.style={scale=1.5}]

\node[scale=0.7,color=red] at (8.45,0.35) {$|D_i^\mathrm{qv}(\mathrm{j}\infty)| = 0$};

\draw [-latex](9.35,-2.05) -- (9.35,2.25);
\draw[-latex] (6.95,0) -- (11.75,0);
\node [scale=0.7,anchor=west] at (9.4,2.15) {$\text{Im}[D_i^\mathrm{qv}(\mathrm{j}\omega)]$};
\node [scale=0.7,anchor=east] at (11.95,-0.3) {$\text{Re}[D_i^\mathrm{qv}(\mathrm{j}\omega)]$};

\node[scale=0.7,color=red] at (9.35,-2.3) {cond (2-i)};



\draw[fill=red,color=red]  (9.35,0) node (v7) {} ellipse (0.06 and 0.06);
\draw [fill=matlab3!30,opacity=0.5,draw=none] (15.55,0) ellipse (1.2 and 1.2);
\draw [color=matlab3,dotted,dash pattern=on 1pt off 1pt,thick] (15.55,0) ellipse (1.2 and 1.2);
\draw[color=matlab3,ultra thin] (15.55,0)  -- (15.75,1.2);
\draw [-latex](14.35,-2.05) -- (14.35,2.25);
\draw[-latex] (11.95,0) -- (16.75,0);
\node [scale=0.7,anchor=west] at (14.4,2.15) {$\text{Im}[D_i^\mathrm{qv}(\mathrm{j}\omega)]$};
\node [scale=0.7,anchor=east] at (16.95,-0.3) {$\text{Re}[D_i^\mathrm{qv}(\mathrm{j}\omega)]$};

\node[scale=0.7,color=matlab3] at (14.35,-2.3) {cond (2-ii)};
\node  [scale=0.7,color=matlab3] at (14.7,-1.5) {$\text{Re}[D_i^\mathrm{qv}(\mathrm{j}\omega)^{-1}]=\tfrac{\text{Re}[D_i^\mathrm{qv}(\mathrm{j}\omega)]}{|D_i^\mathrm{qv}(\mathrm{j}\omega)|^2}>c_i, $};
\node  [scale=0.7,color=matlab3] at (16.4,-2) {$\forall\omega\in[0,\infty)$};

\node [scale=0.7,color=matlab3] at (15.3,0.8) {$\tfrac{1}{2c_i}$};

\node [scale=0.7,color=matlab3] at (16.25,0.3) {$(\tfrac{1}{2c_i},0)$};
\draw [color=matlab3,fill=matlab3] (15.55,0) ellipse (0.05 and 0.05);

\draw[color=matlab4,ultra thin] (19.35,0) node (v4) {} -- (19.1,0.15);
\node[scale=0.7,color=matlab4] at (18.8,0.15) {$\varepsilon_\mathrm{v}$};
\node[scale=0.7,color=matlab4] at (18.1,1.2) {$|D_i^\mathrm{qv}(\mathrm{j}\omega)|\leq \varepsilon_\mathrm{v},$};
\node[scale=0.7,color=matlab4] at (18.55,0.7) {$\forall\omega\geq\omega_\mathrm{bw}$};
\draw [fill=matlab4!30,opacity=0.5,draw=none] (19.35,0) ellipse (0.3 and 0.3);
\draw [color=matlab4,dotted,dash pattern=on 1pt off 1pt,thick] (19.35,0) node (v3) {} ellipse (0.3 and 0.3);
\draw [-latex](19.35,-2.05) -- (19.35,2.25);
\draw[-latex] (16.95,0) -- (21.75,0);
\node [scale=0.7,anchor=west] at (19.4,2.15) {$\text{Im}[D_i^\mathrm{qv}(\mathrm{j}\omega)]$};
\node [scale=0.7,anchor=east] at (21.95,-0.3) {$\text{Re}[D_i^\mathrm{qv}(\mathrm{j}\omega)]$};

\draw [fill=matlab1!20,opacity=0.4,draw=none] (24.35,0) ellipse (1.75 and 1.75);
\draw [color=matlab1,dotted,dash pattern=on 1pt off 1pt,thick] (24.35,0) ellipse (1.75 and 1.75);
\draw [-latex](24.35,-2.05) -- (24.35,2.25);
\draw[-latex] (21.95,0) -- (26.75,0);
\node [scale=0.7,anchor=west] at (24.4,2.15) {$\text{Im}[D_i^\mathrm{qv}(\mathrm{j}\omega)]$};
\node [scale=0.7,anchor=east] at (26.95,-0.3) {$\text{Re}[D_i^\mathrm{qv}(\mathrm{j}\omega)]$};
\node[scale=0.7,color=matlab1] at (24.35,-2.3) {cond (2-iv)};
\draw[color=matlab1,ultra thin] (24.35,0)  -- (24,1.7);
\node [scale=0.7,color=matlab1] at (23.6,0.5) {$\frac{\Delta|v|_\mathrm{max}}{2.5\Delta q_\mathrm{d}^{\tikz[baseline=0] \draw (0.05,0.05) -- (0.15,0.05) -- (0.15,0.15) -- (0.25,0.15);}}$};
\node  [scale=0.7,color=matlab1] at (24.4,-1.05) {$||D_i^\mathrm{qv}(\mathrm{j}\omega)||_\infty\leq \frac{\Delta|v|_\mathrm{max}}{2.5\Delta q_\mathrm{d}^{\tikz[baseline=0] \draw (0.05,0.05) -- (0.15,0.05) -- (0.15,0.15) -- (0.25,0.15);}}$};

\draw [fill=matlab5!20,opacity=0.6,draw=none] (29.35,0) node (v5) {} ellipse (1.1 and 1.1);

\draw [color=matlab5,dotted,dash pattern=on 1pt off 1pt,thick] (29.35,0) node (v5) {} ellipse (1.1 and 1.1);

\draw [-latex](29.35,-2.05) -- (29.35,2.25);
\draw[-latex] (26.95,0) -- (31.75,0);
\node [scale=0.7,anchor=west] at (29.4,2.15) {$\text{Im}[D_i^\mathrm{qv}(\mathrm{j}\omega)]$};
\node [scale=0.7,anchor=east] at (31.95,-0.3) {$\text{Re}[D_i^\mathrm{qv}(\mathrm{j}\omega)]$};
\node[scale=0.7,color=matlab5] at (29.35,-2.3) {cond (2-v)};
\draw[color=matlab5,ultra thin] (29.35,0)  -- (29.55,1.08);
\node [scale=0.7,color=matlab5] at (30.3,0.45) {$\frac{\Delta|v|_\mathrm{ss,\,max}}{\Delta q_\mathrm{d}^{\tikz[baseline=0] \draw (0.05,0.05) -- (0.15,0.05) -- (0.15,0.15) -- (0.25,0.15);}}$};
\node  [scale=0.7,color=matlab5] at (30.3,1.35) {$|D_i^\mathrm{qv}(0)|\leq \frac{\Delta|v|_\mathrm{ss,\,max}}{\Delta q_\mathrm{d}^{\tikz[baseline=0] \draw (0.05,0.05) -- (0.15,0.05) -- (0.15,0.15) -- (0.25,0.15);}}$};

\draw [fill=magenta!30,opacity=0.5,draw=none] (35.35,0) ellipse (1 and 1);

\draw [color=magenta,dotted,dash pattern=on 1pt off 1pt,thick] (35.35,0) ellipse (1 and 1);
\draw[color=magenta,ultra thin] (35.35,0)  -- (35.55,1);
\draw [-latex](34.35,-2.05) -- (34.35,2.25);
\draw[-latex] (31.95,0) -- (36.75,0);
\node [scale=0.7,anchor=west] at (34.4,2.15) {$\text{Im}[D_i^\mathrm{qv}(\mathrm{j}\omega)]$};
\node [scale=0.7,anchor=east] at (36.95,-0.3) {$\text{Re}[D_i^\mathrm{qv}(\mathrm{j}\omega)]$};
\node[scale=0.7,color=magenta] at (34.35,-2.3) {cond (2-vi)};

\node[scale=0.7,color=magenta] at (35,-1.3) {$\text{Re}[D_i^\mathrm{qv}(\mathrm{j}\omega)^{-1}] = \tfrac{\text{Re}[D_i^\mathrm{qv}(\mathrm{j}\omega)]}{|D_i^\mathrm{qv}(\mathrm{j}\omega)|^2}\geq \eta_\mathrm{v},$};
\node[scale=0.7,color=magenta] at (37.1,-1.85) {$\forall\omega\geq0$};

\node [scale=0.7,color=magenta] at (35.2,0.8) {$\tfrac{1}{2\eta_\mathrm{v}}$};

\node [scale=0.7,color=magenta] at (36.1,0.3) {$(\tfrac{1}{2\eta_\mathrm{v}},0)$};
\draw [color=magenta,fill=magenta] (35.35,0) ellipse (0.05 and 0.05);

\node[scale=0.7,color=matlab4] at (19.35,-2.3) {cond (2-iii)};

\end{tikzpicture}  
}
    \vspace{-9mm}
    \caption{Qualitative graphical illustration of the six frequency-domain conditions \ref{cond:2i}-\ref{cond:2vi} for each $D_i^\mathrm{qv}(\mathrm{j}\omega)$ in the complex plane.}
    \label{fig:nyquist_qv_elements}
    \vspace{-4mm}
\end{figure*}
If each device satisfies conditions \ref{cond:1i}, \ref{cond:1ii}, \ref{cond:2i}, and \ref{cond:2ii}, stability of the overall closed-loop system $D \# N$ is ensured. More accurate stability conditions for parametric models derived from the full dynamic network model are provided in \cite{haberle2025decentralized}.

\subsection{Performance Certification Criteria}\label{sec:performance_certificates}
In the previous section, stability conditions were established using the full feedback interconnection to capture inter-node coupling and ensure safety-critical closed-loop stability at each bus. Now, we shift from stability to performance requirements, employing reduced yet accurate representations: the average-mode frequency response for the $\mathrm{pf}$-subsystem (see \cref{sec:avg_mode}) and local decoupled open-loop voltage responses for the $\mathrm{qv}$-subsystem. These performance conditions capture system-level dynamic and steady-state behavior through aggregate, approximate responses to disturbances.

\subsubsection{Frequency Control} 
System operators typically impose performance requirements on the average-mode frequency response of the power grid \cite{ENTSOE_Policy1_2009,ENTSOE_Inertia_RoCoF_2017}, expressed through time-domain metrics after a step disturbance. To evaluate these metrics, we express the average-mode frequency response in \cref{eq:avg_f_response} in the time domain under a predefined worst-case active-power step disturbance $\Delta p_\mathrm{d}^{\sum} (s)=\tfrac{1}{s}\Delta p_\mathrm{d}^{\tikz[baseline=0] \draw (0.05,0.05) -- (0.15,0.05) -- (0.15,0.15) -- (0.25,0.15);}$ (in the global per unit system), as required in grid-code pre-qualification tests\footnote{The value of $\Delta p^{\tikz[baseline=0] \draw (0.05,0.05) -- (0.15,0.05) -- (0.15,0.15) -- (0.25,0.15);}_\mathrm{d}$ is grid-code dependent and typically defined via a reference incident, e.g., a fixed active-power imbalance \cite{ENTSOE_Policy1_2009}.},
\begin{align}\label{eq:time_domain_response_f_avg} \Delta f^{\tikz[baseline=0] \draw (0.05,0.05) -- (0.15,0.05) -- (0.15,0.15) -- (0.25,0.15);}_\mathrm{avg}(t) = \mathcal{L}^{-1}\{D_\mathrm{avg}(s) \tfrac{1}{s} \Delta p_\mathrm{d}^{{\tikz[baseline=0] \draw (0.05,0.05) -- (0.15,0.05) -- (0.15,0.15) -- (0.25,0.15);}} \}, \end{align}
where $\mathcal{L}^{-1}$ is the inverse Laplace transform. The time-domain response \cref{eq:time_domain_response_f_avg} is assessed using standard operational metrics requiring bounded nadir $\Delta f_\mathrm{max}$, steady-state deviation $\Delta f_\mathrm{ss,\,max}$, RoCoF $\Delta\dot{f}_\mathrm{max}$, and adequate oscillation damping, i.e.,\\[-0.5cm]
\begin{subequations}\label{eq:global_f_avg_specs_rigorous}
\begin{align}\label{eq:nadir_spec}
\sup_{t > 0} |\Delta f^{\tikz[baseline=0] \draw (0.05,0.05) -- (0.15,0.05) -- (0.15,0.15) -- (0.25,0.15);}_\mathrm{avg}(t)| &\leq \Delta f_\mathrm{max},\\[-1mm]\label{eq:ss_spec}
|\lim_{t \to \infty} \Delta f^{\tikz[baseline=0] \draw (0.05,0.05) -- (0.15,0.05) -- (0.15,0.15) -- (0.25,0.15);}_\mathrm{avg}(t)| &\leq \Delta f_\mathrm{ss,\,max},\\[-1mm]\label{eq:rocof_spec}
\sup_{t > 0} |\Delta \dot{f}^{\tikz[baseline=0] \draw (0.05,0.05) -- (0.15,0.05) -- (0.15,0.15) -- (0.25,0.15);}_\mathrm{avg}(t)| &\leq \Delta \dot{f}_\mathrm{max},\\[-1mm]\label{eq:osci_spec}
\Delta f^{\tikz[baseline=0] \draw (0.05,0.05) -- (0.15,0.05) -- (0.15,0.15) -- (0.25,0.15);}_\mathrm{avg}(t)
\text{ exhibits sufficiently}& \text{ damped oscillations.}
\end{align}
\end{subequations}
According to the current ENTSO-E and EU regulation framework \cite{ENTSOE_Policy1_2009,ENTSOE_RoCoF_IGD_2018,european2016commission}, typical limit values are $\Delta f_\mathrm{max} = 0.8\,-\,1\,\text{Hz}$, $\Delta f_\mathrm{ss,\,max}= 0.2\,\text{Hz}$, and $\Delta\dot{f}_\mathrm{max}= 2\,-\,2.5\,\text{Hz/s}$. The oscillation-damping requirement reflects standard regulation practice, providing a qualitative assurance that frequency deviations remain well-damped without sustained oscillatory behavior. While these values serve as representative benchmarks, the proposed framework is not tied to any specific regulatory setting and can accommodate arbitrary limit choices.

The \emph{global performance specifications} in \cref{eq:global_f_avg_specs_rigorous} are ensured if each device satisfies the following frequency-domain conditions in addition to conditions \ref{cond:1i} and \ref{cond:1ii}:
\begin{align}
    \tag*{{\color{orange}{(1-iii)}}}\label{cond:1iii}
    & {\color{orange}\angle (D_i^{\mathrm{pf}}(\mathrm{j}\omega))\in [-\tfrac{\pi}{2},\,\tfrac{\pi}{6}],\quad\forall\omega \geq 0,}\\[-1mm]
    \tag*{{\color{matlab4}(1-iv)}}\label{cond:1iv}
    &{\color{matlab4}|D_i^{\mathrm{pf}}(\mathrm{j}\omega)| \le \varepsilon_\mathrm{f}, \quad \forall \omega \ge \omega_{\mathrm{bw}},}\\[-1mm]
    \tag*{{\color{matlab1}(1-v)}}\label{cond:1v}
    &{\color{matlab1} ||D_i^\mathrm{pf}(\mathrm{j}\omega)||_\infty \leq \tfrac{\Delta f_\mathrm{max}}{2.5\Delta p_\mathrm{d}^{\tikz[baseline=0] \draw (0.05,0.05) -- (0.15,0.05) -- (0.15,0.15) -- (0.25,0.15);}},}\\[-1mm]
    \tag*{{\color{matlab5}(1-vi)}}\label{cond:1vi}
    &{\color{matlab5} |D_i^\mathrm{pf}(0)| \leq \tfrac{\Delta f_\mathrm{ss,\,max}}{\Delta p_\mathrm{d}^{\tikz[baseline=0] \draw (0.05,0.05) -- (0.15,0.05) -- (0.15,0.15) -- (0.25,0.15);}},}\\[-1mm]
    \tag*{{\color{cyan}(1-vii)}}\label{cond:1vii}
    &{\color{cyan}| \lim_{\omega\rightarrow \infty} \mathrm{j}\omega D_i^\mathrm{pf}(\mathrm{j}\omega)| \leq \tfrac{\Delta \dot{f}_\mathrm{max}}{\Delta p_\mathrm{d}^{\tikz[baseline=0] \draw (0.05,0.05) -- (0.15,0.05) -- (0.15,0.15) -- (0.25,0.15);}},}\\[-1mm]
    \tag*{{\color{magenta}(1-viii)}}\label{cond:1viii}
    &{\color{magenta}\text{Re}[D_i^\mathrm{pf}(\mathrm{j}\omega)^{-1}]\geq \eta_\mathrm{f},\quad\forall\omega \geq 0.}
\end{align}
Here, $\varepsilon_{\mathrm{f}} <<1$ denotes a sufficiently small bound that ensures proper roll-off for all $\omega \geq \omega_{\mathrm{bw}}$, where $\omega_{\mathrm{bw}}$ is a predefined control bandwidth, typically specified in certain grid codes at around $\omega_{\mathrm{bw}} \approx 5\,\mathrm{Hz}$ \cite{NationalGrid2021}. The constant $\eta_{\mathrm{f}}$ denotes the output feedback passivity index of $D_i^{\mathrm{pf}}(\mathrm{j}\omega)$ \cite{peng2026positive}, which must be chosen sufficiently large to ensure adequate oscillation damping. A graphical illustration of the eight frequency-domain conditions \ref{cond:1i}-\ref{cond:1viii} for each $D_i^\mathrm{pf}(\mathrm{j}\omega)$ is provided in \cref{fig:nyquist_pf_elements}.

Showing that the decentralized conditions in \ref{cond:1i}-\ref{cond:1viii} imply the global performance specifications in \cref{eq:global_f_avg_specs_rigorous} is nontrivial. Specifically, the frequency-nadir bound in \cref{eq:nadir_spec} follows from conditions \ref{cond:1iii}, \ref{cond:1iv}, \ref{cond:1v}; the steady-state bound in \cref{eq:ss_spec} from conditions \ref{cond:1i}, \ref{cond:1iii}, \ref{cond:1vi}; and the RoCoF bound in \cref{eq:rocof_spec} from conditions \ref{cond:1i}, \ref{cond:1iii}, \ref{cond:1vii}. Finally, sufficient oscillation damping in \cref{eq:osci_spec} is guaranteed by condition \ref{cond:1viii}.

The rigorous proof proceeds in two main steps: First, it can be shown (see Appendix \ref{appendix:performance_pf_part1} for the detailed proof) how conditions \ref{cond:1i}-\ref{cond:1viii} ensure the resulting average-mode frequency dynamics $D_\mathrm{avg}(s)$ comes with the following properties\\[-0.45cm]
\begin{subequations}\label{eq:D_avg_characteristics}
\begin{align}\label{eq:avg_1}
    & D_\mathrm{avg}(s)\ \text{is strictly proper and stable},\\[-1mm]\label{eq:avg_2}
    & \mathrm{Re}[D_\mathrm{avg}(\mathrm{j}\omega)] \geq 0,\quad\forall\omega \geq 0,\\[-1mm]\label{eq:avg_3}
    &|D_\mathrm{avg}(\mathrm{j}\omega)| \le \varepsilon_\mathrm{f}, \quad \forall \omega \ge \omega_{\mathrm{bw}},\\[-1mm]\label{eq:avg_4}
    & ||D_\mathrm{avg}(\mathrm{j}\omega)||_\infty \leq \tfrac{\Delta f_\mathrm{max}}{2.5\Delta p_\mathrm{d}^{\tikz[baseline=0] \draw (0.05,0.05) -- (0.15,0.05) -- (0.15,0.15) -- (0.25,0.15);}},\\[-1mm]\label{eq:avg_5}
    & |D_\mathrm{avg}(0)| \leq \tfrac{\Delta f_\mathrm{ss,\,max}}{\Delta p_\mathrm{d}^{\tikz[baseline=0] \draw (0.05,0.05) -- (0.15,0.05) -- (0.15,0.15) -- (0.25,0.15);}},\\[-1mm]\label{eq:avg_6}
    & |\lim_{\omega\rightarrow \infty} \mathrm{j}\omega D_\mathrm{avg}(\mathrm{j}\omega)| \leq \tfrac{\Delta \dot{f}_\mathrm{max}}{\Delta p_\mathrm{d}^{\tikz[baseline=0] \draw (0.05,0.05) -- (0.15,0.05) -- (0.15,0.15) -- (0.25,0.15);}} ,\\[-1mm]\label{eq:avg_7}
    &\text{Re}[D_\mathrm{avg}(\mathrm{j}\omega)^{-1}]\geq n\eta_\mathrm{f},\quad\forall\omega \geq 0.
\end{align}
\end{subequations}

Second, given that $D_\mathrm{avg}(s)$ satisfies the properties in \cref{eq:D_avg_characteristics}, the global performance specifications of bounded nadir, steady-state deviation, RoCoF, and oscillation damping in \cref{eq:global_f_avg_specs_rigorous} can be derived (the detailed proof can be found in Appendix \ref{appendix:performance_pf_part2}). 

\begin{remark}
The properties \cref{eq:avg_3} -\cref{eq:avg_6} follow from the bound\\[-0.35cm]
\begin{align*}
\begin{split}
|D_\mathrm{avg}(\mathrm{j}\omega)|&=\left| \left(\textstyle\sum_{i=1}^nD_i^\mathrm{pf}(\mathrm{j}\omega)^{-1} \right)\right|\\[-0.7mm]
&\leq{\hspace{-0.1mm} \tfrac{1}{2}\hspace{-0.1mm}\left( \textstyle\sum_{i=2}^n(\tfrac{1}{2})^{n-i}|D_i^\mathrm{pf}(\mathrm{j}\omega)|\hspace{-0.5mm}+\hspace{-0.5mm}(\tfrac{1}{2})^{n-2}|D_1^\mathrm{pf}(\mathrm{j}\omega)|\right)}\hspace{-0.5mm}.\hspace{-2.5mm}
\end{split}
\end{align*}
given in \cref{eq:D_avg_magnitude_upper_bound} (Appendix~\ref{appendix:performance_pf_part1}). The bound is conservative (particularly in steady state) and could be tightened for \cref{eq:avg_5} by directly summing all $D_i^\mathrm{pf}(0)^{-1}>0$; however, it is retained for consistency across all properties.
\end{remark}

\begin{remark}
\fontdimen2\font=0.6ex As a direct consequence of the conditions in \ref{cond:1i}-\ref{cond:1viii}, the individual bus frequency responses (i.e., the open-loop mapping from $\Delta p_i$ to $\Delta \omega_i$ shown in the bottom-left of \cref{fig:power_sys_model_orig}) also exhibit a bounded nadir, RoCoF, steady-state deviation, and oscillation damping capability under worst-case active power disturbances. This follows from a proof analogous to that in Appendix \ref{appendix:performance_pf_part2}, with $D_\mathrm{avg}(s)$ replaced by $D_i^\mathrm{pf}(s)$.  
\end{remark}

\begin{remark}
    \fontdimen2\font=0.6ex The damping indicated by the output passivity index is only qualitative and cannot generally be mapped to a specific damping coefficient. Explicit damping-ratio bounds are available only for special cases, such as second-order systems.
\end{remark}

\subsubsection{Voltage Control}

In contrast to the $\mathrm{pf}$-subsystem, which exhibits an average-mode frequency response, the $\mathrm{qv}$-subsystem admits no analogous average-mode voltage response. Instead, consistent with standard power-system insight, the voltage magnitude deviation $\Delta |v|_i$ at each bus is predominantly influenced by the local reactive power disturbance $\Delta q_{\mathrm{d},i}$, i.e., 
\begin{align}\label{eq:local_v_approx}
    \Delta |v|_i(s) \approx D_i^\mathrm{qv}(s)\,\Delta q_{\mathrm{d},i}(s), \quad \forall\, i \in \{1,\ldots,n\}.
\end{align}
This motivates imposing performance conditions directly at the device level, such that the locally approximated\footnote{This local open-loop approximation is particularly accurate when $\lvert D_i^{\mathrm{qv}}(s)N^{\mathrm{vq}}\rvert < 1$ holds over the frequency range of interest.} open-loop voltage responses satisfy prescribed performance specifications.

More specifically, for a predefined worst-case reactive-power step disturbance $\Delta q_{\mathrm{d},i}(s)=\tfrac{1}{s}\Delta q_\mathrm{d}^{{\tikz[baseline=0] \draw (0.05,0.05) -- (0.15,0.05) -- (0.15,0.15) -- (0.25,0.15);}}$ (in the global per unit system), as required in the corresponding grid-code pre-qualification test, the resulting time-domain response of the approximated bus voltage at bus $i$ in \cref{eq:local_v_approx} is given by
\begin{align}
    \Delta |v|^{\tikz[baseline=0] \draw (0.05,0.05) -- (0.15,0.05) -- (0.15,0.15) -- (0.25,0.15);}_i(t) = \mathcal{L}^{-1}\{D_i^\mathrm{qv} (s)\tfrac{1}{s} \Delta q_\mathrm{d}^{{\tikz[baseline=0] \draw (0.05,0.05) -- (0.15,0.05) -- (0.15,0.15) -- (0.25,0.15);}}\}.
\end{align}
We require that each bus voltage response satisfies bounds on the maximum and the steady-state deviation, and comes with sufficient oscillation damping, i.e.,
\begin{subequations}\label{eq:local_volt_spec_rigorous}
\begin{align}
    \sup_{t > 0} |\Delta |v|^{\tikz[baseline=0] \draw (0.05,0.05) -- (0.15,0.05) -- (0.15,0.15) -- (0.25,0.15);}_i(t)| &\leq \Delta |v|_{\mathrm{max}},\\[-1mm]
|\lim_{t \to \infty} \Delta |v|^{\tikz[baseline=0] \draw (0.05,0.05) -- (0.15,0.05) -- (0.15,0.15) -- (0.25,0.15);}_i(t)| &\leq \Delta |v|_\mathrm{ss,\,max},\\[-1mm]
\Delta |v|^{\tikz[baseline=0] \draw (0.05,0.05) -- (0.15,0.05) -- (0.15,0.15) -- (0.25,0.15);}_i(t)
\text{ exhibits sufficiently}& \text{ damped oscillations,}
\end{align}
\end{subequations}
where $\Delta |v|_{\mathrm{max}}$ denotes the maximum acceptable voltage peak and $\Delta |v|_\mathrm{ss,\,max}$ the maximum acceptable steady-state voltage magnitude deviation from the current operating point. ENTSO-E and related regulatory frameworks specify absolute steady-state voltage limits of $0.9$-$1.1~\text{pu}$ \cite{european2016commission}. Consequently, the permissible deviations $\Delta |v|_{\mathrm{ss,max}}$ and $\Delta |v|_{\mathrm{max}}$ depend on the operating point $|v|_{0,i}$, and should either be adapted accordingly or chosen sufficiently small to prevent any violation of the absolute limits.

\begin{figure}[t!]
    \centering
    \vspace{-2mm}
    \resizebox{0.5\textwidth}{!}{
\begin{tikzpicture}[scale=1,every node/.style={scale=0.6}]
\draw [fill=matlab3!30,color=matlab3!30,opacity=0.5] (-0.75,2.1) node (v6) {} rectangle (1.4,-2.05) node (v1) {};
\draw[fill=orange!20, color=orange!20,opacity = 0.6]  (-0.8,0) node (v2) {} rectangle  (1.4,-2.05);

\draw[fill=orange!20, color=orange!20,opacity = 0.6] (-0.8,0) -- (1.4,1.5) node (v8) {} -- (1.4,0) --(-0.8,0);

\draw [fill=matlab1!20,opacity=0.4,draw=none] (-0.8,0) ellipse (1.75 and 1.75);
\draw [fill=matlab5!20,opacity=0.6,draw=none] (-0.8,0) node (v5) {} ellipse (1.1 and 1.1);
\draw [fill=magenta!30,opacity=0.4,draw=none] (0.2,0) ellipse (1 and 1);
\draw [fill=matlab4!30,opacity=0.5,draw=none] (-0.8,0) node (v3) {} ellipse (0.3 and 0.3);

\draw [-latex](-0.8,-2.05) -- (-0.8,2.25);
\draw[-latex] (-3.2,0) -- (1.6,0);
\node [scale=0.8]at (-0.25,2.25) {$\text{Im}[D_i^\mathrm{pf}(\mathrm{j}\omega)]$};
\node [scale=0.8]at (1.5,-0.2) {$\text{Re}[D_i^\mathrm{pf}(\mathrm{j}\omega)]$};
\node[scale=0.6,color=red,     anchor=west] at (-3.1,-1.075) {cond (1-i)};
\node[scale=0.6,color=matlab3, anchor=west] at (-3.1,-1.2)   {cond (1-ii)};
\node[scale=0.6,color=orange,  anchor=west] at (-3.1,-1.325) {cond (1-iii)};
\node[scale=0.6,color=matlab4, anchor=west] at (-3.1,-1.45)  {cond (1-iv)};
\node[scale=0.6,color=matlab1,anchor=west] at (-3.1,-1.575) {cond (1-v)};
\node[scale=0.6,color=matlab5,anchor=west] at (-3.1,-1.7)   {cond (1-vi)};
\node[scale=0.6,color=cyan,anchor=west] at (-3.1,-1.825) {cond (1-vii)};
\node[scale=0.6,color=magenta, anchor=west] at (-3.1,-1.95)  {cond (1-viii)};

\node[scale=0.6,color=red] at (-1.25,-0.12) {$|D_i^\mathrm{pf}(\mathrm{j}\infty)| = 0$};

\node [scale=0.6,color=matlab3] at (0.5,1.9) {$\text{Re}[D_i^\mathrm{pf}(\mathrm{j}\omega)]>0,\,\forall \omega\in[0,\infty)$};

\node  [scale=0.6,color=orange] at (0.45,-1.8) {$\angle (D_i^\mathrm{pf}(\mathrm{j}\omega))\in[-\frac{\pi}{2},\frac{\pi}{6}],\,\forall\omega\geq 0$};

\draw[color=matlab4,ultra thin, opacity = 0.5] (-0.8,0) node (v4) {} -- (-1.05,0.15);
\node[scale=0.6,color=matlab4] at (-0.9,0.15) {$\varepsilon_\mathrm{f}$};
\node[scale=0.6,color=matlab4] at (-1.6,0.37) {$|D_i^\mathrm{pf}(\mathrm{j}\omega)|\leq \varepsilon_\mathrm{f},\,\forall\omega\geq\omega_\mathrm{bw}$};

\draw[color=matlab1,ultra thin,opacity = 0.5] (-0.8,0)  -- (-1.15,1.7);
\node [scale=0.6,color=matlab1] at (-1.3,1.45) {$\frac{\Delta f_\mathrm{max}}{2.5\Delta p_\mathrm{d}^{\tikz[baseline=0] \draw (0.05,0.05) -- (0.15,0.05) -- (0.15,0.15) -- (0.25,0.15);}}$};
\node  [scale=0.6,color=matlab1] at (-1.75,1.9) {$||D_i^\mathrm{pf}(\mathrm{j}\omega)||_\infty\leq \frac{\Delta f_\mathrm{max}}{2.5\Delta p_\mathrm{d}^{\tikz[baseline=0] \draw (0.05,0.05) -- (0.15,0.05) -- (0.15,0.15) -- (0.25,0.15);}}$};

\draw[color=matlab5,ultra thin,opacity=0.5] (-0.8,0)  -- (-0.6,1.08);
\node [scale=0.6,color=matlab5] at (-0.45,0.6) {$\frac{\Delta f_\mathrm{ss,\,max}}{\Delta p_\mathrm{d}^{\tikz[baseline=0] \draw (0.05,0.05) -- (0.15,0.05) -- (0.15,0.15) -- (0.25,0.15);}}$};

\node  [scale=0.6,color=matlab5] at (-0.2,1.2) {$|D_i^\mathrm{pf}(0)|\leq \frac{\Delta f_\mathrm{ss,\,max}}{\Delta p_\mathrm{d}^{\tikz[baseline=0] \draw (0.05,0.05) -- (0.15,0.05) -- (0.15,0.15) -- (0.25,0.15);}}$};


\draw [color=magenta,fill=magenta] (0.2,0) ellipse (0.02 and 0.02);

\node [scale=0.6,color=cyan] at (-1.55,-1.9) {$\&|\underset{{\omega\rightarrow\infty}}{\text{lim}}\mathrm{j}\omega D_i^\mathrm{pf}(\mathrm{j}\omega)|\hspace{-0.7mm}\leq\hspace{-0.7mm}\frac{\Delta\dot{f}_\mathrm{max}}{\Delta p_\mathrm{d}^{\tikz[baseline=0] \draw (0.05,0.05) -- (0.15,0.05) -- (0.15,0.15) -- (0.25,0.15);}}$};

\draw [ultra thin] (-3.15,-0.95) rectangle (-2.35,-2.05);

\draw [fill=black!40,opacity=0.4,draw=none] plot[smooth, tension=.7] coordinates { (-0.8,0) (-0.7,0.08) (-0.46,0.24) (-0.06,0.52) (0.25,0.72) (0.51,0.9)(0.56,0.94) (0.62,0.92) (0.68,0.89) (0.73,0.85) (0.75,0.81) (0.81,0.69) (0.88,0.48) (0.95,0.18) (0.95,-0.08) (0.92,-0.32) (0.83,-0.63) (0.77,-0.78) (0.73,-0.85) (0.67,-0.89) (0.49,-0.95) (0.14,-0.99) (-0.18,-0.92) (-0.48,-0.71) (-0.66,-0.48) (-0.75,-0.28) (-0.79,-0.12) (-0.79,0)};

\draw [pattern=dots, pattern color=black!40,draw=none] plot[smooth, tension=.7] coordinates { (-0.8,0) (-0.7,0.08) (-0.46,0.24) (-0.06,0.52) (0.25,0.72) (0.51,0.9)(0.56,0.94) (0.62,0.92) (0.68,0.89) (0.73,0.85) (0.75,0.81) (0.81,0.69) (0.88,0.48) (0.95,0.18) (0.95,-0.08) (0.92,-0.32) (0.83,-0.63) (0.77,-0.78) (0.73,-0.85) (0.67,-0.89) (0.49,-0.95) (0.14,-0.99) (-0.18,-0.92) (-0.48,-0.71) (-0.66,-0.48) (-0.75,-0.28) (-0.79,-0.12) (-0.79,0)};

\node[scale=0.6,color=magenta] at (0.45,-1.15) {$\text{Re}[D_i^\mathrm{pf}(\mathrm{j}\omega)^{-1}]\geq \eta_\mathrm{f},\,\forall\omega\geq0$};

\node [scale=0.6,color=magenta]at (0.5,0.1) {$(\frac{1}{2\eta_\mathrm{f}},0)$};
\node [scale=0.6,color=magenta]at (0.55,0.7) {$\frac{1}{2\eta_\mathrm{f}}$};
\draw[color=magenta,ultra thin,opacity=0.5] (0.2,0)  -- (0.5,0.97);
\node [scale=0.6,color=orange] at (1.25,1.25) {$\tfrac{\pi}{6}$};

\draw [dotted,dash pattern=on 1pt off 1pt,color=matlab3] (-0.75,2.1) -- (-0.75,-2.05);

\draw [color=matlab5,dotted,dash pattern=on 1pt off 1pt] (-0.8,0) node (v5) {} ellipse (1.1 and 1.1);

\draw [color=matlab4,dotted,dash pattern=on 1pt off 1pt] (-0.8,0) node (v3) {} ellipse (0.3 and 0.3);

\draw [color=matlab1,dotted,dash pattern=on 1pt off 1pt] (-0.8,0) ellipse (1.75 and 1.75);

\draw[color=orange,dotted,dash pattern=on 1pt off 1pt](-0.8,0) node (v10) {} --(1.4,1.5);

\draw [color=magenta,dotted,dash pattern=on 1pt off 1pt] (0.2,0) ellipse (1 and 1);

\draw[
  black,
  line width=0.4pt,
  line cap=round,
  line join=round,
  use Hobby shortcut,
  tension=1.2
]
(0.05,0)
.. (0.2,0.3)
.. (0.5,0.4)
.. (0.75,0.25)
.. (0.85,0.05)
.. (0.8,-0.2)
.. (0.55,-0.5)
.. (-0.05,-0.7)
.. (-0.5,-0.52)
.. (-0.65,-0.32)
.. (-0.737,-0.16)
.. (-0.8,0);

\node [scale=0.6]at (0.05,-0.1) {$\omega=0$};

\draw [fill=black] (0.05,0) ellipse (0.02 and 0.02);

\draw [fill=black] (-0.675,-0.27) ellipse (0.02 and 0.02);
\node  [scale=0.6] at (-0.37,-0.27) {$\omega=\omega_\mathrm{bw}$};
\node [scale=0.6,color=red] at (-0.5,-0.07) {$\omega=\infty$};
\draw[fill=red,color=red]  (-0.8,0) ellipse (0.02 and 0.02);


\end{tikzpicture}  
}
    \vspace{-9mm}
    \caption{Qualitative graphical illustration of the superimposed NGGC 1 conditions \ref{cond:1i}-\ref{cond:1viii} in \cref{fig:nyquist_pf_elements}. The gray shaded region indicates the feasible set for all $\omega\in(0,\omega_\mathrm{bw})$, and the black curve depicts an exemplary feasible Nyquist plot of $D_i^\mathrm{pf}(\mathrm{j}\omega)$ that satisfies all conditions simultaneously.}
    \label{fig:nyquist_pf}
    \vspace{-4mm}
\end{figure}

These \emph{local performance requirements} are guaranteed if each device locally satisfies the following frequency-domain conditions in addition to conditions \ref{cond:2i} and \ref{cond:2ii}:
\begin{align}
    \tag*{{\color{matlab4}(2-iii)}}\label{cond:2iii}
    &{\color{matlab4}|D_i^{\mathrm{qv}}(\mathrm{j}\omega)| \le \varepsilon_\mathrm{v}, \quad \forall \omega \ge \omega_{\mathrm{bw}},}\\[-1mm]
    \tag*{{\color{matlab1}(2-iv)}}\label{cond:2iv}
    & {\color{matlab1}||D_i^\mathrm{qv}(\mathrm{j}\omega)||_\infty \leq \tfrac{\Delta |v|_\mathrm{max}}{2.5\Delta q_\mathrm{d}^{\tikz[baseline=0] \draw (0.05,0.05) -- (0.15,0.05) -- (0.15,0.15) -- (0.25,0.15);}},}\\[-1mm]
    \tag*{{\color{matlab5}(2-v)}}\label{cond:2v}
    &{\color{matlab5}|D_i^\mathrm{qv}(0)| \leq \tfrac{\Delta |v|_\mathrm{ss,\,max}}{\Delta q_\mathrm{d}^{\tikz[baseline=0] \draw (0.05,0.05) -- (0.15,0.05) -- (0.15,0.15) -- (0.25,0.15);}},}\\[-1mm]
    \tag*{{\color{magenta}(2-vi)}}\label{cond:2vi}&{\color{magenta}\text{Re}[D_i^\mathrm{qv}(\mathrm{j}\omega)^{-1}]\geq \eta_\mathrm{v},\quad\forall\omega \geq 0.}
\end{align}
Here, $\varepsilon_{\mathrm{v}} <<1$ denotes a sufficiently small bound that ensures proper roll-off for all $\omega \geq \omega_{\mathrm{bw}}$, where $\omega_{\mathrm{bw}}$ is a predefined control bandwidth, typically specified in certain grid codes at around $\omega_{\mathrm{bw}} \approx 5\,\mathrm{Hz}$ \cite{NationalGrid2021}. The constant $\eta_{\mathrm{v}}$ denotes the output feedback passivity index of $D_i^{\mathrm{qv}}(\mathrm{j}\omega)$, which must be chosen sufficiently large to ensure adequate oscillation damping. A graphical illustration of the six frequency-domain conditions \ref{cond:2i}-\ref{cond:2vi} for each $D_i^\mathrm{qv}(\mathrm{j}\omega)$ is provided in \cref{fig:nyquist_qv_elements}.

The conditions closely mirror those for the $\mathrm{pf}$-subsystem in \ref{cond:1iv}, \ref{cond:1v}, \ref{cond:1vi} and \ref{cond:1viii}, and the proof follows the same steps as in Appendix \ref{appendix:performance_pf_part2}, with $D_\mathrm{avg}(s)$ replaced by $D_i^\mathrm{qv}(s)$.

\section{Next Generation Grid Code}\label{sec:NGGC}
Based on the preceding stability and performance criteria, we propose the following next-generation grid code (NGGC) specifications for fast, reliable frequency and voltage regulation.
\begin{gridcode}[Frequency Regulation]
    \fontdimen2\font=0.6ex The dynamic mapping of a generation unit $i$ from the local active power measurement deviation $\Delta p_i$ to the local frequency deviation $\Delta f_i$ shall be implemented via a transfer function $D_i^\mathrm{pf}(s)$ necessarily satisfying the frequency-domain conditions in \ref{cond:1i}-\ref{cond:1viii}.
\end{gridcode}

\begin{figure}[t!]
    \centering
    \vspace{-2mm}
    \resizebox{0.5\textwidth}{!}{
\begin{tikzpicture}[scale=1,every node/.style={scale=0.6}]

\draw [fill=matlab1!20,opacity=0.4,draw=none] (-0.8,0) ellipse (1.75 and 1.75);
\draw [fill=matlab5!20,opacity=0.6,draw=none] (-0.8,0) node (v5) {} ellipse (1.1 and 1.1);
\draw [fill=matlab3!30,opacity=0.5,draw=none] (0.25,0) ellipse (1.05 and 1.05);
\draw [fill=magenta!30,opacity=0.4,draw=none] (0.15,0) ellipse (0.95 and 0.95);
\draw [fill=matlab4!30,opacity=0.5,draw=none] (-0.8,0) node (v3) {} ellipse (0.3 and 0.3);
\draw [-latex](-0.8,-2.05) -- (-0.8,2.25);
\draw[-latex] (-3.2,0) -- (1.6,0);
\node [scale=0.8]at (-0.25,2.25) {$\text{Im}[D_i^\mathrm{qv}(\mathrm{j}\omega)]$};
\node [scale=0.8]at (1.5,-0.2) {$\text{Re}[D_i^\mathrm{qv}(\mathrm{j}\omega)]$};
\node[scale=0.6,color=red,     anchor=west] at (-3.1,-1.325) {cond (2-i)};
\node[scale=0.6,color=matlab3,  anchor=west] at (-3.1,-1.45) {cond (2-ii)};
\node[scale=0.6,color=matlab4, anchor=west] at (-3.1,-1.575)  {cond (2-iii)};
\node[scale=0.6,color=matlab1, anchor=west] at (-3.1,-1.7) {cond (2-iv)};
\node[scale=0.6,color=matlab5, anchor=west] at (-3.1,-1.825)   {cond (2-v)};
\node[scale=0.6,color=magenta, anchor=west] at (-3.1,-1.95)  {cond (2-vi)};

\node[scale=0.6,color=red] at (-1.25,-0.12) {$|D_i^\mathrm{qv}(\mathrm{j}\infty)| = 0$};

\node  [scale=0.6,color=matlab3] at (0.75,-1.18) {$\text{Re}[D_i^\mathrm{qv}(\mathrm{j}\omega)^{-1}]>c_i,\,\forall\omega\in[0,\infty) $};


\draw[color=matlab4,ultra thin, opacity = 0.5] (-0.8,0) node (v4) {} -- (-1.05,0.15);
\node[scale=0.6,color=matlab4] at (-0.9,0.15) {$\varepsilon_\mathrm{v}$};
\node[scale=0.6,color=matlab4] at (-1.73,0.3) {$|D_i^\mathrm{qv}(\mathrm{j}\omega)|\leq \varepsilon_\mathrm{v}, \forall \omega\geq\omega_\mathrm{bw}$};

\draw[color=matlab1,ultra thin,opacity = 0.5] (-0.8,0)  -- (-1.15,1.7);
\node [scale=0.6,color=matlab1] at (-1.3,1.45) {$\frac{\Delta|v|_\mathrm{max}}{2.5\Delta q_\mathrm{d}^{\tikz[baseline=0] \draw (0.05,0.05) -- (0.15,0.05) -- (0.15,0.15) -- (0.25,0.15);}}$};
\node  [scale=0.6,color=matlab1] at (-1.75,1.9) {$||D_i^\mathrm{qv}(\mathrm{j}\omega)||_\infty\leq \frac{\Delta|v|_\mathrm{max}}{2.5\Delta q_\mathrm{d}^{\tikz[baseline=0] \draw (0.05,0.05) -- (0.15,0.05) -- (0.15,0.15) -- (0.25,0.15);}}$};

\draw[color=matlab5,ultra thin,opacity=0.5] (-0.8,0)  -- (-0.6,1.08);

\node  [scale=0.6,color=matlab5] at (-0.2,1.2) {$|D_i^\mathrm{qv}(0)|\leq \frac{\Delta|v|_\mathrm{ss,\,max}}{\Delta q_\mathrm{d}^{\tikz[baseline=0] \draw (0.05,0.05) -- (0.15,0.05) -- (0.15,0.15) -- (0.25,0.15);}}$};


\draw [ultra thin] (-3.15,-1.2) rectangle (-2.35,-2.05);


\draw [fill=magenta,color=magenta] (0.15,0) ellipse (0.02 and 0.02);

\draw [color=black!40,fill=black!40,opacity=0.4,draw=none]plot[smooth, tension=.7] coordinates { (-0.8,0)(-0.76,0.28) (-0.57,0.62) (-0.25,0.86) (0.13,0.95)(0.51,0.88) (0.72,0.77) (0.79,0.71) (0.86,0.55) (0.93,0.22) (0.94,-0.21) (0.86,-0.55) (0.82,-0.67) (0.74,-0.74) (0.52,-0.87) (0.12,-0.95) (-0.32,-0.83) (-0.6,-0.57) (-0.77,-0.24) (-0.8,0.03)};

\draw [pattern=dots, pattern color=black!40,draw=none] plot[smooth, tension=.7] coordinates { (-0.8,0)(-0.76,0.28) (-0.57,0.62) (-0.25,0.86) (0.13,0.95)(0.51,0.88) (0.72,0.77) (0.79,0.71) (0.86,0.55) (0.93,0.22) (0.94,-0.21) (0.86,-0.55) (0.82,-0.67) (0.74,-0.74) (0.52,-0.87) (0.12,-0.95) (-0.32,-0.83) (-0.6,-0.57) (-0.77,-0.24) (-0.8,0.03)};
\node [scale=0.6,color=matlab5] at (-0.95,0.7) {$\frac{\Delta|v|_\mathrm{ss,\,max}}{\Delta q_\mathrm{d}^{\tikz[baseline=0] \draw (0.05,0.05) -- (0.15,0.05) -- (0.15,0.15) -- (0.25,0.15);}}$};

\draw [color=matlab3,fill=matlab3] (0.25,0) ellipse (0.02 and 0.02);
\node [scale=0.6,color=matlab3!100]at (0.55,0.1) {$(\frac{1}{2c_i},0)$};
\node [scale=0.6,color=matlab3!100]at (0.75,1.05) {$\frac{1}{2c_i}$};
\draw[color=matlab3!120,ultra thin,opacity=0.8] (0.25,0)  -- (0.72,0.94);

\node[scale=0.6,color=magenta] at (-1.25,-0.85) {$\text{Re}[D_i^\mathrm{qv}(\mathrm{j}\omega)^{-1}]\geq \eta_\mathrm{v},\,\forall\omega\geq 0$};
\node [scale=0.6,color=magenta]at (-0.15,0.1) {$(\frac{1}{2\eta_\mathrm{v}},0)$};
\node [scale=0.6,color=magenta]at (0.4,0.6) {$\frac{1}{2\eta_\mathrm{v}}$};
\draw[color=magenta,ultra thin,opacity=0.5] (0.15,0)  -- (0.39,0.92);

\draw [color=matlab3,dotted,dash pattern=on 1pt off 1pt] (0.25,0) ellipse (1.05 and 1.05);

\draw [color=matlab4,dotted,dash pattern=on 1pt off 1pt] (-0.8,0) node (v3) {} ellipse (0.3 and 0.3);

\draw [color=matlab5,dotted,dash pattern=on 1pt off 1pt] (-0.8,0) node (v5) {} ellipse (1.1 and 1.1);

\draw [color=matlab1,dotted,dash pattern=on 1pt off 1pt] (-0.8,0) ellipse (1.75 and 1.75);
\draw [color=magenta,dotted,dash pattern=on 1pt off 1pt] (0.15,0) ellipse (0.95 and 0.95);
\draw[
  black,
  line width=0.4pt,
  line cap=round,
  line join=round,
  use Hobby shortcut,
  tension=1.2
]
(0.05,0)
.. (0.2,0.3)
.. (0.5,0.4)
.. (0.75,0.25)
.. (0.85,0.05)
.. (0.8,-0.2)
.. (0.55,-0.5)
.. (-0.05,-0.7)
.. (-0.5,-0.52)
.. (-0.65,-0.32)
.. (-0.737,-0.16)
.. (-0.8,0);
\draw [fill=black](0.05,0) ellipse (0.02 and 0.02);
\node [scale=0.6,color=red] at (-0.5,-0.07) {$\omega=\infty$};
\draw[fill=red,color=red]  (-0.8,0) ellipse (0.02 and 0.02);
\node [scale=0.6]at (0,-0.1) {$\omega=0$};

\draw [fill=black] (-0.68,-0.27) ellipse (0.02 and 0.02);
\node  [scale=0.6] at (-0.35,-0.27) {$\omega=\omega_\mathrm{bw}$};

\end{tikzpicture}  
}
    \vspace{-8mm}
    \caption{Qualitative graphical illustration of the superimposed NGGC 2 conditions \ref{cond:2i}-\ref{cond:2vi} in \cref{fig:nyquist_qv_elements}. The gray shaded region indicates the feasible set for all $\omega\in(0,\omega_\mathrm{bw})$, and the black curve depicts an exemplary feasible Nyquist plot of $D_i^\mathrm{qv}(\mathrm{j}\omega)$ that satisfies all conditions simultaneously.}
    \label{fig:nyquist_qv}
    \vspace{-4mm}
\end{figure}

\begin{gridcode}[Voltage Regulation]
    \fontdimen2\font=0.6ex The dynamic mapping of a generation unit $i$ from the local reactive power measurement deviation $\Delta q_i$ to the local voltage magnitude deviation $\Delta |v|_i$ shall be implemented via a transfer function $D_i^\mathrm{qv}(s)$ necessarily satisfying the frequency-domain conditions in \ref{cond:2i}-\ref{cond:2vi}.
\end{gridcode}

Compliance with these NGGC specifications by all generation units in the power system ensures overall closed-loop stability, while simultaneously satisfying the global and local performance metrics defined in \cref{eq:global_f_avg_specs_rigorous,eq:local_volt_spec_rigorous}. A graphical illustration of the superimposed grid-code specifications is provided in \cref{fig:nyquist_pf,fig:nyquist_qv}. Representative Nyquist plots of transfer functions that satisfy the associated frequency-domain conditions will be presented in the case studies in \cref{sec:case_study}.

\section{Case Studies}\label{sec:case_study}
To validate the proposed NGGC framework and to substantiate its stability and performance guarantees, we conduct a series of numerical case studies in MATLAB/Simulink.

The first two case studies are tutorial examples based on a simple two-node system with two devices interconnected through an $RL$ line. The implementation follows the NGGC modeling assumptions, using a quasistationary phasor-domain network model and including only the outer-loop control dynamics represented in the framework, while neglecting inner converter loops and detailed synchronous machine dynamics.

The \emph{first case study} examines the monotonic relationship between the local NGGC frequency-domain conditions and the system-wide time-domain stability and performance metrics. Two grid-forming VSCs are interconnected and their control parameters are progressively varied beyond the admissible NGGC bounds to assess the resulting dynamic behavior.

The \emph{second case study} investigates conventional grid-forming VSC and SG control concepts within the NGGC framework. This setup enables a systematic evaluation of the feasibility and applicability of both converter-based and synchronous generation technologies under the NGGC framework.
\begin{figure*}[t!]
    \centering
    \scalebox{0.45}{\includegraphics[]{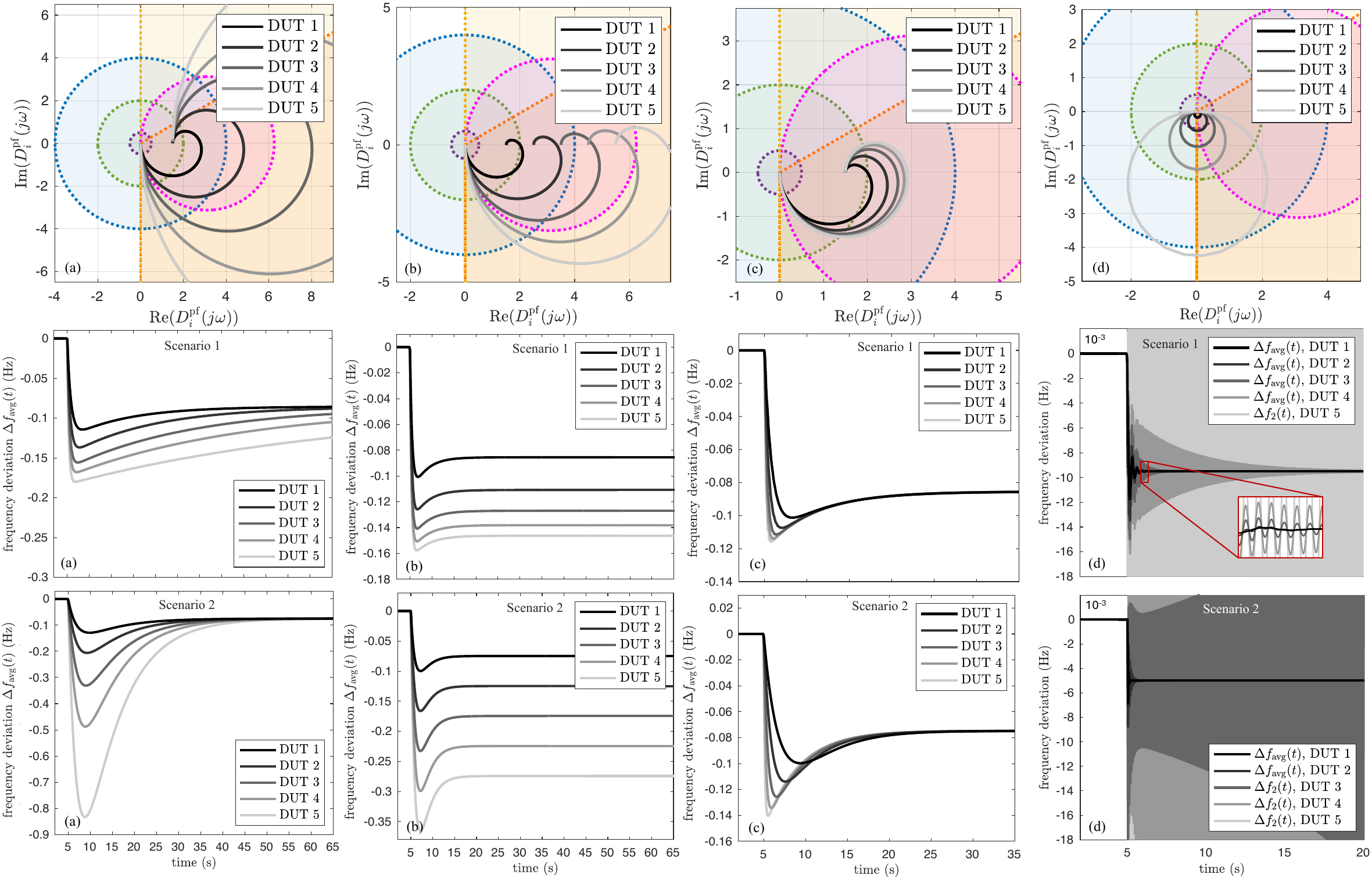}}
    \vspace{-2mm}
    \caption{Simulation results of Case Study I: The \emph{top} row shows the Nyquist plots of the DUT controller $D_2^\mathrm{pf}(s)$ for progressive variations (a)–(d) of the NGGC 1 boundaries, using the same colors as in \cref{fig:nyquist_pf}. The gray region indicates the feasible set for $\omega\in (0,\omega_\mathrm{bw})$. The \emph{middle} and \emph{bottom} row display the average-node frequency responses $\Delta f_\mathrm{avg}(t)$ of the two-node system after a load increase for Scenario 1 in \cref{fig:2Node1} (middle row) and Scenario 2 in \cref{fig:2Node2} (bottom row), under the same NGGC boundary variations (a)–(d).}
    \vspace{-4mm}
    \label{fig:monotonicity_CSI_nyquist_and_avg_response}
\end{figure*}

Finally, the \emph{third case study} considers a more realistic transmission system based on the IEEE nine-bus network with three grid-forming VSCs. We use a full nonlinear electromagnetic transient (EMT) model with detailed converter representations, including inner control loops. We perform Monte Carlo simulations by varying the outer-loop control parameters of all three converters within the feasible region defined by the NGGC framework and evaluate the resulting dynamic responses with respect to system stability and performance metrics.

\subsection{Case Study I: Monotonicity of NGGC Conditions}\label{sec:CSI}
We consider the two-node system shown in \cref{fig:2Node} (parameters in \cref{tab:2node_parameters}), consisting of two grid-forming VSCs interconnected through a $RL$ line. The setup strictly follows the modeling assumptions of the NGGC framework: the network is represented by a quasi-stationary phasor model, and only outer-loop device dynamics are implemented via the device transfer functions $D_i^\mathrm{pf}(s)$ and $D_i^\mathrm{qv}(s)$, $i=\{1,2\}$, while inner converter control loops, filter dynamics, and PWM switching are neglected.

We focus on the active-power–frequency dynamics of the $\mathrm{pf}$-subsystem and fix the nodal voltages at 1~pu (i.e., $D_i^\mathrm{qv}(s) =0$). Two interconnection scenarios are analyzed. In the first scenario, an ideal VSC with a feasible controller $D_1^\mathrm{pf}(s)$ satisfying all NGGC 1 conditions \ref{cond:1i}–\ref{cond:1viii} is connected at node 1, whereas the device under test (DUT) at node 2 is assigned different outer-loop control laws $D_2^\mathrm{pf}(s)$ (cf. \cref{fig:2Node1}). In the second scenario, the DUT is connected at both buses, i.e., $D_1^\mathrm{pf}(s)=D_2^\mathrm{pf}(s)$, and varied simultaneously (cf. \cref{fig:2Node2}).

For both scenarios, we examine the monotonicity of the local frequency-domain NGGC 1 conditions \ref{cond:1i}–\ref{cond:1viii} with respect to time-domain stability and performance metrics of the average-mode frequency response defined in \cref{eq:global_f_avg_specs_rigorous}. To this end, individual local conditions are progressively violated. First, the $\mathcal{H}_\infty$-norm $||D_2^\mathrm{pf}(s)||_\infty$ is increased beyond the bound in condition \ref{cond:1v} (Nyquist plots in \cref{fig:monotonicity_CSI_nyquist_and_avg_response}(a)). Second, the DC gain $|D_2^\mathrm{pf}(0)|$ is increased beyond the bound in condition \ref{cond:1vi} (subfigure \cref{fig:monotonicity_CSI_nyquist_and_avg_response}(b)). Third, the high-frequency derivative gain $|\lim_{\omega\rightarrow\infty} \mathrm{j}\omega D_2^\mathrm{pf}(\mathrm{j}\omega)|$ is increased beyond the bound in condition \ref{cond:1vii} (subfigure \cref{fig:monotonicity_CSI_nyquist_and_avg_response}(c)). Finally, the real part of the inverse transfer function, $\mathrm{Re}[D_2^\mathrm{pf}(\mathrm{j}\omega)^{-1}]$, is varied to progressively violate the oscillation condition in \ref{cond:1viii} and the stability boundary associated with the imaginary axis in condition \ref{cond:1ii} (subfigure \cref{fig:monotonicity_CSI_nyquist_and_avg_response}(d)).

\begin{table}[b!]
    \centering
    \vspace{-2mm}
    \begin{tabular}{c||c|c}
    \toprule
         Parameter&Symbol & Value  \\ \hline
         $\hspace{-2mm}$ Base power, voltage, frequency $\hspace{-2mm}$ & $\hspace{-2mm}$ $S_\mathrm{b}$, $V_\mathrm{b}$, $f_\mathrm{b}$ $\hspace{-2mm}$& $\hspace{-2mm}$ 100 MVA, 13.8 kV, 50 Hz$\hspace{-2mm}$\\
         $RL$ line components & $r_{12}$, $l_{12}$& 0.02, 0.2 pu\\
         Converter power ratings & $\hspace{-2mm}$ $S_\mathrm{b,1}$, $S_\mathrm{b,2}$ $\hspace{-2mm}$ & 100, 100 MVA\\ 
         $\hspace{-2mm}$Converter steady-state voltages$\hspace{-2mm}$& $\hspace{-1mm}$$|v|_{0,1}$, $|v|_{0,2}$ $\hspace{-2mm}$&1, 1 pu\\\bottomrule
    \end{tabular}
            \vspace{-1mm}
    \caption{Parameters of the two-node test system in Case Studies I and II.}
        \vspace{-2mm}
    \label{tab:2node_parameters}
\end{table}

\begin{figure}[b!]
\centering
\begin{subfigure}[b]{0.23\textwidth}
    \centering
    \resizebox{0.93\textwidth}{!}{
\begin{tikzpicture}[scale=0.4, every node/.style={scale=0.65}]
	
	\draw(-7,19.4) -- (-2.5,19.4) node (v1) {};
	\draw [ultra thick](-3.1,20) -- (-3.1,18.8);
	\draw [ultra thick](-6.4,20) -- (-6.4,18.8);

	\fill[black] (-6.4,19.4)circle (0.7 mm); 

	\fill[black] (-3.1,19.4)circle (0.7 mm);

	\node at (-6.4,20.5) {1};

	\node at (-3.1,20.5) {2};

	\draw [-latex,thick](-3.1,19.1) -- (-3.6,19.1) -- (-3.6,18.1);
	\fill[black] (-3.1,19.1) circle (0.7 mm);


	\node at (-7.8,18.1) {ideal VSC};
	\node at (0.4,18.2) {SG};

	\draw [rounded corners = 3,fill=gray!30] (-2.2,19.1) rectangle (-0.1,17.3);
\draw (-1.15,18.18) ellipse (0.8 and 0.8);
\draw (-1.3,17.9) node (v4) {} -- (-1.3,18.45) node (v3) {};
\draw (-1,18.45) node (v5) {} -- (-1,17.9) node (v6) {};
\draw  (-1.3,18.45) -- (-1.55,18.45);
\draw  (-1.3,17.9) -- (-1.55,17.9); 
\draw(-1,18.45)-- (-0.75,18.45); 
\draw  (-1,17.9) -- (-0.75,17.9); 
\draw (-0.75,18.45) arc (0:180:0.4);
\draw (-0.75,17.9) arc (0:-180:0.4);

\draw[rounded corners = 3,fill=gray!30]  (-9,20.3) rectangle (-6.9,18.5);
\draw (-8.7,19.4) -- (-8.3,19.4); 
\draw (-8.3,19.8) -- (-8.3,19); 
\draw (-8.2,19.8) -- (-8.2,19); 
\draw (-8.2,19.65) -- (-7.95,19.75) -- (-7.95,20.15); 

\draw (-8.2,19.2) -- (-7.95,19.05) node (v8) {} -- (-7.95,18.65);
\draw (-7.95,20.05) -- (-7.5,20.05) -- (-7.5,19.65) node (v7) {}; 
\draw (-7.95,18.8) -- (-7.5,18.8) -- (-7.5,19.25);
\draw(-7.5,19.65)-- (-7.7,19.25) -- (-7.3,19.25) -- (-7.5,19.65);
\draw (-7.7,19.65) -- (-7.3,19.65);
\draw (-8.01,19.18) -- (-7.95,19.05) node (v9) {} ;
\draw (-8.1,19.05) -- (-7.95,19.05) ;
\draw [fill=white] (-5.6,19.6) rectangle (-3.9,19.2);

\draw[rounded corners = 3,fill=gray!30]  (-2.2,21.6) rectangle (-0.1,19.8);
\draw (-1.9,20.7) -- (-1.5,20.7); 
\draw (-1.5,21.1) -- (-1.5,20.3); 
\draw (-1.4,21.1) -- (-1.4,20.3); 
\draw (-1.4,20.95) -- (-1.15,21.05) -- (-1.15,21.45); 

\draw (-1.4,20.5) -- (-1.15,20.35) node (v8) {} -- (-1.15,19.95);
\draw (-1.15,21.35) -- (-0.7,21.35) -- (-0.7,20.95) node (v7) {}; 
\draw (-1.15,20.1) -- (-0.7,20.1) -- (-0.7,20.55);
\draw(-0.7,20.95)-- (-0.9,20.55) -- (-0.5,20.55) -- (-0.7,20.95);
\draw (-0.9,20.95) -- (-0.5,20.95);
\draw (-1.21,20.48) -- (-1.15,20.35) node (v9) {} ;
\draw (-1.3,20.35) -- (-1.15,20.35) ;

\draw [dashed, rounded corners = 2] (-2.5,21.9) rectangle (1.3,17.1);
\node at (-1.1,19.4) {or};
\node at (0.6,20.7) {VSC};
\node at (-0.4,16.75) {DUT};
\end{tikzpicture}

}
    \vspace{-9mm}
    \caption{Scenario 1: Two-node system with an ideal grid-forming VSC at node~1 using fixed, feasible controllers, and a DUT (VSC or SG) at node~2 on which various $D_2^{\mathrm{pf}}(s)$ and $D_2^{\mathrm{qv}}(s)$ controls are evaluated.}
    \label{fig:2Node1}
\end{subfigure}
\hfill
\begin{subfigure}[b]{0.24\textwidth}
    \centering
    \resizebox{1.05\textwidth}{!}{
\begin{tikzpicture}[scale=0.4, every node/.style={scale=0.65}]
	
	\draw(-7,19.4) -- (-2.5,19.4) node (v1) {};
	\draw [ultra thick](-3.1,20) -- (-3.1,18.8);
	\draw [ultra thick](-6.4,20) -- (-6.4,18.8);
	
\draw [fill=white] (-5.6,19.6) rectangle (-3.9,19.2);

	\fill[black] (-6.4,19.4)circle (0.7 mm); 

	\fill[black] (-3.1,19.4)circle (0.7 mm);

	\node at (-6.4,20.5) {1};

	\node at (-3.1,20.5) {2};

	\draw [-latex,thick](-3.1,19.1) -- (-3.6,19.1) -- (-3.6,18.1);
	\fill[black] (-3.1,19.1) circle (0.7 mm);


	\node at (-7.9,18.2) {SG};

	\draw [rounded corners = 3,fill=gray!30] (-10.5,19.1) rectangle (-8.4,17.3);
\draw (-9.45,18.18) ellipse (0.8 and 0.8);
\draw (-9.6,17.9) node (v4) {} -- (-9.6,18.45) node (v3) {};
\draw (-9.3,18.45) node (v5) {} -- (-9.3,17.9) node (v6) {};
\draw  (-9.6,18.45) -- (-9.85,18.45);
\draw  (-9.6,17.9) -- (-9.85,17.9); 
\draw(-9.3,18.45)-- (-9.05,18.45); 
\draw  (-9.3,17.9) -- (-9.05,17.9); 
\draw (-9.05,18.45) arc (0:180:0.4);
\draw (-9.05,17.9) arc (0:-180:0.4);

\draw[rounded corners = 3,fill=gray!30]  (-10.5,21.6) rectangle (-8.4,19.8);
\draw (-10.2,20.7) -- (-9.8,20.7); 
\draw (-9.8,21.1) -- (-9.8,20.3); 
\draw (-9.7,21.1) -- (-9.7,20.3); 
\draw (-9.7,20.95) -- (-9.45,21.05) -- (-9.45,21.45); 

\draw (-9.7,20.5) -- (-9.45,20.35) node (v8) {} -- (-9.45,19.95);
\draw (-9.45,21.35) -- (-9,21.35) -- (-9,20.95) node (v7) {}; 
\draw (-9.45,20.1) -- (-9,20.1) -- (-9,20.55);
\draw(-9,20.95)-- (-9.2,20.55) -- (-8.8,20.55) -- (-9,20.95);
\draw (-9.2,20.95) -- (-8.8,20.95);
\draw (-9.51,20.48) -- (-9.45,20.35) node (v9) {} ;
\draw (-9.6,20.35) -- (-9.45,20.35) ;

\draw [dashed, rounded corners = 2] (-10.8,21.9) rectangle (-7,17.1);
\node at (-9.4,19.4) {or};
\node at (-7.7,20.7) {VSC};
\node at (-8.7,16.75) {DUT};

	\draw [rounded corners = 3,fill=gray!30] (-2.2,19.1) rectangle (-0.1,17.3);
\draw (-1.15,18.18) ellipse (0.8 and 0.8);
\draw (-1.3,17.9) node (v4) {} -- (-1.3,18.45) node (v3) {};
\draw (-1,18.45) node (v5) {} -- (-1,17.9) node (v6) {};
\draw  (-1.3,18.45) -- (-1.55,18.45);
\draw  (-1.3,17.9) -- (-1.55,17.9); 
\draw(-1,18.45)-- (-0.75,18.45); 
\draw  (-1,17.9) -- (-0.75,17.9); 
\draw (-0.75,18.45) arc (0:180:0.4);
\draw (-0.75,17.9) arc (0:-180:0.4);

\draw[rounded corners = 3,fill=gray!30]  (-2.2,21.6) rectangle (-0.1,19.8);
\draw (-1.9,20.7) -- (-1.5,20.7); 
\draw (-1.5,21.1) -- (-1.5,20.3); 
\draw (-1.4,21.1) -- (-1.4,20.3); 
\draw (-1.4,20.95) -- (-1.15,21.05) -- (-1.15,21.45); 

\draw (-1.4,20.5) -- (-1.15,20.35) node (v8) {} -- (-1.15,19.95);
\draw (-1.15,21.35) -- (-0.7,21.35) -- (-0.7,20.95) node (v7) {}; 
\draw (-1.15,20.1) -- (-0.7,20.1) -- (-0.7,20.55);
\draw(-0.7,20.95)-- (-0.9,20.55) -- (-0.5,20.55) -- (-0.7,20.95);
\draw (-0.9,20.95) -- (-0.5,20.95);
\draw (-1.21,20.48) -- (-1.15,20.35) node (v9) {} ;
\draw (-1.3,20.35) -- (-1.15,20.35) ;

\draw [dashed, rounded corners = 2] (-2.5,21.9) rectangle (1.3,17.1);
\node at (-1.1,19.4) {or};
\node at (0.6,20.7) {VSC};
\node at (-0.4,16.75) {DUT};
\node at (0.4,18.2) {SG};
\end{tikzpicture}

}
    \vspace{-9mm}
    \caption{Scenario 2: Two-node system with the DUT (VSC or SG) connected at both node~1 and node~2, on which various $D_i^{\mathrm{pf}}(s)$ and $D_i^{\mathrm{qv}}(s)$ controls are evaluated under identical test conditions.}
    \label{fig:2Node2}
\end{subfigure}
\vspace{-1mm}
\caption{Two-node test system for NGGC evaluation in Case Studies I and II.}
\label{fig:2Node}
    \vspace{-1mm}
\end{figure}

To evaluate the resulting system-wide time-domain behavior, a 0.1~pu active-power step disturbance, relative to the total system rating, is applied at node~2. The corresponding average-mode frequency responses are shown in \cref{fig:monotonicity_CSI_nyquist_and_avg_response} for both interconnection scenarios. The plotted trajectories represent the average frequency obtained by averaging the measured signals of the individual devices. The results clearly demonstrate a monotonic correspondence between the local frequency-domain NGGC 1 conditions and the global average-mode frequency behavior in the time domain. In particular, progressive violations of the frequency-domain boundaries in conditions \ref{cond:1i}–\ref{cond:1viii} lead to a systematic degradation of the time-domain performance, reflected in increasing exceedances of the prescribed metrics $\Delta f_\mathrm{max} = 0.8\,-\,1\,\text{Hz}$, $\Delta f_\mathrm{ss,\,max}= 0.2\,\text{Hz}$, and $\Delta\dot{f}_\mathrm{max}= 2\,-\,2.5\,\text{Hz/s}$, as well as in a reduction of the oscillation damping capability, ultimately resulting in instability.

\subsection{Case Study II: Feasibility of Classical VSC and SG Controls}\label{sec:CSII}
Building on the same two-node setup introduced in Case Study~I, we now assess the compatibility of classical grid-forming VSC and SG controls within the NGGC framework. Rather than varying abstract controller properties as in Case Study I, this study evaluates local control implementations commonly used in practice and examines their compliance with the NGGC stability and performance conditions.

\textit{Active-Power–Frequency Control:}
We first consider the $\mathrm{pf}$-subsystem while maintaining constant nodal voltages of 1 pu. The same interconnection scenarios defined in \cref{fig:2Node} are used to distinguish between the influence of a single DUT and the case where identical DUTs are deployed at both nodes.

\begin{figure}[t]
    \centering
\scalebox{0.32}{\includegraphics[]{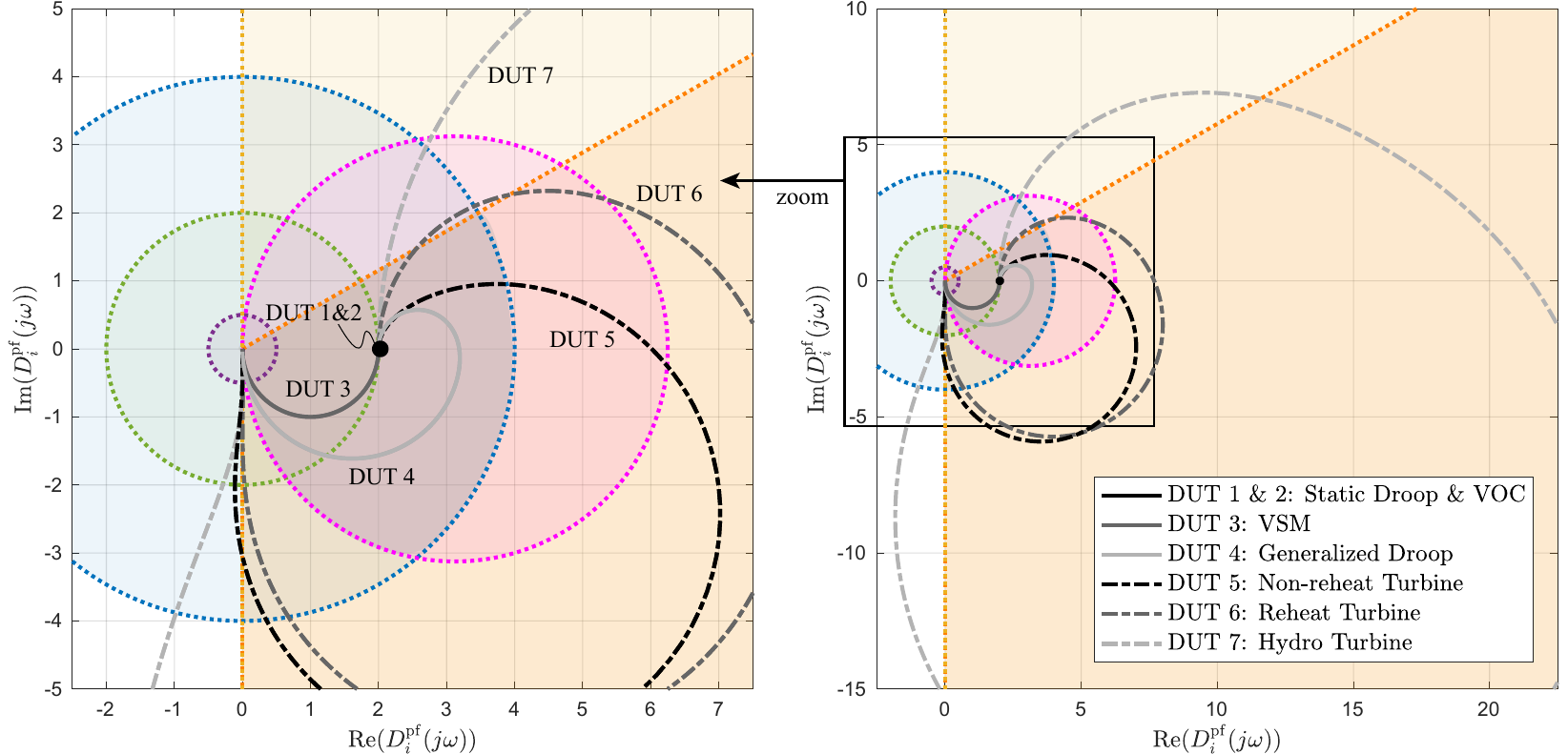}}
    \vspace{-6mm}
    \caption{Nyquist plots of the DUT controllers $D_2^{\mathrm{pf}}(s)$ in Case Study II, with NGGC 1 conditions indicated using the same colors as in \cref{fig:nyquist_pf}. The gray shaded region marks the feasible set for all $\omega\in(0,\omega_{\mathrm{bw}})$.}
    \label{fig:2NodeNyquist_PF}
    \vspace{-3mm}
\end{figure}

The Nyquist plots in \cref{fig:2NodeNyquist_PF} show different grid-forming VSC and SG control implementations used as $D_2^\mathrm{pf}(s)$:
\begin{itemize}
\item \textbf{DUT~1}: static droop control \cite{tayyebi2020frequency}, 
\item \textbf{DUT~2}: virtual oscillator control (VOC), using the linear approximation in \cite{he2024quantitative}, 
\item \textbf{DUT~3}: virtual synchronous machine (VSM) control (i.e., first-order filtered droop control) \cite{tayyebi2020frequency}, 
\item \textbf{DUT~4}: generalized second-order droop control \cite{ordono2025transfer},
\item \textbf{DUT~5}: SG with non-reheat steam turbine,
\item \textbf{DUT~6}: SG with reheat steam turbine,
\item \textbf{DUT~7}: SG with hydro turbine,
\end{itemize}
where the SG governor-turbine models and parameters are taken from \cite{kundur2007power}. The ideal grid-forming VSC at node 1 in Scenario 1 (\cref{fig:2Node1}) remains strictly compliant with all NGGC 1 conditions \ref{cond:1i}-\ref{cond:1viii}. The NGGC 1 compliance of the DUT controllers $D_2^\mathrm{pf}(s)$ at node 2 is summarized in \cref{tab:CSI_conditions_pf}. The control laws are evaluated using representative parameter values. Although different parameters may alter the quantitative results, the aim is to illustrate the qualitative behavior.

\begin{figure}[t!]
\centering
\begin{subfigure}[b]{0.23\textwidth}
    \centering
    \scalebox{0.43}{\includegraphics[]{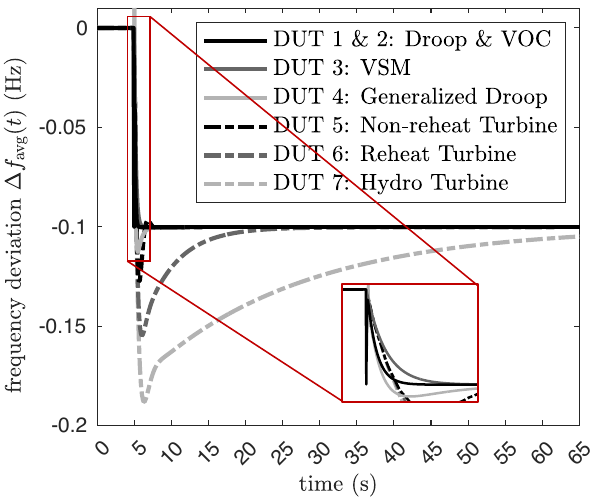}}
    \vspace{-6mm}
    \caption{Average-mode frequency response $\Delta f_\mathrm{avg}(t)$ after a load increase for Scenario 1 in \cref{fig:2Node1}.}
    \label{fig:2Node_nominal_freq_response_1DUT}
\end{subfigure}
\hfill
\begin{subfigure}[b]{0.24\textwidth}
    \centering
    \scalebox{0.43}{\includegraphics[]{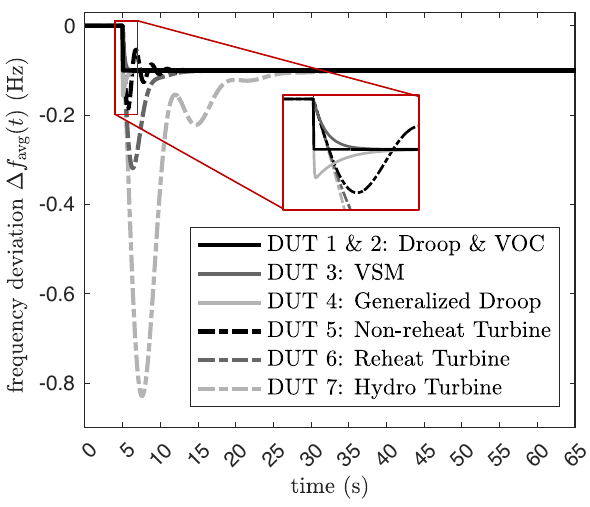}}
    \vspace{-6mm}
    \caption{Average-mode frequency response $\Delta f_\mathrm{avg}(t)$ after a load increase for Scenario 2 in \cref{fig:2Node2}.}
    \label{fig:2Node_nominal_freq_response_2DUT}
\end{subfigure}
\vspace{-1mm}
\caption{Average-mode frequency responses $\Delta f_\mathrm{avg}(t)$ of the two-node test system in Case Study II.}
\label{fig:2Node_nominal_freq_response}
\vspace{-4mm}
\end{figure}

A 0.1~pu active-power step disturbance is applied to evaluate the system-wide time-domain behavior. The resulting average-mode frequency response dynamics for the single-DUT configuration in Scenario 1 (\cref{fig:2Node1}) are shown in \cref{fig:2Node_nominal_freq_response_1DUT}. All investigated controllers produce stable responses, consistent with approximate satisfaction of the strict passivity conditions \ref{cond:1i} and \ref{cond:1ii} (except DUT 7, which may become unstable under specific grid conditions). Furthermore, the steady-state frequency deviation remains within the ENTSO-E limit of $\Delta f_\mathrm{ss,max}=0.2$~Hz for all cases, reflecting compliance with condition~\ref{cond:1vi}. The dynamic time-domain characteristics of $\Delta f_\mathrm{avg}(t)$ correlate closely with the local NGGC 1 frequency-domain conditions. In particular, the frequency nadir increases with the $\mathcal{H}_\infty$-norm of $D_2^\mathrm{pf}(s)$, confirming the performance implications of condition~\ref{cond:1v}. Nevertheless, although SG-based DUTs 5-7 violate condition~\ref{cond:1v}, the ideal VSC at node 1 limits the nadir well below the ENTSO-E bound $\Delta f_{\mathrm{max}}=0.8$-$1$~Hz. RoCoF behavior and oscillation damping are consistent with conditions \ref{cond:1vii} and \ref{cond:1viii}, respectively, where DUT 1\&2 and DUT 4, violating condition \ref{cond:1vii}, exhibit the highest RoCoF values, even exceeding the ENTSO-E limit of $\Delta\dot{f}_\mathrm{max}=2$ Hz/s.

When both nodes are equipped with identical DUT controllers in Scenario 2 (\cref{fig:2Node2}), the resulting average-mode frequency responses (\cref{fig:2Node_nominal_freq_response_2DUT}) become more demanding from an operational perspective: the nadir exceeds the ENTSO-E limit for DUT 7 (violation of conditions \ref{cond:1iii} and \ref{cond:1v}), RoCoF constraints are violated for DUT 1\&2 and DUT 4 (violation of condition \ref{cond:1vii}), and pronounced oscillations appear for SG-based DUTs 5-7 (violation of condition \ref{cond:1viii}). Overall, in alignment with Case Study I, these results demonstrate that the proposed NGGC conditions capture key stability and performance trends and provide meaningful guidance for device-level control design.

\begin{table}[]
    \centering
    \begin{tabular}{c||c|c|c|c|c|c|c|c}
    \toprule
     & $\hspace{-1mm}$\ref{cond:1i} $\hspace{-2mm}$& $\hspace{-1mm}$\ref{cond:1ii} $\hspace{-2mm}$& $\hspace{-1mm}$\ref{cond:1iii}$\hspace{-1mm}$ & $\hspace{-1mm}$\ref{cond:1iv} $\hspace{-2mm}$&$\hspace{-1mm}$ \ref{cond:1v} $\hspace{-2mm}$& $\hspace{-1mm}$\ref{cond:1vi} $\hspace{-2mm}$&$\hspace{-1mm}$ \ref{cond:1vii} $\hspace{-2mm}$& $\hspace{-1mm}$\ref{cond:1viii}$\hspace{-2mm}$   \\ \hline 
      DUT 1 \& 2 & $\times$&$\checkmark$&$\checkmark$&$\times$&$\checkmark$&$\checkmark$&$\times$&$\checkmark$\\
      DUT 3 &$\checkmark$&$\checkmark$&$\checkmark$&$\checkmark$&$\checkmark$&$\checkmark$&$\checkmark$&$\checkmark$\\
      DUT 4 &$\checkmark$&$\checkmark$&$\checkmark$&$\times$&$\checkmark$&$\checkmark$&$\times$&$\checkmark$\\
      DUT 5 &$\checkmark$&$\times$&$\times$&$\checkmark$&$\times$&$\checkmark$&$\checkmark$&$\times$\\
      DUT 6 &$\checkmark$&$\checkmark$&$\times$&$\checkmark$&$\times$&$\checkmark$&$\checkmark$&$\times$\\
      DUT 7 &$\checkmark$&$\times$&$\times$&$\checkmark$&$\times$&$\checkmark$&$\checkmark$&$\times$\\
      \bottomrule
    \end{tabular}
    \vspace{-1mm}
    \caption{NGGC 1 compliance of the different DUT control laws $D_2^{\mathrm{pf}}(s)$ in Case Study II.}
    \vspace{-4mm}
    \label{tab:CSI_conditions_pf}
\end{table}

\begin{figure}[t!]
\centering
\begin{subfigure}[b]{0.23\textwidth}
    \centering
    \vspace{-1mm}
    \scalebox{0.42}{\includegraphics[]{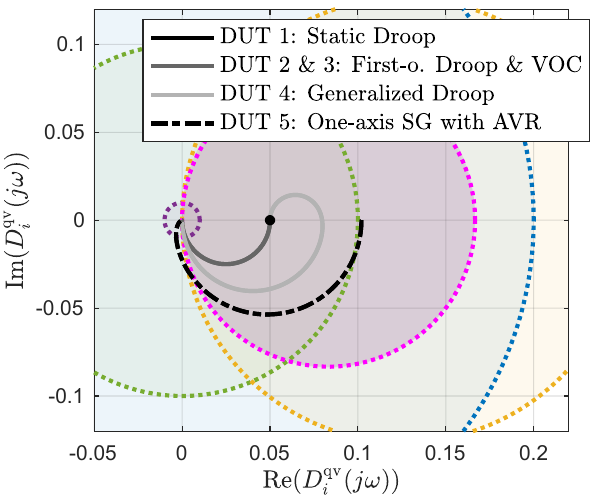}}
    \vspace{-6mm}
    \caption{Nyquist plots of the DUT controllers $D_2^{\mathrm{qv}}(s)$ in Case Study II, with NGGC 2 conditions indicated using the same colors as in \cref{fig:nyquist_qv}. The gray shaded region marks the feasible set for all $\omega\in(0,\omega_{\mathrm{bw}})$.}
    \label{fig:2NodeNyquist_QV}
\end{subfigure}
\hfill
\begin{subfigure}[b]{0.24\textwidth}
    \centering
        \vspace{-2.5mm}
    \scalebox{0.42}{\includegraphics[]
    {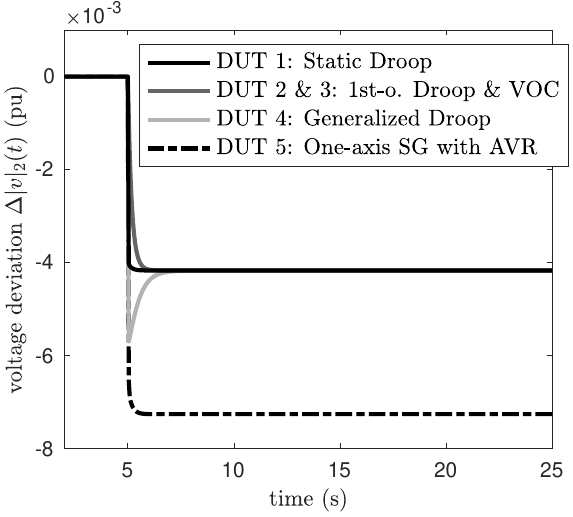}}
    \vspace{-2mm}
    \caption{Local voltage response $\Delta |v|_2(t)$ at node 2 after a small load increase in the two-node system in \cref{fig:2Node1}.\\
    {\color{white}Local voltage response  at bus 2 after a small load increase in the two-node system in}}
    \label{fig:2Node_nominal_volt_response_1DUT}
\end{subfigure}
\vspace{-1mm}
\caption{Nyquist plots and local voltage responses in Case Study II.}
\label{fig:2Node_nominal_volt_response}
\vspace{-4mm}
\end{figure}

\textit{Reactive-Power-Voltage Control:}  We next investigate the dynamic behavior of the $\mathrm{qv}$-subsystem, while keeping the frequency at each bus fixed at 50 Hz. We adopt a robust approach and choose $\Delta |v|_{\mathrm{ss,max}}=0.01\text{ pu}$ and $\Delta |v|_{\mathrm{max}}=0.05\text{ pu}$ sufficiently small to ensure that the ENTSO-E absolute voltage limits of 0.9-1.1 pu are never violated for any operating points $|v|_{0,i}\in[0.95,1.05]\text{ pu}$, thereby also ensuring $|D_i^\mathrm{qv}(s)N^\mathrm{vq}|<1$ to be satisfied. We consider the two-node system under Scenario 1 in \cref{fig:2Node1}, where an ideal VSC with a feasible device control $D_1^\mathrm{qv}(s)$ is connected at node 1 (satisfying all NGGC 2 conditions \ref{cond:2i}-\ref{cond:2vi}), and the DUT at node 2 is varied by assigning different control laws $D_2^\mathrm{qv}(s)$. The Nyquist plots in \cref{fig:2NodeNyquist_QV} illustrate several conventional grid-forming VSC and SG $\mathrm{qv}$-control designs used as $D_2^\mathrm{qv}(s)$:
\begin{itemize}
    \item \textbf{DUT 1}: static droop control \cite{pan2019transient},
    \item \textbf{DUT 2}: first-order filtered droop control \cite{schiffer2014conditions},
    \item \textbf{DUT 3}: linearized VOC control \cite{he2024quantitative}, and
    \item \textbf{DUT 4}: generalized 2nd-order droop control \cite{ordono2025transfer}
    \item \textbf{DUT 5}: one-axis SG model with 1st-order AVR \cite{kundur2007power,machowski2020power}.
\end{itemize}
\cref{tab:CSI_conditions_qv} summarizes NGGC 2 compliance of the different DUT control laws $D_2^\mathrm{qv}(s)$. The control laws are evaluated with representative parameters to illustrate qualitative behavior.

\begin{table}[t!]
    \centering
    \begin{tabular}{c||c|c|c|c|c|c}
    \toprule
     & $\hspace{-1mm}$\ref{cond:2i} $\hspace{-2mm}$& $\hspace{-1mm}$\ref{cond:2ii} $\hspace{-2mm}$& $\hspace{-1mm}$\ref{cond:2iii}$\hspace{-1mm}$ & $\hspace{-1mm}$\ref{cond:2iv} $\hspace{-2mm}$&$\hspace{-1mm}$ \ref{cond:2v} $\hspace{-2mm}$& $\hspace{-1mm}$\ref{cond:2vi} $\hspace{-2mm}$\\ \hline 
    {DUT 1} & $\times$ & $\checkmark$ & $\times$ & $\checkmark$ & $\checkmark$ & $\checkmark$\\
        {DUT 2 \& 3} & $\checkmark$ & $\checkmark$ & $\checkmark$ & $\checkmark$ & $\checkmark$ & $\checkmark$\\
            {DUT 4} & $\checkmark$ & $\checkmark$ & $\times$ & $\checkmark$ & $\checkmark$ & $\checkmark$\\
                {DUT 5} & $\checkmark$ & $\times$ & $\times$ & $\checkmark$ & $\times$ & $\times$\\
      \bottomrule
    \end{tabular}
    \vspace{-1mm}
    \caption{NGGC 2 compliance of the different DUT control laws $D_2^{\mathrm{qv}}(s)$ in Case Study II.}
    \vspace{-3mm}
    \label{tab:CSI_conditions_qv}
\end{table}

To evaluate overall system stability and the time-domain voltage behavior at each DUT, we apply a 0.1 pu reactive-power step (relative to the total system rating) at node 2. The resulting local-voltage responses are shown in \cref{fig:2Node_nominal_volt_response_1DUT}. All DUTs exhibit stable behavior, consistent with their control laws remaining largely in the right-half plane and therefore approximately satisfying conditions \ref{cond:2i} and \ref{cond:2ii} (except DUT 5, which may become unstable under specific grid conditions). The imposed peak-voltage constraint $\Delta|v|_{\max}=0.05\text{ pu}$ is also met for all DUTs because condition \ref{cond:2iv} holds. DUTs 1–4 share the same feasible DC gain and therefore satisfy the steady-state requirement in condition \ref{cond:2v}; their steady-state voltage deviations remain within $\Delta|v|_{\mathrm{ss,max}}=0.01\text{ pu}$. DUT 5 slightly violates this condition, resulting in a larger steady-state deviation close to the imposed limit. Finally, since the Nyquist plots of all DUTs lie predominantly within the pink region (condition \ref{cond:2vi}), their voltage responses do not exhibit pronounced oscillations.

Overall, the results of the first two case studies confirm that the proposed NGGC performance conditions are valid, although generally conservative. However, given the safety-critical nature of power-system operation as well as the modelling uncertainties inherent in grid-code design, such conservatism is both reasonable and desirable. Nevertheless, future work should investigate opportunities to reduce this conservatism where appropriate. Beyond validating NGGC performance, the proposed criteria also provide guidance for rethinking device-level control design, which is a topic of independent interest.

\subsection{Case Study III: Validation in IEEE Nine-Bus System}\label{sec:CSIII}
\begin{figure}
    \centering
    \vspace{-2mm}
    \resizebox{0.43\textwidth}{!}{
\begin{tikzpicture}[scale=0.4, every node/.style={scale=0.65}]
	
	\draw(-7.3,19.4) -- (0.5,19.4);
	\draw [ultra thick](-3.4,20) -- (-3.4,18.8);
	\draw [ultra thick](-9.5,20) -- (-9.5,18.8);
	
	\draw [ultra thick](-6.4,20) -- (-6.4,18.8);
	
	\draw [ultra thick](2.7,20) -- (2.7,18.8);
	
	\draw [ultra thick](-0.4,20) -- (-0.4,18.8);
	\draw[ultra thick] (-5.6,17.9) -- (-4.2,17.9);
	
	\draw [ultra thick](-2.6,17.9) -- (-1.2,17.9);
	\draw [ultra thick](-4.1,16.1) -- (-2.7,16.1);
	\draw [ultra thick](-4.1,13.5) -- (-2.7,13.5);
	\fill[black] (-9.5,19.4)circle (0.7 mm); 
	\fill[black]  (-6.4,19.4)circle (0.7 mm); 
	\fill[black] (-3.4,19.4)circle (0.7 mm); 
	\fill[black]  (-0.4,19.4)circle (0.7 mm); 
	\fill[black] (2.7,19.4) circle (0.7 mm);

	\fill[black] (-1.5,17.9)circle (0.7 mm); 
	\fill[black] (-2.3,17.9)circle (0.7 mm); 
	\fill[black] (-4.5,17.9)circle (0.7 mm); 
	\fill[black] (-5.3,17.9)circle (0.7 mm); 
	\fill[black] (-3.8,16.1)circle (0.7 mm); 
	\fill[black] (-3,16.1)circle (0.7 mm); 
	\fill[black] (-3.4,16.1)circle (0.7 mm); 
	\fill[black]  (-3.4,13.5) circle (0.7 mm); 
	
	\node at (-9.5,20.5) {2};
	\node at (-6.4,20.5) {7};
	\node at (-3.4,20.5) {8};
	\node at (-0.4,20.5) {9};
	\node at (2.7,20.5) {3};
	\node at (-6,17.9) {5};
	\node at (-0.8,17.9) {6};
	\node at (-2.3,16.1) {4};
	\node at (-2.3,13.5) {1};

	\node (v2) at (-6.4,19.1) {};
	\node at (2.7,19.1) {};
	\draw (-6.4,19.1) -- (-5.3,19.1) -- (-5.3,17.9);
	\fill[black] (-6.4,19.1) circle (0.7 mm); 
	\draw (-4.5,17.9) -- (-4.5,17.2) -- (-3.8,16.8) -- (-3.8,16.1);
	\draw (-2.3,17.9) -- (-2.3,17.2) -- (-3,16.8) -- (-3,16.1);
	
	\draw (-3.4,16.1) -- (-3.4,15.4) node (v1) {};
	\draw (-0.4,19.1) -- (-1.5,19.1) -- (-1.5,17.9);
	\fill[black] (-0.4,19.1) circle (0.7 mm); 
	\draw [-latex,thick](-3.4,19.1) -- (-2.9,19.1) -- (-2.9,18.3);
	\fill[black] (-3.4,19.1) circle (0.7 mm); 
	\draw [-latex,thick](-4.9,17.9) -- (-4.9,16.7);
	\fill[black](-4.9,17.9)circle (0.7 mm); 
	\draw [-latex, thick](-1.9,17.9) -- (-1.9,16.7);
	\fill[black](-1.9,17.9)circle (0.7 mm);


	\draw(-8.2,19.4)  circle (4 mm); 
	\draw(-7.7,19.4)  circle (4 mm); 

	\draw(-3.4,15)  circle (4 mm); 
	\draw(-3.4,14.5)  circle (4 mm); 

	\draw(1.4,19.4)  circle (4 mm); 
	\draw(0.9,19.4)  circle (4 mm); 
	\draw (-10,19.4) -- (-8.6,19.4);
	\draw (3.15,19.4) -- (1.8,19.4);
	\draw (-3.4,13.05) -- (-3.4,14.1);

	\node at (-10.9,18.15) {VSC 2};
	\node at (4.2,18.15) {VSC 3};
	\node at (-1.45,12.1) {VSC 1};
	
	\draw [rounded corners = 3,fill=gray!30] (3.15,20.3) rectangle (5.25,18.5);

\draw[rounded corners = 3,fill=gray!30]  (-4.45,13.05) rectangle (-2.35,11.25);
\draw (3.45,19.4) -- (3.85,19.4); 
\draw (3.85,19.8) -- (3.85,19); 
\draw (3.95,19.8) -- (3.95,19); 
\draw (3.95,19.65) -- (4.2,19.75) -- (4.2,20.15); 

\draw (3.95,19.2) -- (4.2,19.05) node (v8) {} -- (4.2,18.65);
\draw (4.2,20.05) -- (4.65,20.05) -- (4.65,19.65) node (v7) {}; 
\draw (4.2,18.8) -- (4.65,18.8) -- (4.65,19.25);
\draw(4.65,19.65)-- (4.45,19.25) -- (4.85,19.25) -- (4.65,19.65);
\draw (4.45,19.65) -- (4.85,19.65);
\draw (4.14,19.18) -- (4.2,19.05) node (v9) {} ;
\draw (4.05,19.05) -- (4.2,19.05) ;

\draw (-4.15,12.15) -- (-3.75,12.15); 
\draw (-3.75,12.55) -- (-3.75,11.75); 
\draw (-3.65,12.55) -- (-3.65,11.75); 
\draw (-3.65,12.4) -- (-3.4,12.5) -- (-3.4,12.9); 

\draw (-3.65,11.95) -- (-3.4,11.8) node (v8) {} -- (-3.4,11.4);
\draw (-3.4,12.8) -- (-2.95,12.8) -- (-2.95,12.4) node (v7) {}; 
\draw (-3.4,11.55) -- (-2.95,11.55) -- (-2.95,12);
\draw(-2.95,12.4)-- (-3.15,12) -- (-2.75,12) -- (-2.95,12.4);
\draw (-3.15,12.4) -- (-2.75,12.4);
\draw (-3.46,11.93) -- (-3.4,11.8) node (v9) {} ;
\draw (-3.55,11.8) -- (-3.4,11.8) ;

\draw[rounded corners = 3,fill=gray!30]  (-12.1,20.35) rectangle (-10,18.55);
\draw (-11.8,19.45) -- (-11.4,19.45); 
\draw (-11.4,19.85) -- (-11.4,19.05); 
\draw (-11.3,19.85) -- (-11.3,19.05); 
\draw (-11.3,19.7) -- (-11.05,19.8) -- (-11.05,20.2); 

\draw (-11.3,19.25) -- (-11.05,19.1) node (v8) {} -- (-11.05,18.7);
\draw (-11.05,20.1) -- (-10.6,20.1) -- (-10.6,19.7) node (v7) {}; 
\draw (-11.05,18.85) -- (-10.6,18.85) -- (-10.6,19.3);
\draw(-10.6,19.7)-- (-10.8,19.3) -- (-10.4,19.3) -- (-10.6,19.7);
\draw (-10.8,19.7) -- (-10.4,19.7);
\draw (-11.11,19.23) -- (-11.05,19.1) node (v9) {} ;
\draw (-11.2,19.1) -- (-11.05,19.1) ;
\end{tikzpicture}

}
        \vspace{-7mm}
    \caption{IEEE nine-bus test system for NGGC validation during Monte Carlo Simulations in Case Study III.}
    \label{fig:9bus}
        \vspace{-4mm}
\end{figure}
\begin{figure}[t!]
    \centering
\scalebox{0.65}{\includegraphics[]{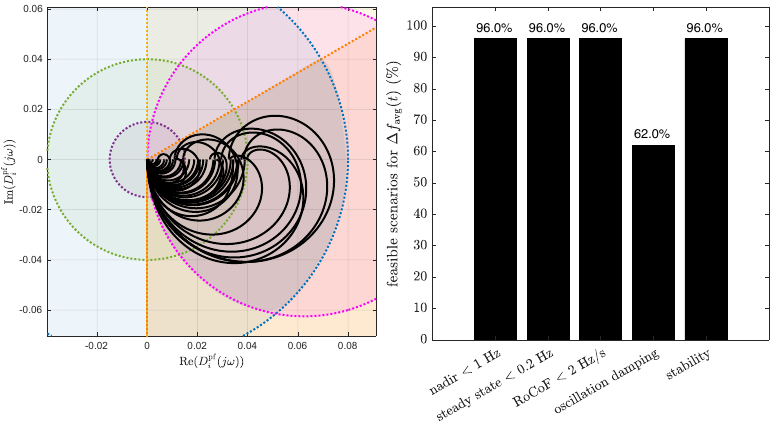}}
\vspace{-3mm}
    \caption{Nyquist plots of local control laws $D_i^\mathrm{pf}(s),\,i=1,2,3$ (left), and bar charts indicating the percentage of feasible stability and performance metric evaluations during Monte Carlo simulations in Case Study~III (right).}
    \label{fig:monte_carlo_frequency}
        \vspace{-2mm}
\end{figure}
\begin{figure}[t!]
    \centering
\scalebox{0.65}{\includegraphics[]{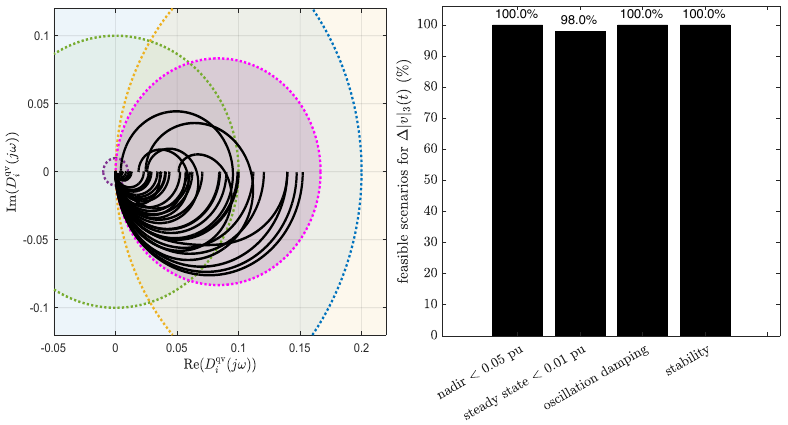}}
\vspace{-2mm}
\caption{Nyquist plots of the local control law $D_3^\mathrm{qv}(s)$ (left), and bar charts indicating the percentage of feasible stability and performance metric evaluations during Monte Carlo simulations in Case Study~III (right).}
    \label{fig:monte_carlo_voltage}
    \vspace{-4mm}
\end{figure}

To validate the NGGC framework under realistic conditions, we consider the IEEE nine-bus system with three grid-forming VSCs (\cref{fig:9bus}) and perform detailed nonlinear EMT simulations, including network dynamics and inner converter-control loops. Network and device parameters are provided in Appendix \ref{appendix:ninebus_parameters}. A Monte Carlo analysis is conducted by randomly varying the outer-loop controllers of all three VSCs within the NGGC feasible region. In particular, 50 feasible realizations of the frequency controllers $D_i^\mathrm{pf}(s)$, $i=\{1,2,3\}$, and 50 realizations of the voltage controller $D_3^\mathrm{qv}(s)$ are generated, as illustrated by the Nyquist plots on the left in \cref{fig:monte_carlo_frequency,fig:monte_carlo_voltage}. For each realization, system-level responses following small active- and reactive-power disturbances are evaluated.

The resulting average-mode frequency response to an active-power disturbance at bus~6 and the local voltage response to a reactive-power disturbance at bus~3 are evaluated against the NGGC stability and performance metrics. As shown on the right in \cref{fig:monte_carlo_frequency,fig:monte_carlo_voltage}, voltage stability and performance requirements are satisfied in nearly all cases. Similarly, frequency stability, nadir, steady-state deviation, and RoCoF constraints are met with close to 100\% reliability. The oscillation damping criterion, in turn, only achieves a lower acceptance rate of 62\%. The reduced damping performance is attributed to higher-order dynamics, including inner-loop and electromagnetic network effects, which are not captured by the considered NGGC modelling. Nevertheless, the results confirm that the NGGC framework provides reliable guarantees in detailed nonlinear EMT simulations, while indicating the potential for improved damping prediction by including higher-order dynamics.

\section{Conclusion}\label{sec:conclusion}
This paper introduced the Next-Generation Grid Code (NGGC) framework for dynamic ancillary service provision in future power systems. The proposed approach establishes system-wide stability through decentralized frequency-domain certificates and provides explicit performance bounds within the same analytical structure. Owing to its model-agnostic and non-parametric formulation, the framework is broadly applicable and does not require explicit device parameterization, thereby enabling the unified integration and assessment of grid-forming controls and synchronous machines.

Future research will focus on systematic controller design methods that explicitly satisfy the NGGC conditions. In addition, the framework will be extended to incorporate dynamic and lossy network models and to include grid-following devices. Finally, data-driven and identification-based representations of grid elements will be investigated to enable a more comprehensive characterization and enhancement of oscillation damping in practical future power systems.\\[-0.2cm]

\renewcommand{\baselinestretch}{0.92}
\bibliographystyle{IEEEtran}
\fontdimen2\font=0.6ex
\bibliography{IEEEabrv,mybibliography}

\renewcommand{\baselinestretch}{0.96}
\appendices
\section{Derivation of Average-Mode Frequency Response}\label{appendix:avg_mode}
To derive the average-mode frequency response in \cref{eq:cl_w_3}, we start from the closed-loop transfer function in \cref{eq:cl_w_2}, obtained after eigenvalue decomposition of the Laplacian matrix $L^{\mathrm{fp}}$, and apply the following algebraic manipulations:
\begin{align}\label{eq:avg_mode_derivation1}
\begin{split}
    \scriptstyle\Delta f(s)
    &\scriptstyle= \mathrm{diag}( D_i^\mathrm{pf}(s))
    (I+\frac{1}{s}
    V \Lambda V^\top
    \mathrm{diag}( D_i^\mathrm{pf}(s))
    )^{-1}\Delta p_\mathrm{d}(s)\\
    &\scriptstyle=( \mathrm{diag}( D_i^\mathrm{pf}(s)^{-1}) + \frac{1}{s}
    V \Lambda V^\top)^{-1}
    \Delta p_\mathrm{d}(s)\\
    &\scriptstyle= ( I+\frac{1}{s}
    V \Lambda V^\top-\mathrm{diag}( 1-D_i^\mathrm{pf}(s)^{-1}))^{-1}
    \Delta p_\mathrm{d}(s)\\
    &\scriptstyle= ( V(I+\frac{1}{s}\Lambda)V^\top-\mathrm{diag}( 1-D_i^\mathrm{pf}(s)^{-1}))^{-1}
    \Delta p_\mathrm{d}(s).
\end{split}
\end{align}
By introducing the transfer function $H(s)^{-1}:=I+\tfrac{1}{s}\Lambda$, we can rewrite \cref{eq:avg_mode_derivation1} accordingly as
\begin{align}\label{eq:avg_mode_derivation2}
\begin{split}
    \hspace{-2mm}\scriptstyle\Delta f(s)
    &\scriptstyle=( VH(s)^{-1}V^\top-\mathrm{diag}( 1-D_i^\mathrm{pf}(s)^{-1}))^{-1}
    \Delta p_\mathrm{d}(s)\\
    &\scriptstyle=(I-VH(s)V^\top\mathrm{diag}(1-D_i^\mathrm{pf}(s)^{-1}) )^{-1} VH(s)V^\top
    \Delta p_\mathrm{d}(s)\\
    &\scriptstyle= V(I-H(s)V^\top\mathrm{diag}(1-D_i^\mathrm{pf}(s)^{-1})V )^{-1} H(s)V^\top
    \Delta p_\mathrm{d}(s).\hspace{-2mm}
    \end{split}
\end{align}
The average-mode frequency response represents the dominant behavior on long time scales and is obtained by considering the quasi–steady-state limit $s \to 0$ of \cref{eq:avg_mode_derivation2}. To this end, we exploit the fact that $\lim_{s \to 0} H(s) = e_1 e_1^\top$, reflecting the dominant zero eigenvalue $\lambda_1 = 0$ of $L^\mathrm{pf}$, where $e_1 = [1,0,0,\ldots,0]^\top$ denotes the first standard basis vector in $\mathbb{R}^n$.
\begin{align*}
    \hspace{-2mm}\scriptstyle\underset{s\to 0}{\text{lim }}\Delta f(s)&\scriptstyle=\underset{s\to 0}{\text{lim }}\hspace{-0.5mm} V(I-H(s)V^\top\mathrm{diag}(1-D_i^\mathrm{pf}(s)^{-1})V )^{-1} H(s)V^\top
    \Delta p_\mathrm{d}(s)\hspace{-2mm}\\
     &\scriptstyle= V(I-e_1e_1^\top V^\top\mathrm{diag}(1-D_i^\mathrm{pf}(s\to 0)^{-1})V )^{-1} e_1e_1^\top V^\top
    \Delta p_\mathrm{d}(s\to 0)\hspace{-2mm}\\
    &\scriptstyle=V(I-e_1v_0^\top\mathrm{diag}(1-D_i^\mathrm{pf}(s\to 0)^{-1})V )^{-1} e_1e_1^\top V^\top
    \Delta p_\mathrm{d}(s\to 0)\hspace{-2mm}\\
    &\scriptstyle=Ve_1e_1^\top(e_1e_1^\top-e_1v_0^\top\mathrm{diag}(1-D_i^\mathrm{pf}(s\to 0)^{-1})Ve_1e_1^\top )^{-1} e_1e_1^\top V^\top
    \Delta p_\mathrm{d}(s\to 0)\hspace{-2mm}\\
    &\scriptstyle=v_0e_1^\top(e_1e_1^\top-e_1\frac{1}{n}\sum(1-D_i^\mathrm{pf}(s\to 0)^{-1})e_1^\top )^{-1} e_1e_1^\top V^\top
    \Delta p_\mathrm{d}(s\to 0)\hspace{-2mm}\\
    &\scriptstyle=v_0e_1^\top(e_1(1-\frac{1}{n}\sum(1-D_i^\mathrm{pf}(s\to 0)^{-1}))e_1^\top )^{-1} e_1e_1^\top V^\top
    \Delta p_\mathrm{d}(s\to 0)\hspace{-2mm}\\
    &\scriptstyle=v_0e_1^\top(e_1(\frac{1}{n}\sum D_i^\mathrm{pf}(s\to 0)^{-1})e_1^\top )^{-1} e_1e_1^\top V^\top
    \Delta p_\mathrm{d}(s\to 0)\hspace{-2mm}\\
    &\scriptstyle=v_0e_1^\top V^\top(\frac{1}{n}\sum D_i^\mathrm{pf}(s\to 0)^{-1}))^{-1}
    \Delta p_\mathrm{d}(s\to 0)\hspace{-2mm}\\
    &\scriptstyle=v_0v_0^\top n(\sum_{i=1}^{n}(D_i^\mathrm{pf}(s\to 0))^{-1})^{-1}
    \Delta p_\mathrm{d}(s\to 0)\hspace{-2mm}\\
&\scriptstyle=(\sum_{i=1}^{n} (D_i^\mathrm{pf}(s\to 0))^{-1})^{-1} 
    \mathbf{1}_n \mathbf{1}_n^\top    \Delta p_\mathrm{d}(s\to 0)
\end{align*}
which corresponds to \cref{eq:cl_w_3}, and $v_0=\tfrac{1}{\sqrt{n}}\mathbf{1}_n$ is the eigenvector associated with the zero eigenvalue $\lambda_1=0$ of $L^\mathrm{pf}$.

\section{Proof of $\mathrm{qv}$-Stability \fontdimen2\font=0.6ex Certification Criteria}\label{appendix:stability_qv}
For the $\mathrm{qv}$ control, strict passivity of $D_i'(s)$ in \cref{eq:D_dash} requires
\begin{align}\label{eq:D_i_qv_dash}
    \scriptstyle D_{i}^{\mathrm{qv}\,\prime}(s)=\frac{D_i^\mathrm{qv}(s)}{1-c_iD_i^\mathrm{qv}(s)}
\end{align}
to be strictly passive. According to the standard criteria for strictly passive transfer functions \cite{haberle2025decentralized}, $D_{i}^{\mathrm{qv}\,\prime}(s)$ must be stable and satisfy $\mathrm{Re}[D_{i}^{\mathrm{qv}\,\prime}(\mathrm{j}\omega)] > 0$. 

Stability of $D_{i}^{\mathrm{qv}\,\prime}(s)$ is ensured if the roots of $1-c_iD_i^\mathrm{qv}(s) = 0$ lie in the open left-half plane. Applying the Argument Principle \cite{feedback2015franklin}, for $D_i^\mathrm{qv}(s)$ stable (i.e., with no poles in the closed right-half plane; cf. condition \ref{cond:2i}), it is required that the Nyquist plot of $D_i^\mathrm{qv}(s)$ does \emph{not} encircle the point $(1/c_i,0)$ in the clockwise direction. This is guaranteed by condition \ref{cond:2ii}.

For $\mathrm{Re}[D_{i}^{\mathrm{qv},\prime}(\mathrm{j}\omega)] > 0$ to hold for all $\omega\in(-\infty,\infty)$, we can equivalently write
\begin{align}
\begin{split}
    \scriptstyle\mathrm{Re}[D_{i}^{\mathrm{qv}\,\prime}(\mathrm{j}\omega)] > 0 &\scriptstyle\Leftrightarrow \angle(D_{i}^{\mathrm{qv}\,\prime}(\mathrm{j}\omega))\in (-\frac{\pi}{2},\frac{\pi}{2}) \\
    &\scriptstyle\Leftrightarrow\angle \left(\frac{D_i^\mathrm{qv}(\mathrm{j}\omega)}{1 - c_i D_i^\mathrm{qv}(\mathrm{j}\omega)}  \right)\in (-\frac{\pi}{2},\frac{\pi}{2})\\
    &\scriptstyle\Leftrightarrow\angle\left(\frac{1}{D_i^\mathrm{qv}(\mathrm{j}\omega)^{-1} - c_i }  \right)\in (-\frac{\pi}{2},\frac{\pi}{2})\\
    &\scriptstyle\Leftrightarrow-\angle(D_i^\mathrm{qv}(\mathrm{j}\omega)^{-1} - c_i ) \in (-\frac{\pi}{2},\frac{\pi}{2})\\
    &\scriptstyle\Leftrightarrow c_i < \mathrm{Re}[D_i^{\mathrm{qv}}(\mathrm{j}\omega)^{-1}],
    \end{split}
\end{align}
which corresponds to condition \ref{cond:2ii}.

Finally, for $\bar{\sigma}(N'(\mathrm{j}\infty))\bar{\sigma}(D'(\mathrm{j}\infty))<1$, we require $D_i^\mathrm{qv}(s)$ to be be strictly proper which ensures that $\bar{\sigma}(D_{i}^{\mathrm{qv}\,\prime}(\mathrm{j}\infty))=0$.

\section{Proof of $\mathrm{pf}$-Performance Certification Criteria}\label{appendix:performance_pf}
\fontdimen2\font=0.6ex The proof proceeds in two steps. First, we show that the decentralized conditions in \ref{cond:1i}-\ref{cond:1viii} imply the properties of the average-mode frequency dynamics $D_\mathrm{avg}(s)$ stated in \cref{eq:D_avg_characteristics}. Second, assuming these properties of $D_\mathrm{avg}(s)$, we demonstrate that the global performance specifications in \cref{eq:global_f_avg_specs_rigorous} are satisfied.

\subsection{Conditions \ref{cond:1i}-\ref{cond:1viii}  $\Rightarrow$ Characteristics \cref{eq:D_avg_characteristics}}\label{appendix:performance_pf_part1}
\subsubsection{Property \cref{eq:avg_1}} Stability of $D_\mathrm{avg}(s)$ follows directly from the stability of the overall closed-loop system, which is guaranteed by conditions \ref{cond:1i} and \ref{cond:1ii}. Strict properness of $D_\mathrm{avg}(s)$ can be shown by expressing the rational device transfer functions $D_i^\mathrm{pf}(s)=\tfrac{n_i^\mathrm{pf}(s)}{d_i^\mathrm{pf}(s)}$ in terms of their numerator and denominator polynomials, where $\text{deg}(n_i^\mathrm{pf}(s))=m_i\in\mathbb{Z}$ and $\text{deg}(d_i^\mathrm{pf}(s))=n_i\in\mathbb{Z}$ with $m_i<n_i,\,\forall i\in\{ 1,...,n\}$ by condition \ref{cond:1i}. Based on this, $D_\mathrm{avg}(s)$ can be written as
\begin{align}
    \scriptstyle D_\mathrm{avg}(s)=\frac{1}{\sum_{i=1}^n(D_i^\mathrm{pf}(s))^{-1}}=\frac{\prod_{i=1}^n n_i(s)}{\sum_{i=1}^n d_i(s)\prod_{j\ne i}^n n_j(s)},
\end{align}
with the numerator degree $\text{deg}(\prod_{i=1}^n n_i(s))\hspace{-1mm}=\hspace{-1mm} \textstyle\sum_{i=1}^n m_i$ and the denominator degree $\text{deg}(\sum_{i=1}^n d_i(s)\prod_{j\ne i}^n n_j(s))=\text{sup}_{i\in\{1,...,n\}}(\textstyle\sum_{j\ne i}m_j+n_i) > \textstyle \sum_{i=1}^n m_i$, rendering $D_\mathrm{avg}(s)$ strictly proper.
\subsubsection{Property \cref{eq:avg_2}} We consider the phase of $D_\mathrm{avg}(\mathrm{j}\omega)$ and apply condition \ref{cond:1iii}, i.e.,
\begin{align}
\begin{split}
    \scriptstyle\angle\,(D_\mathrm{avg}(\mathrm{j}\omega)) &\scriptstyle= \angle \left( \frac{1}{\sum_{i=1}^n(D_i^\mathrm{pf}(s))^{-1}} \right)\\
    & \scriptstyle=\scriptstyle - \angle (\sum_{i=1}^n(D_i^\mathrm{pf}(s))^{-1}) \in [-\frac{\pi}{2},\frac{\pi}{6}],
    \end{split}
\end{align}
which implies $\text{Re}[D_\mathrm{avg}(\mathrm{j}\omega)]\geq 0$ for all $\omega\in [-\infty,\infty]$.

\subsubsection{Property \cref{eq:avg_3}}\label{sec:induction_proof} We start by considering $|D_\mathrm{avg}(\mathrm{j}\omega)|$, which can be upper bounded by
\begin{align}\label{eq:D_avg_magnitude_upper_bound}
\begin{split}
\scriptstyle|D_\mathrm{avg}(\mathrm{j}\omega)|&\scriptstyle=\left| \frac{1}{\sum_{i=1}^nD_i^\mathrm{pf}(\mathrm{j}\omega)^{-1}}\right|\\
&\scriptstyle\leq{\scriptstyle\hspace{-0.1mm} \frac{1}{2}\hspace{-0.1mm}\left( \sum_{i=2}^n(\frac{1}{2})^{n-i}|D_i^\mathrm{pf}(\mathrm{j}\omega)|\hspace{-0.5mm}+\hspace{-0.5mm}(\frac{1}{2})^{n-2}|D_1^\mathrm{pf}(\mathrm{j}\omega)|\right)}\hspace{-0.5mm}.\hspace{-2.5mm}
\end{split}
\end{align}
The inequality in \cref{eq:D_avg_magnitude_upper_bound} follows from an induction argument: For the \emph{Induction Start ($n=2$),} we can show that 
    \begin{align}\label{eq:induction_start}
        \scriptstyle\left| \frac{1}{D_1^\mathrm{pf}(\mathrm{j}\omega)^{-1}+D_2^\mathrm{pf}(\mathrm{j\omega})^{-1}}\right| \leq \frac{1}{2}\left(|D_1^\mathrm{pf}(\mathrm{j}\omega)|+|D_2^\mathrm{pf}(\mathrm{j}\omega)|  \right).
    \end{align}
In particular, to show \cref{eq:induction_start}, we consider the polar coordinate representation $D_i^\mathrm{pf}(\mathrm{j}\omega)=r_i(\omega)e^{\mathrm{j}\varphi_i(\omega)}$, where $r_i(\omega):= |D_i^\mathrm{pf}(\mathrm{j}\omega)|$ and $\varphi_i(\omega):=\angle(D_i^\mathrm{pf}(\mathrm{j}\omega))$, and rewrite \cref{eq:induction_start} as
\begin{align}\label{eq:induction_start2}
   \scriptstyle \left| \frac{1}{\frac{1}{r_1(\omega)}e^{-\mathrm{j}\varphi_1(\omega)}+\frac{1}{r_2(\omega)}e^{-\mathrm{j}\varphi_2(\omega)}} \right| \leq \frac{1}{2}(r_1(\omega)+r_2(\omega)).
\end{align}
To show \cref{eq:induction_start2}, we can equivalently show
\begin{align}\label{eq:induction_start3}
    \scriptstyle\left| \frac{1}{r_1(\omega)}e^{-\mathrm{j}\varphi_1(\omega)}+\frac{1}{r_2(\omega)}e^{-\mathrm{j}\varphi_2(\omega)} \right| \geq \frac{2}{r_1(\omega)+r_2(\omega)}
\end{align}
Indeed, the left side of \cref{eq:induction_start3} will be smallest if the phase difference $|-\varphi_1(\omega)+\varphi_2(\omega)|=|\tfrac{\pi}{2}+\tfrac{\pi}{6}|$ is largest, by using condition \ref{cond:1iii}. We can thus lower bound the left side as
\begin{align}\label{eq:induction_start4}
    \scriptstyle \left| \frac{1}{r_1(\omega)}e^{-\mathrm{j}\varphi_1(\omega)} + \frac{1}{r_2(\omega)}e^{-\mathrm{j}\varphi_2(\omega)} \right|  \geq \left| \frac{1}{r_1(\omega)}e^{\mathrm{j}\frac{\pi}{2}} + \frac{1}{r_2(\omega)}e^{-\mathrm{j}\frac{\pi}{6}} \right| \hspace{-1.5mm}
\end{align}
So it remains to show that
\begin{align}\label{eq:induction_start5}
   \scriptstyle \left| \frac{1}{r_1(\omega)}e^{\mathrm{j}\frac{\pi}{2}}+\frac{1}{r_2(\omega)}e^{-\mathrm{j}\frac{\pi}{6}} \right|\geq \frac{2}{r_1(\omega)+r_2(\omega)}
\end{align}
Using the law of cosines, we can write
\begin{align}\label{eq:induction_start6}
    \scriptstyle\hspace{-1.5mm}\left| \frac{1}{r_1(\omega)}e^{\mathrm{j}\frac{\pi}{2}}+\frac{1}{r_2(\omega)}e^{-\mathrm{j}\frac{\pi}{6}} \right|^2 = \frac{1}{r_1(\omega)^2}+\frac{1}{r_2(\omega)^2}-\frac{1}{r_1(\omega)r_2(\omega)}\hspace{-1.5mm}
\end{align}
Next, using \cref{eq:induction_start6}, we define the positive scaling factor $x(\omega)=\tfrac{r_1(\omega)}{r_2(\omega)} \ge 0$ at the frequency $\omega$ and square \cref{eq:induction_start5}, i.e.,
\begin{align}
\begin{split}
    &\scriptstyle\frac{1}{r_2(\omega)^2}\left(\frac{1}{x(\omega)^2}+1-\frac{1}{x(\omega)} \right)\geq \frac{4}{r_2(\omega)^2(1+x(\omega))^2}\\
     \scriptstyle\Leftrightarrow\, &\scriptstyle\left(\frac{1}{x(\omega)^2}+1-\frac{1}{x(\omega)} \right) \geq \frac{4}{(1+x(\omega))^2}\\
     \scriptstyle\Leftrightarrow\, &\scriptstyle-1\geq \frac{4x(\omega)}{(1+x(\omega))^2}-x(\omega)-\frac{1}{x(\omega)},
\end{split}
\end{align}
where the right side is maximized for $x(\omega)=1$, and therefore proves the inequality in \cref{eq:induction_start}.

In the \emph{Induction Hypothesis ($n=k$)}, we assume that
    \begin{align}\label{eq:induction_hypothesis}
        \scriptstyle\left| \frac{1}{\sum_{i=1}^kD_i^\mathrm{pf}(\mathrm{j}\omega)^{-1}}\right|\leq{\scriptstyle\frac{1}{2}\hspace{-0.75mm}\left(\sum_{i=2}^k(\frac{1}{2})^{k-i}|D_i^\mathrm{pf}(\mathrm{j}\omega)|\hspace{-0.5mm}+\hspace{-0.5mm}(\frac{1}{2})^{k-2}|D_1^\mathrm{pf}(\mathrm{j}\omega)|\hspace{-0.5mm}\right)}\hspace{-1mm}
    \end{align}
    holds true. In the \emph{Induction Step ($n=k+1$)}, we can show that 
    \begin{align}\label{eq:induction_step}
        \scriptstyle\hspace{-3mm}\left| \frac{1}{\sum_{i=1}^{k+1}D_i^\mathrm{pf}(\mathrm{j}\omega)^{-1}}\right|\leq{\scriptstyle\frac{1}{2}\hspace{-0.75mm}\left( \sum_{i=2}^{k+1}(\frac{1}{2})^{k+1-i}|D_i^\mathrm{pf}(\mathrm{j}\omega)|\hspace{-0.5mm}+\hspace{-0.5mm}(\frac{1}{2})^{k-1}|D_1^\mathrm{pf}(\mathrm{j}\omega)|\hspace{-0.5mm}\right)}\hspace{-2mm}
    \end{align}
holds true. Namely, by defining $\tilde{D}(\mathrm{j}\omega)^{-1}:=\textstyle\sum_{i=1}^{k}D_i^\mathrm{pf}(\mathrm{j}\omega)^{-1}$, and by using the induction start in \cref{eq:induction_start}, we can rewrite \cref{eq:induction_step} as
\begin{align}\label{eq:induction_step2}
\begin{split}
    \scriptstyle\hspace{-1mm}\left| \frac{1}{\sum_{i=1}^{k+1}D_i^\mathrm{pf}(\mathrm{j}\omega)^{-1}}\right|&\scriptstyle=\left| \frac{1}{\tilde{D}(\mathrm{j}\omega)^{-1}+D_{k+1}^\mathrm{pf}(\mathrm{j}\omega)^{-1}}\right|\\
    &\scriptstyle\leq \frac{1}{2}\left( |\tilde{D}(\mathrm{j}\omega)|+|D_{k+1}^\mathrm{pf}(\mathrm{j}\omega)|\right).
    \end{split}
\end{align}
Furthermore, by the induction hypothesis in \cref{eq:induction_hypothesis}, we have
\begin{align}\label{eq:induction_step3}
\begin{split}
   \scriptstyle |\tilde{D}(\mathrm{j}\omega)|&\scriptstyle= \left| \frac{1}{\sum_{i=1}^kD_i^\mathrm{pf}(\mathrm{j}\omega)^{-1}}\right|\\
    & \scriptstyle\leq {\scriptstyle\frac{1}{2}\hspace{-0.75mm}\left(\sum_{i=2}^k(\frac{1}{2})^{k-i}|D_i^\mathrm{pf}(\mathrm{j}\omega)|\hspace{-0.5mm}+\hspace{-0.5mm}(\frac{1}{2})^{k-2}|D_1^\mathrm{pf}(\mathrm{j}\omega)|\hspace{-0.5mm}\right)}\hspace{-0.5mm},
\end{split}
\end{align}
based on which \cref{eq:induction_step2} can be upper bounded as
\begin{align}\notag
    \scriptstyle\hspace{-1mm}\left| \frac{1}{\sum_{i=1}^{k+1}D_i^\mathrm{pf}(\mathrm{j}\omega)^{-1}}\right|\hspace{-0.5mm}&\leq 
    {\scriptstyle\frac{1}{4}\hspace{-0.75mm}\left(\sum_{i=2}^k(\frac{1}{2})^{k-i}|D_i^\mathrm{pf}(\mathrm{j}\omega)|\hspace{-0.5mm}+\hspace{-0.5mm}(\frac{1}{2})^{k-2}|D_1^\mathrm{pf}(\mathrm{j}\omega)|\hspace{-0.5mm}\right)}\\
     \scriptstyle& \scriptstyle\quad\,+{\scriptstyle\frac{1}{2}|D_{k+1}^\mathrm{pf}(\mathrm{j}\omega)|}\\\notag
     \scriptstyle& \scriptstyle=  {\scriptstyle\frac{1}{2}\hspace{-0.75mm}\left( \sum_{i=2}^{k+1}(\frac{1}{2})^{k+1-i}|D_i^\mathrm{pf}(\mathrm{j}\omega)|\hspace{-0.5mm}+\hspace{-0.5mm}(\frac{1}{2})^{k-1}|D_1^\mathrm{pf}(\mathrm{j}\omega)|\hspace{-0.5mm}\right)}\hspace{-0.5mm},\hspace{-1mm}
\end{align}
which corresponds to \cref{eq:induction_step} and proves the inequality in \cref{eq:D_avg_magnitude_upper_bound}.

Finally, for $\omega\geq\omega_\mathrm{bw}$, we can rewrite \cref{eq:D_avg_magnitude_upper_bound} using \ref{cond:1iv}, and applying the geometric series expansion, i.e., 
\begin{align}\label{eq:simplify_D_avg_bound}
    \begin{split}
         \scriptstyle\hspace{-2.5mm}|D_\mathrm{avg}(\mathrm{j}\omega)|& \scriptstyle\leq {\scriptstyle \frac{1}{2}\hspace{-0.75mm}\left( \sum_{i=2}^n(\frac{1}{2})^{n-i}|D_i^\mathrm{pf}(\mathrm{j}\omega)|\hspace{-0.5mm}+\hspace{-0.5mm}(\frac{1}{2})^{n-2}|D_1^\mathrm{pf}(\mathrm{j}\omega)|\right)}\\
        & \scriptstyle\leq {\scriptstyle\frac{1}{2}\hspace{-0.75mm}\left( \sum_{i=2}^n(\frac{1}{2})^{n-i}\varepsilon_\mathrm{f}\hspace{-0.5mm}+\hspace{-0.5mm}(\frac{1}{2})^{n-2}\varepsilon_\mathrm{f}\right)}\\
        & \scriptstyle= {\scriptstyle\frac{1}{2}\varepsilon_\mathrm{f}\left (\sum_{i=2}^n(\frac{1}{2})^{n-i}+(\frac{1}{2})^{n-2}\right)}\\
        & \scriptstyle={\scriptstyle\frac{1}{2}\varepsilon_\mathrm{f}\left (\sum_{k=0}^{n-2}(\frac{1}{2})^{k}+(\frac{1}{2})^{n-2}\right)}\\
        & \scriptstyle={\scriptstyle\frac{1}{2}\varepsilon_\mathrm{f} \left(\frac{1-(\frac{1}{2})^{n-1}}{1-\frac{1}{2}} +(\frac{1}{2})^{n-2}\right)}= {\scriptstyle\varepsilon_\mathrm{f}},
    \end{split}
\end{align}
which proves the property in \cref{eq:avg_3}.

\subsubsection{Property \cref{eq:avg_4}} We can rewrite \cref{eq:D_avg_magnitude_upper_bound} using \ref{cond:1v}, and applying the same steps as in \cref{eq:simplify_D_avg_bound}, i.e., 
\begin{align}\nonumber
    \scriptstyle\hspace{-2.5mm}||D_\mathrm{avg}(\mathrm{j}\omega)||_\infty& \scriptstyle\leq {\scriptstyle \frac{1}{2}\hspace{-0.75mm}\left( \sum_{i=2}^n(\frac{1}{2})^{n-i}||D_i^\mathrm{pf}(\mathrm{j}\omega)||_\infty\hspace{-0.5mm}+\hspace{-0.5mm}(\frac{1}{2})^{n-2}||D_1^\mathrm{pf}(\mathrm{j}\omega)||_\infty\right)}\\\nonumber
                & \scriptstyle\hspace{-0.5mm}\leq {\scriptstyle \frac{1}{2}\hspace{-0.75mm}\left( \sum_{i=2}^n(\frac{1}{2})^{n-i} \frac{\Delta f_\mathrm{max}}{2.5\Delta p_\mathrm{d}^{\tikz[baseline=0] \draw (0.05,0.05) -- (0.15,0.05) -- (0.15,0.15) -- (0.25,0.15);}}\hspace{-0.5mm}+\hspace{-0.5mm}(\frac{1}{2})^{n-2}\frac{\Delta f_\mathrm{max}}{2.5\Delta p_\mathrm{d}^{\tikz[baseline=0] \draw (0.05,0.05) -- (0.15,0.05) -- (0.15,0.15) -- (0.25,0.15);}}\right)}\\
                & \scriptstyle={\scriptstyle\frac{1}{2}\frac{\Delta f_\mathrm{max}}{2.5\Delta p_\mathrm{d}^{\tikz[baseline=0] \draw (0.05,0.05) -- (0.15,0.05) -- (0.15,0.15) -- (0.25,0.15);}}\left (\sum_{i=2}^n(\frac{1}{2})^{n-i}+(\frac{1}{2})^{n-2}\right)}\\\nonumber
                & \scriptstyle= {\scriptstyle\frac{\Delta f_\mathrm{max}}{2.5\Delta p_\mathrm{d}^{\tikz[baseline=0] \draw (0.05,0.05) -- (0.15,0.05) -- (0.15,0.15) -- (0.25,0.15);}}},
\end{align}
which corresponds to the property in \cref{eq:avg_4}.

\subsubsection{Property \cref{eq:avg_5}} We can evaluate \cref{eq:D_avg_magnitude_upper_bound} at $\omega = 0$ using \ref{cond:1vi}, and apply the same steps as in \cref{eq:simplify_D_avg_bound}, i.e., 
\begin{align}
    \nonumber
                 \scriptstyle\hspace{-2.5mm}|D_\mathrm{avg}(0)|& \scriptstyle\leq {\scriptstyle \frac{1}{2}\hspace{-0.75mm}\left( \sum_{i=2}^n(\frac{1}{2})^{n-i}|D_i^\mathrm{pf}(0)|\hspace{-0.5mm}+\hspace{-0.5mm}(\frac{1}{2})^{n-2}|D_1^\mathrm{pf}(0)|\right)}\\\nonumber
                & \scriptstyle\hspace{-0.5mm}\leq {\scriptstyle \frac{1}{2}\hspace{-0.75mm}\left( \sum_{i=2}^n(\frac{1}{2})^{n-i} \frac{\Delta f_\mathrm{ss,\,max}}{\Delta p_\mathrm{d}^{\tikz[baseline=0] \draw (0.05,0.05) -- (0.15,0.05) -- (0.15,0.15) -- (0.25,0.15);}}\hspace{-0.5mm}+\hspace{-0.5mm}(\frac{1}{2})^{n-2}\frac{\Delta f_\mathrm{ss,\,max}}{\Delta p_\mathrm{d}^{\tikz[baseline=0] \draw (0.05,0.05) -- (0.15,0.05) -- (0.15,0.15) -- (0.25,0.15);}}\right)}\\
                & \scriptstyle={\scriptstyle\frac{1}{2}\frac{\Delta f_\mathrm{ss,\,max}}{\Delta p_\mathrm{d}^{\tikz[baseline=0] \draw (0.05,0.05) -- (0.15,0.05) -- (0.15,0.15) -- (0.25,0.15);}}\left (\sum_{i=2}^n(\frac{1}{2})^{n-i}+(\frac{1}{2})^{n-2}\right)}\\\nonumber
                & \scriptstyle= {\scriptstyle\frac{\Delta f_\mathrm{ss,\,max}}{\Delta p_\mathrm{d}^{\tikz[baseline=0] \draw (0.05,0.05) -- (0.15,0.05) -- (0.15,0.15) -- (0.25,0.15);}}},
\end{align}
which corresponds to the property in \cref{eq:avg_5}.

\subsubsection{Property \cref{eq:avg_6}} We first realize that
\begin{align}\label{eq:sDavg_rewritten}
     \scriptstyle\mathrm{j}\omega D_\mathrm{avg}(\mathrm{j}\omega)=\frac{\mathrm{j}\omega}{\sum_{i=1}^n D_i^\mathrm{pf}(\mathrm{j}\omega)^{-1}}=\frac{1}{\sum_{i=1}^n(\mathrm{j}\omega D_i^\mathrm{pf}(\mathrm{j}\omega))^{-1}}.
\end{align}
Next, by taking the magnitude of \cref{eq:sDavg_rewritten}, we can show that 
\begin{align}\label{eq:sD_avg_magnitude_upper_bound}
\begin{split}
 \scriptstyle \hspace{-3.5mm}|\mathrm{j}\omega D_\mathrm{avg}(\mathrm{j}\omega)| & \scriptstyle=\left| \frac{1}{\sum_{i=1}^n(\mathrm{j}\omega D_i^\mathrm{pf}(\mathrm{j}\omega))^{-1}}\right|\\
 & \scriptstyle\leq{\scriptstyle \frac{1}{2}\left( \sum_{i=2}^n(\frac{1}{2})^{n-i}|(\mathrm{j}\omega D_i^\mathrm{pf}(\mathrm{j}\omega)|\hspace{-0.5mm}+\hspace{-0.5mm}(\frac{1}{2})^{n-2}|\mathrm{j}\omega D_1^\mathrm{pf}(\mathrm{j}\omega)|\right)}\hspace{-0.5mm},\hspace{-2.5mm}
 \end{split}
\end{align}
which follows from an induction argument analogous to that presented in \cref{sec:induction_proof}. The only modification is the incorporation of an additional phase offset of $+\tfrac{\pi}{2}$ in equation \cref{eq:induction_start4}, arising from the multiplication of $D_i^\mathrm{pf}(\mathrm{j}\omega)$ by $\mathrm{j}\omega$. 

Finally, we can evaluate \cref{eq:sD_avg_magnitude_upper_bound} at $\omega \rightarrow \infty$ using \ref{cond:1vii}, and apply the same steps as in \cref{eq:simplify_D_avg_bound}, i.e., 
\begin{align}
\begin{split}
     \scriptstyle|\underset{\omega \rightarrow \infty}{\text{lim}}\,\mathrm{j}\omega D_\mathrm{avg}(\mathrm{j}\omega)| & \scriptstyle\leq {\scriptstyle \frac{1}{2}\left( \sum_{i=2}^n(\frac{1}{2})^{n-i}|\underset{\omega \rightarrow \infty}{\text{lim}}\,\mathrm{j}\omega D_i^\mathrm{pf}(\mathrm{j}\omega)| + \right.}\\
    & \scriptstyle\,\,\,\quad\quad\quad\quad{\left.\scriptstyle +(\frac{1}{2})^{n-2}|\underset{\omega \rightarrow \infty}{\text{lim}}\,\mathrm{j}\omega D_1^\mathrm{pf}(\mathrm{j}\omega)|\right)}\\\nonumber
    & \scriptstyle={\scriptstyle\frac{1}{2}\frac{\Delta\dot{f}_\mathrm{max}}{\Delta p_\mathrm{d}^{\tikz[baseline=0] \draw (0.05,0.05) -- (0.15,0.05) -- (0.15,0.15) -- (0.25,0.15);}}\left (\sum_{i=2}^n(\frac{1}{2})^{n-i}+(\frac{1}{2})^{n-2}\right)}= {\scriptstyle\frac{\Delta\dot{f}_\mathrm{max}}{\Delta p_\mathrm{d}^{\tikz[baseline=0] \draw (0.05,0.05) -- (0.15,0.05) -- (0.15,0.15) -- (0.25,0.15);}}},
    \end{split}
\end{align}
which corresponds to the property in \cref{eq:avg_6}.

\subsubsection{Property \cref{eq:avg_7}} We can express $\text{Re}[D_\mathrm{avg}(\mathrm{j}\omega)^{-1}]$ in terms of $\text{Re}[D_i^\mathrm{pf}(\mathrm{j}\omega)^{-1}]$ and use \ref{cond:1viii} as
\begin{align}
\begin{split}
     \scriptstyle\text{Re}[D_\mathrm{avg}(\mathrm{j}\omega)^{-1}] & \scriptstyle= \sum_{i=1}^n\text{Re}[D_i^\mathrm{pf}(\mathrm{j}\omega]^{-1}\geq n\eta_\mathrm{f}
    \end{split}
\end{align}
which corresponds to the property in \cref{eq:avg_7}.

\subsection{Characteristics \cref{eq:D_avg_characteristics} $\Rightarrow$ Global Performance Specs \cref{eq:global_f_avg_specs_rigorous}}\label{appendix:performance_pf_part2} 
\subsubsection{Bounded nadir} To show that the nadir bound in \cref{eq:nadir_spec} is satisfied, we recall from \cite{papoulis1962The,franchek1998direct,jayasuriya1992qft} that the step response of a stable and causal transfer function (cf. \cref{eq:avg_1}) can be uniquely expressed in terms of $\text{Re}[D_\mathrm{avg}(\mathrm{j}\omega)]$, i.e., 
\begin{align}\label{eq:step_response_integral1}
     \scriptstyle\Delta {f}^{\tikz[baseline=0] \draw (0.05,0.05) -- (0.15,0.05) -- (0.15,0.15) -- (0.25,0.15);}_\mathrm{avg}(t)=\frac{2}{\pi}\Delta p_\mathrm{d}^{{\tikz[baseline=0] \draw (0.05,0.05) -- (0.15,0.05) -- (0.15,0.15) -- (0.25,0.15);}}\hspace{-1mm}\int_0^\infty \frac{\text{Re}[D_\mathrm{avg}(\mathrm{j}\omega)]}{\omega} \sin({\omega t})d\omega,\quad t > 0.
\end{align}
By applying \cref{eq:avg_3} for some $\varepsilon_\mathrm{f}\approx 0$, we have that $\text{Re}[D_\mathrm{avg}(\mathrm{j}\omega)]\approx 0,\, \forall \omega\geq \omega_\mathrm{bw}$ and thus rewrite \cref{eq:step_response_integral1} as
\begin{align}\label{eq:step_response_integral2}
    \scriptstyle \Delta {f}^{\tikz[baseline=0] \draw (0.05,0.05) -- (0.15,0.05) -- (0.15,0.15) -- (0.25,0.15);}_\mathrm{avg}(t)=\frac{2}{\pi}\Delta p_\mathrm{d}^{{\tikz[baseline=0] \draw (0.05,0.05) -- (0.15,0.05) -- (0.15,0.15) -- (0.25,0.15);}}\hspace{-1mm}\int_0^{\omega_\mathrm{bw}} \frac{\text{Re}[D_\mathrm{avg}(\mathrm{j}\omega)]}{\omega} \sin({\omega t})d\omega,\quad t > 0.
\end{align}
To derive an upper bound on the time-domain nadir, similar as in \cite{jayasuriya1992qft}, we take the absolute value of \cref{eq:step_response_integral2} and use the upper bound $|\text{Re}[D_\mathrm{avg}(\mathrm{j}\omega)]|\leq ||D_\mathrm{avg}(\mathrm{j}\omega)||_\infty$, i.e., 
\begin{align}\label{eq:step_response_integral3}
    \begin{split}
         \scriptstyle|\Delta {f}^{\tikz[baseline=0] \draw (0.05,0.05) -- (0.15,0.05) -- (0.15,0.15) -- (0.25,0.15);}_\mathrm{avg}(t)|& \scriptstyle= \scriptstyle\left| \frac{2}{\pi}\Delta p_\mathrm{d}^{{\tikz[baseline=0] \draw (0.05,0.05) -- (0.15,0.05) -- (0.15,0.15) -- (0.25,0.15);}}\hspace{-1mm}\int_0^{\omega_\mathrm{bw}} \frac{\text{Re}[D_\mathrm{avg}(\mathrm{j}\omega)]}{\omega} \sin({\omega t})d\omega\right|\\
        & \scriptstyle\leq\frac{2}{\pi}\Delta p_\mathrm{d}^{{\tikz[baseline=0] \draw (0.05,0.05) -- (0.15,0.05) -- (0.15,0.15) -- (0.25,0.15);}}\hspace{-1mm}\int_0^{\omega_\mathrm{bw}} \left|\frac{\text{Re}[D_\mathrm{avg}(\mathrm{j}\omega)]}{\omega} \sin({\omega t})\right|d\omega\\
        & \scriptstyle= \scriptstyle \frac{2}{\pi}\Delta p_\mathrm{d}^{{\tikz[baseline=0] \draw (0.05,0.05) -- (0.15,0.05) -- (0.15,0.15) -- (0.25,0.15);}}\hspace{-1mm}\int_0^{\omega_\mathrm{bw}} \left|\text{Re}[D_\mathrm{avg}(\mathrm{j}\omega)] \right| \hspace{-1mm}\left|\frac{\sin({\omega t})}{\omega}\right|d\omega\\
        & \scriptstyle\leq \frac{2}{\pi}\Delta p_\mathrm{d}^{{\tikz[baseline=0] \draw (0.05,0.05) -- (0.15,0.05) -- (0.15,0.15) -- (0.25,0.15);}}\hspace{-1mm}\int_0^{\omega_\mathrm{bw}} ||D_\mathrm{avg}(\mathrm{j}\omega)||_\infty \hspace{-1mm}\left|\frac{\sin({\omega t})}{\omega}\right|d\omega\\
        & \scriptstyle= \frac{2}{\pi}\Delta p_\mathrm{d}^{{\tikz[baseline=0] \draw (0.05,0.05) -- (0.15,0.05) -- (0.15,0.15) -- (0.25,0.15);}}||D_\mathrm{avg}(\mathrm{j}\omega)||_\infty \hspace{-1mm}\int_0^{\omega_\mathrm{bw}}\left|\frac{\sin({\omega t})}{\omega}\right|d\omega,\quad t > 0.
    \end{split}
\end{align}
Next, we perform the coordinate change $u:=\omega t$ which transforms the integral in \cref{eq:step_response_integral3} as
\begin{align}
\begin{split}
 \scriptstyle\int_0^{\omega_\mathrm{bw}}\left|\frac{\sin({\omega t})}{\omega}\right|d\omega & \scriptstyle= \int_0^{\omega_\mathrm{bw}t}\left|\frac{\sin({u})}{u}\right|du\\
    & \scriptstyle\leq\int_0^{\omega_\mathrm{bw}t_\mathrm{max}}\left|\frac{\sin({u})}{u}\right|du\approx 4,
    \end{split}
\end{align}
where $t_\mathrm{max}$ denotes the maximum time of interest for the dynamic response of the average-mode frequency before higher-level secondary control actions are activated. For typical values $\omega_{\mathrm{bw}} \approx 2\pi \cdot 5\mathrm{rad/s}$ and $t_{\mathrm{max}} \approx 5\mathrm{s}$, as commonly specified in grid codes \cite{NationalGrid2021}, the integral can be empirically approximated as 4. Finally, by \cref{eq:avg_4}, we can obtain the nadir bound in \cref{eq:nadir_spec}
\begin{align}
      \scriptstyle\sup_{t > 0} |\Delta f^{\tikz[baseline=0] \draw (0.05,0.05) -- (0.15,0.05) -- (0.15,0.15) -- (0.25,0.15);}_\mathrm{avg}(t)| \lesssim 2.5 \Delta p_\mathrm{d}^{\tikz[baseline=0] \draw (0.05,0.05) -- (0.15,0.05) -- (0.15,0.15) -- (0.25,0.15);}||D_\mathrm{avg}(\mathrm{j}\omega)||_\infty\leq \Delta f_\mathrm{max}.
\end{align}

\subsubsection{Bounded steady-state deviation} To show that the steady-state bound in \cref{eq:ss_spec} is satisfied, we first note that since $D_\mathrm{avg}(s)$ is strictly proper and stable by \cref{eq:avg_1}, $D_\mathrm{avg}(0)$ is finite. We then apply the final value theorem, which, together with \cref{eq:avg_5}, yields
\begin{align}
\begin{split}
   \scriptstyle|\underset{t\rightarrow \infty}{\text{lim}}\Delta f^{\tikz[baseline=0] \draw (0.05,0.05) -- (0.15,0.05) -- (0.15,0.15) -- (0.25,0.15);}_\mathrm{avg}(t)| & \scriptstyle=|\underset{s\rightarrow 0}{\text{lim}}s D_\mathrm{avg}(s) \frac{1}{s} \Delta p_\mathrm{d}^{{\tikz[baseline=0] \draw (0.05,0.05) -- (0.15,0.05) -- (0.15,0.15) -- (0.25,0.15);}}|\\
   & \scriptstyle=|D_\mathrm{avg}(0)\Delta p_\mathrm{d}^{{\tikz[baseline=0] \draw (0.05,0.05) -- (0.15,0.05) -- (0.15,0.15) -- (0.25,0.15);}}|\leq\Delta f_\mathrm{ss,\,max},
   \end{split}
\end{align}
which corresponds to the steady-state bound in \cref{eq:ss_spec}.

\subsubsection{Bounded RoCoF} To show that the RoCoF bound in \cref{eq:rocof_spec} is satisfied, we first realize that given \cref{eq:avg_2}, the maximum derivative $\sup_{t > 0} |\Delta \dot{f}^{\tikz[baseline=0] \draw (0.05,0.05) -- (0.15,0.05) -- (0.15,0.15) -- (0.25,0.15);}_\mathrm{avg}(t)|$ will happen in the limit $t\rightarrow0^+$: More specifically, the derivative of the step response in \cref{eq:avg_f_response} is equivalent to the impulse response of $D_\mathrm{avg}(s)$, i.e., 
\begin{align*}
     \scriptstyle\Delta \dot{f}^{\tikz[baseline=0] \draw (0.05,0.05) -- (0.15,0.05) -- (0.15,0.15) -- (0.25,0.15);}_\mathrm{avg}(t)=\mathcal{L}^{-1}\{sD_\mathrm{avg}(s)\tfrac{1}{s}\Delta p_\mathrm{d}^{{\tikz[baseline=0] \draw (0.05,0.05) -- (0.15,0.05) -- (0.15,0.15) -- (0.25,0.15);}} \} = \mathcal{L}^{-1}\{D_\mathrm{avg}(s)\Delta p_\mathrm{d}^{{\tikz[baseline=0] \draw (0.05,0.05) -- (0.15,0.05) -- (0.15,0.15) -- (0.25,0.15);}}\}.
\end{align*}
Next, as derived in \cite{papoulis1962The,franchek1998direct}, the impulse response of a stable and causal transfer function (cf. \cref{eq:avg_1}) can be uniquely expressed in terms of $\text{Re}[D_\mathrm{avg}(\mathrm{j}\omega)]$:
\begin{align}\label{eq:int_qv_rocof}
     \scriptstyle\Delta \dot{f}^{\tikz[baseline=0] \draw (0.05,0.05) -- (0.15,0.05) -- (0.15,0.15) -- (0.25,0.15);}_\mathrm{avg}(t)= \frac{2}{\pi}\Delta p_\mathrm{d}^{{\tikz[baseline=0] \draw (0.05,0.05) -- (0.15,0.05) -- (0.15,0.15) -- (0.25,0.15);}}\hspace{-1mm}\int_0^\infty \text{Re}[D_\mathrm{avg}(\mathrm{j}\omega)] \cos({\omega t})d\omega,\quad t > 0.
\end{align}
Given \cref{eq:avg_2}, it can be seen from \cref{eq:int_qv_rocof} how the maximum derivative happens in the limit $t\rightarrow0^+$, i.e., $\sup_{t > 0} |\Delta \dot{f}^{\tikz[baseline=0] \draw (0.05,0.05) -- (0.15,0.05) -- (0.15,0.15) -- (0.25,0.15);}_\mathrm{avg}(t)| = |\underset{t\rightarrow 0^+}{\text{lim}}\,\Delta \dot{f}^{\tikz[baseline=0] \draw (0.05,0.05) -- (0.15,0.05) -- (0.15,0.15) -- (0.25,0.15);}_\mathrm{avg}(t)|$. Since $D_\mathrm{avg}(s)$ is strictly proper and stable by \cref{eq:avg_1}, $sD_\mathrm{avg}(s)$ is finite at $s=\mathrm{j}\infty$. We can therefore apply the initial value theorem and use \cref{eq:avg_6} as
\begin{align}
\begin{split}
 \scriptstyle\sup_{t > 0} |\Delta \dot{f}^{\tikz[baseline=0] \draw (0.05,0.05) -- (0.15,0.05) -- (0.15,0.15) -- (0.25,0.15);}_\mathrm{avg}(t)| &  \scriptstyle=|\underset{t\rightarrow 0^+}{\text{lim}}\,\Delta \dot{f}^{\tikz[baseline=0] \draw (0.05,0.05) -- (0.15,0.05) -- (0.15,0.15) -- (0.25,0.15);}_\mathrm{avg}(t)|\\
& \scriptstyle= |\underset{s\rightarrow \infty}{\text{lim}}\,sD_\mathrm{avg}(s)\Delta p_\mathrm{d}^{{\tikz[baseline=0] \draw (0.05,0.05) -- (0.15,0.05) -- (0.15,0.15) -- (0.25,0.15);}}|\\
& \scriptstyle=|\underset{\omega\rightarrow \infty}{\text{lim}}\,\mathrm{j}\omega D_\mathrm{avg}(\mathrm{j}\omega)|\leq \Delta \dot{f}_\mathrm{max},
\end{split}
\end{align}
which corresponds to the RoCoF bound in \cref{eq:rocof_spec}.

\subsubsection{Sufficient oscillation damping} To ensure sufficient oscillation damping as required in \cref{eq:osci_spec}, we recall that, by \cref{eq:avg_1,eq:avg_7}, $D_\mathrm{avg}(s)$ is output feedback passive with index $\eta:=n\eta_\mathrm{f}>0$, i.e., $\text{Re}(D_\mathrm{avg}(\mathrm{j}\omega)^{-1}) \geq \eta$. Notice that stability of $D_\mathrm{avg}(s)/(1-\eta D_\mathrm{avg}(s))$ is guaranteed by stability of $D_\mathrm{avg}(s)$, following the same arguments as in the proof in \cref{appendix:stability_qv}. Hence, there exists a nonnegative storage function $V(x)$ such that the dissipation inequality \cite{bao2007process}
\begin{align}
     \scriptstyle\dot V \le {\Delta p_\mathrm{d}^{\sum}(t)}^\top \Delta f_\mathrm{avg}(t) - \eta|\Delta f_\mathrm{avg}(t)|^2
\end{align}
holds. Namely, for the zero-input response $\Delta p_\mathrm{d}^{\sum}\equiv 0$, one has
\begin{align}
    \scriptstyle \dot V \le -\eta|\Delta f_\mathrm{avg}(t)|^2\quad\Longrightarrow
\quad\scriptstyle \int_0^\infty |\Delta f_\mathrm{avg}(t)|^2\,dt \le \frac{V(0)}{\eta} < \infty.
\end{align}
Hence, the output belongs to $L_2[0,\infty)$ and converges to zero as $t \to \infty$. This shows that output-feedback passivity with index $\eta>0$ induces a strictly dissipative output channel, thereby guaranteeing finite average output energy. Moreover, when combined with the $\mathcal{H}_\infty$ bound in \cref{eq:avg_4}, which constrains the worst-case energy amplification, we conclude that all observable modes are sufficiently damped, as characterized in \cref{eq:osci_spec}. However, the passivity index $\eta$ alone does not permit inference of an explicit exponential decay rate.

\section{IEEE Nine-Bus System and Device Parameters}\label{appendix:ninebus_parameters}
To validate the NGGC framework under realistic conditions, we consider the IEEE nine-bus system with three grid-forming VSCs (cf. \cref{fig:9bus}) and perform detailed EMT simulations. We adopt the network parameters from \cite{anderson}, and provide the device parameters in \cref{tab:9bus_parameters} below.
\begin{table}[h!]
    \centering
    \begin{tabular}{c||c|c}
    \toprule
         Parameter&Symbol & Value  \\ \hline
 {$\hspace{-3mm}$Base power, voltage, frequency$\hspace{-3mm}$}&{$S_\mathrm{b}$, $V_\mathrm{b}$, $f_\mathrm{b}$}&{$\hspace{-2mm}$100 MVA, 230 kV, 50 Hz$\hspace{-2mm}$}\\ 
         {Bus base voltages}&{$\hspace{-3mm}$$V_{\mathrm{b},1}$, $V_{\mathrm{b},2}$, $V_{\mathrm{b},3}$}$\hspace{-3mm}$&{16.5, 18, 13.8 kV}\\ 
         {Constant impedance loads}&{$p_{\mathrm{l},i}$, $q_{\mathrm{l},i}$}&{1, 0.275 pu}\\\hline
{GFM base power}&{$S_1$, $S_2$, $S_3$}&{100, 100, 100 MVA}\\
         {GFM active power setpoints}$\hspace{-2mm}$&$\hspace{-1mm}${$p_\mathrm{0,1}$, $p_\mathrm{0,2}$, $p_\mathrm{0,3}$}&{1.13, 0.94, 0.92 pu}$\hspace{-2mm}$\\
         {GFM reactive power setpoints}$\hspace{-2mm}$&$\hspace{-1mm}${$q_\mathrm{0,1}$, $q_\mathrm{0,2}$, $q_\mathrm{0,3}$}&{$\hspace{-2mm}$ 0.01, -0.02, 0.01 pu$\hspace{-1mm}$}\\
         GFM steady-state voltages &$\hspace{-2mm}$$|v|_{0,1}$,$|v|_{0,2}$,$|v|_{0,3}$$\hspace{-2mm}$& 1, 1, 1 pu\\
{GFM $RLC$ filter components}&{$r_{\mathrm{f},i}$, $l_{\mathrm{f},i}$, $c_{\mathrm{f},i}$}&{0.01, 0.1, 0.1 pu}\\ \bottomrule
    \end{tabular}
            \vspace{-1mm}
    \caption{Nine-bus system parameters (in local per unit system).}
    \label{tab:9bus_parameters}
\end{table}
\end{document}